\documentclass[useAMS,usenatbib]{mn2e}
\usepackage{graphicx,epsfig,longtable}
\usepackage{epstopdf}
\usepackage{color}
\usepackage{amsmath}
\usepackage{tabularx} 
\usepackage{hyperref}

\newcommand{\Msun}{$M_{\sun}$}
\newcommand{\aap}{A\&A}
\newcommand{\apjl}{ApJL}
\newcommand{\apjs}{ApJS}
\newcommand{\apj}{ApJ}
\newcommand{\aanda}{A\&A}
\newcommand{\prd}{Phys. Rev. D}
\newcommand{\prc}{Phys. Rev. C}
\newcommand{\prl}{Phys. Rev. Lett.}
\newcommand{\mnras}{MNRAS}
\newcommand{\nat}{Nature}
\newcommand{\nucpha}{Nucl. Phys. A}

\newcommand{\physrep}{Phys. Rep.}

\newcommand{\araa}{ARA\& A}

\newcommand{\pasa}{PASA}
\newcommand{\jqsrt}{J.~Quant.~Spec.~Radiat.~Transf.}

\def\ga{\,\,\raise0.14em\hbox{$>$}\kern-0.76em\lower0.28em\hbox
{$\sim$}\,\,}
\def\la{\,\,\raise0.14em\hbox{$<$}\kern-0.76em\lower0.28em\hbox
{$\sim$}\,\,}
\def\Msun{$M_{\odot}$}

\title[Nucleosynthesis from compact binary mergers]
  {Comprehensive nucleosynthesis analysis for ejecta of compact binary mergers }
\author[O.~Just et al.]
  {O.~Just,$^{1,2}$ A.~Bauswein,$^3$ 
    R.~Ardevol Pulpillo,$^{1,4}$ S.~Goriely$^5$ and H.-T.~Janka$^1$ \\
    $^1$Max-Planck-Institut f\"ur Astrophysik, Postfach 1317, 85741 Garching, Germany \\
    $^2$Max-Planck/Princeton Center for Plasma Physics (MPPC) \\
    $^3$Department of Physics, Aristotle University of Thessaloniki, 54124 Thessaloniki, Greece \\
    $^4$Physik Department, Technische Universit\"at M\"unchen, James-Franck-Stra\ss e 1, 85748 Garching, Germany \\
    $^5$Institut d'Astronomie et d'Astrophysique, CP-226, Universit\'e Libre de Bruxelles, 
    1050 Brussels, Belgium
}
\date{Released 2014 Xxxxx XX}

\pagerange{\pageref{firstpage}--\pageref{lastpage}} \pubyear{2014}

\begin{document}

\label{firstpage}

\maketitle

\begin{abstract}
  We present the first comprehensive study of r-process element nucleosynthesis in the ejecta of
  compact binary mergers (CBMs) and their relic black-hole (BH)-torus systems.  The evolution of the
  BH-accretion tori is simulated for seconds with a Newtonian hydrodynamics code including viscosity
  effects, pseudo-Newtonian gravity for rotating BHs, and an energy-dependent two-moment closure
  scheme for the transport of electron neutrinos and antineutrinos. The investigated cases are
  guided by relativistic double neutron star (NS-NS) and NS-BH merger models, producing
  $\sim$3--6\,$M_\odot$ BHs with rotation parameters of $A_\mathrm{BH}\sim 0.8$ and tori of
  0.03--0.3\,$M_\odot$.  Our nucleosynthesis analysis includes the dynamical (prompt) ejecta
  expelled during the CBM phase and the neutrino and viscously driven outflows of the relic BH-torus
  systems.  While typically $\sim$20--25\% of the initial accretion-torus mass are lost by viscously
  driven outflows, neutrino-powered winds contribute at most another $\sim$1\%, but neutrino heating
  enhances the viscous ejecta significantly. Since BH-torus ejecta possess a wide distribution of
  electron fractions (0.1--0.6) and entropies, they produce heavy elements from $A\sim 80$ up to the
  actinides, with relative contributions of $A \ga 130$ nuclei being subdominant and sensitively
  dependent on BH and torus masses and the exact treatment of shear viscosity. The combined ejecta
  of CBM and BH-torus phases can reproduce the solar abundances amazingly well for $A \ga
  90$. Varying contributions of the torus ejecta might account for observed variations of lighter
  elements with $40\le Z\le 56$ relative to heavier ones, and a considerable reduction of the prompt
  ejecta compared to the torus ejecta, e.g.\ in highly asymmetric NS-BH mergers, might explain the
  composition of heavy-element deficient stars.
\end{abstract}


\section{Introduction}

Double neutron star (NS-NS) and neutron star-black hole (NS-BH) binaries radiate gravitational waves
and their orbits shrink due to the associated angular momentum and energy loss until, after millions
to hundreds of millions of years, a catastrophic merger event terminates the evolution of these
binary systems.  The frequency of such events can be estimated on grounds of the known double NS
systems in the solar neighborhood \citep[e.g.,][]{Kalogera2004} and theoretical population synthesis
studies \citep[e.g.,][]{Belczynski2008} to be of the order of one NS-NS merger in some $10^5$ years
for Milky Way-like galaxies and possibly up to several times this rate for NS-BH mergers, but these
numbers contain considerable uncertainties \citep{Postnov2014}.

Compact binary mergers are among the most promising extragalactic sources of gravitational waves
(GWs) in the $\la$100 to $\ga$1000\,Hz range to be measured by the upcoming advanced interferometer
antennas (advLIGO, advVIRGO, KAGRA). Strong arguments suggest that their remnants are good
candidates for the still enigmatic central engines of short gamma-ray bursts (GRBs; see, e.g.,
\citealp{Berger2014, Nakar2007} for reviews). The merger rate, reduced by the beaming factor
measuring the probability that the Earth is hit by the collimated, ultrarelativistic GRB beam, can
well account for the number of detected short GRBs.

\begin{figure}
  \includegraphics[width=84mm]{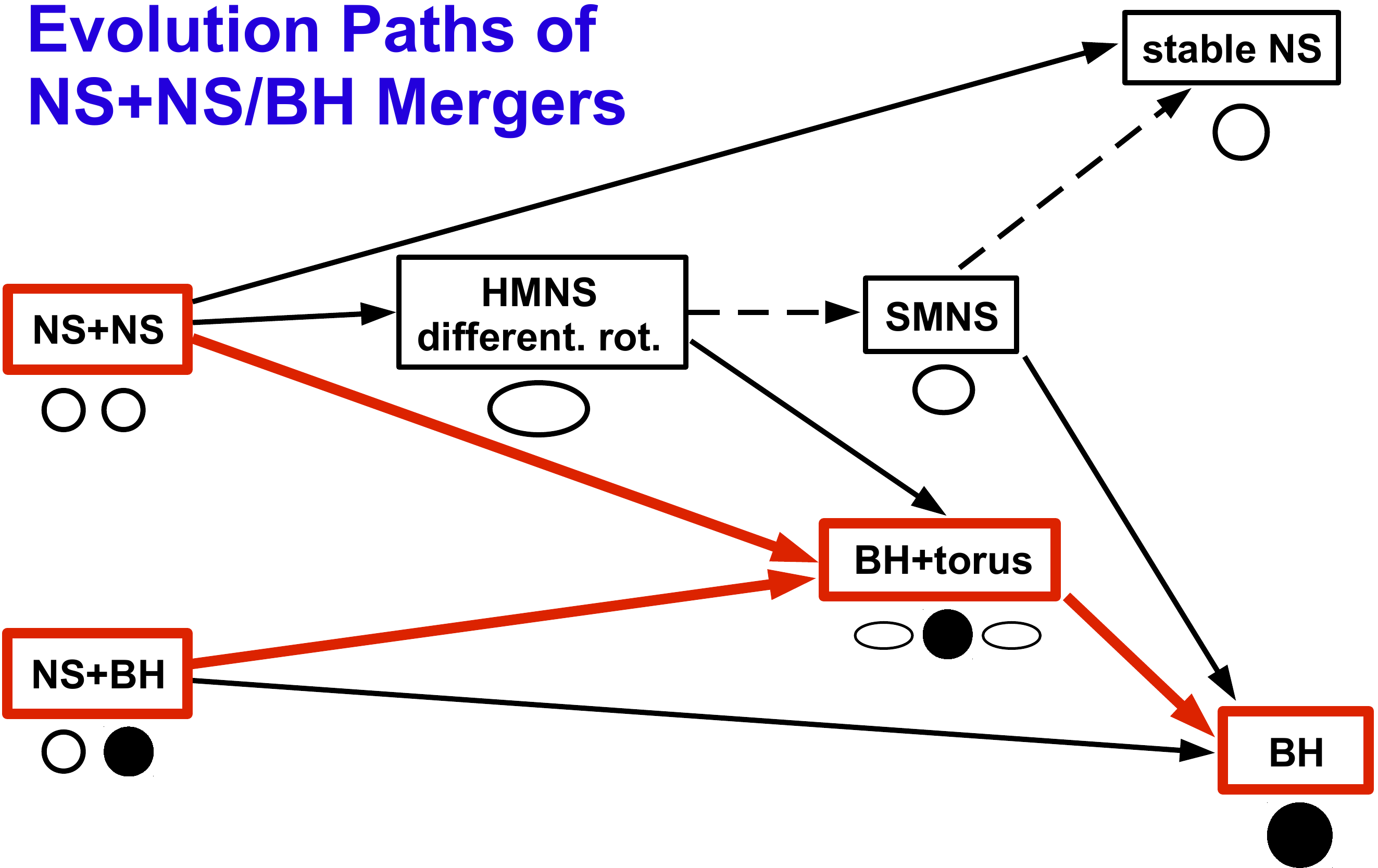}
  \vspace{5pt}
  \\
  \includegraphics[width=84mm]{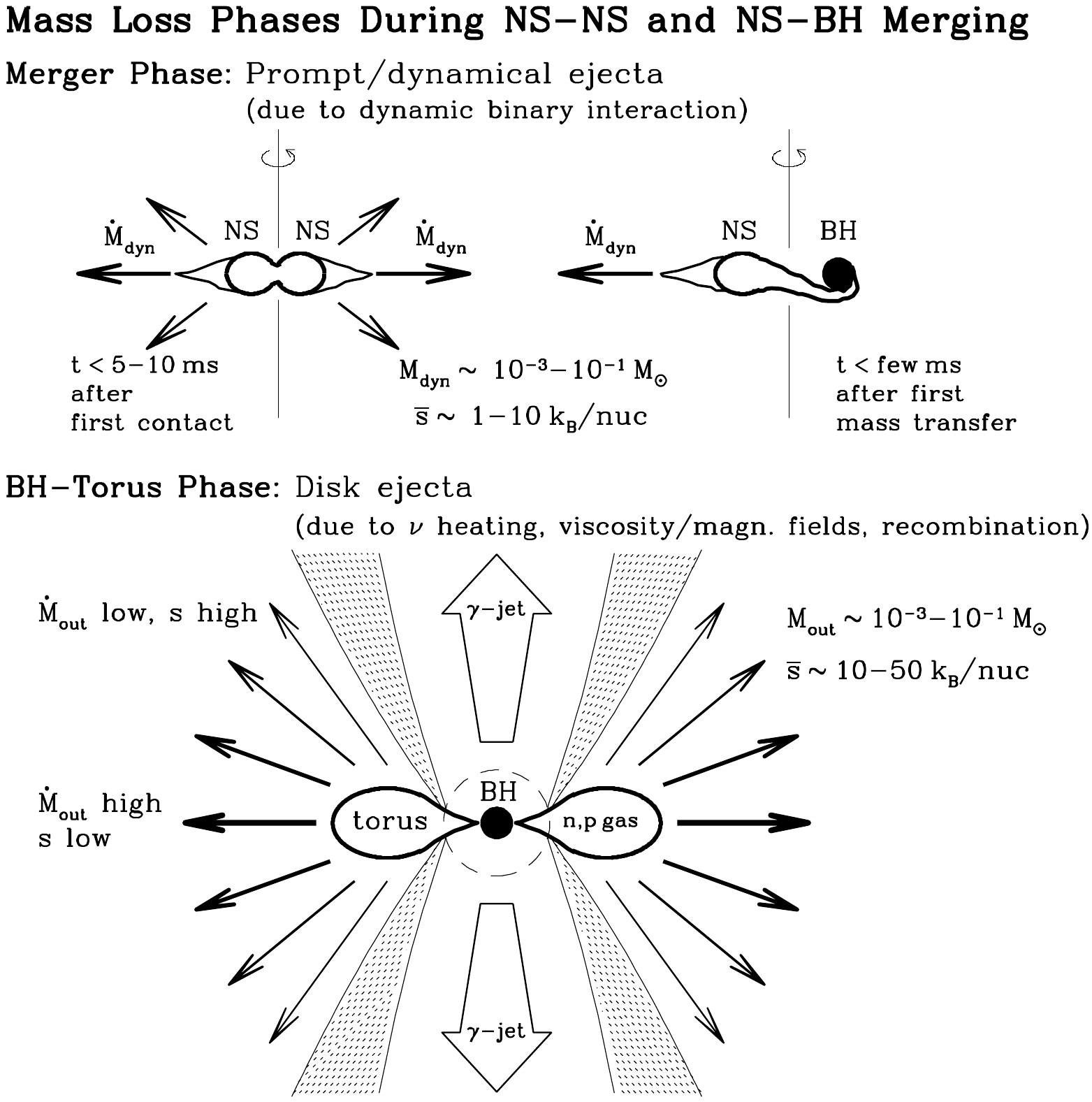}
  \caption{{\em Top panel:} Evolution paths of NS-NS and NS-BH mergers. Depending on
    the binary parameters and the properties of the nuclear equation of state, binary
    NS mergers can lead to the formation of a stable NS, a transient hypermassive NS
    (HMNS; stabilized by differential rotation) and supermassive NS (SMNS, stabilized
    by rigid rotation), or a BH plus accretion-torus system. The last scenario is also
    the outcome of a NS-BH merger if the BH/NS mass ratio is not too large. Transitions
    between the different evolution stages can be accompanied by mass loss. 
    In the present work we focus exclusively on the evolution tracks and stages
    highlighted by thick, solid red lines.
    {\em Bottom panel:} Mass-loss phases during the dynamical interaction of NS-NS
    and NS-BH binaries and the subsequent secular evolution of a relic BH-torus system
    (corresponding to the evolution paths indicated by red lines in the upper panel).
    The dynamical mass ejection takes place within a few milliseconds when the two
    binary components merge with each other. Typical ejecta masses are around
    0.01\,$M_\odot$ and the average entopies of the ejecta are low. BH-torus systems
    eject matter mainly in viscously driven outflows and to a smaller extent also in
    neutrino-driven winds. Baryon-poor polar funnels may provide suitable conditions
    for neutrino or magnetohydrodynamically powered, ultrarelativistic, collimated
    outflows, which are likely to produce short gamma-ray bursts. The image is adapted
    from \citet{Ruffert1999a, Janka2002}.}
  \label{fig:merger_channels}
\end{figure}

Moreover, a small fraction of the NS matter can be expelled during and after the coalescence of the
binary components. Because of its extremely neutron-rich initial state, this matter has long been
speculated to be a possible site for the formation of r-process nuclei \citep{Lattimer1974,
  Lattimer1976, Eichler1989}. Indeed, Newtonian \citep[e.g.,][]{Ruffert1997, Ruffert1999a,
  Ruffert2001, Janka1999, Rosswog1999, Freiburghaus1999, Rosswog2005b, Korobkin2012} and conformally
flat general relativistic \citep[e.g.,][]{Oechslin2007, Goriely2011,Bauswein2013} as well as fully
relativistic \citep[e.g.,][]{Kyutoku2011, Kyutoku2013, Hotokezaka2013, Hotokezaka2013a, Deaton2013,
  Wanajo2014a, Foucart2014} hydrodynamical simulations of NS-NS and NS-BH mergers with microphysical
equations of state (EOSs) have demonstrated that typically some $10^{-3}\,M_\odot$ up to more than
0.1\,$M_\odot$ can become gravitationally unbound on roughly dynamical timescales due to shock
acceleration and tidal stripping. Also the relic object, either a hot, transiently stable
hypermassive NS (HMNS; stabilized by differential rotation and thermal pressure;
\citealp{Baumgarte2000}) and later supermassive NS (SMNS; stabilized by rigid rotation and thermal
pressure) or a BH-torus system (Fig.~\ref{fig:merger_channels}, upper panel), can lose mass along
with its secular evolution in outflows that are driven by viscous energy dissipation and turbulent
angular momentum transport, magnetic pressure, nucleon-recombination heating and neutrino-energy
deposition (Fig.~\ref{fig:merger_channels}, lower panel; e.g., \citealp{Popham1999, Ruffert1999a,
  Dessart2009, Fernandez2013, Siegel2014, Metzger2014, Perego2014a}. Such ejecta could be
interesting environments for r-process, p-process, and nickel nucleosynthesis \citep[e.g.,][and
references therein]{Surman2008, Caballero2012, Wanajo2012, Malkus2014, Surman2014}.

The presence of significant mass fractions of radioactive material (nickel, r-nuclei,....) was
pointed out to lead to long-term decay heating of the material ejected by compact binary mergers and
thus to thermal radiation that can potentially be observed as electromagnetic transient
\citep{Li1998} termed ``macro-nova'' \citep{Kulkarni2005} or ``kilonova'' \citep{Metzger2010c,
  Metzger2012}. Indeed, a near-infrared data point at the position of the short-hard GRB\,130603B
about 7 days (rest-frame time) after the burst was interpreted as possible first detection of such
an r-process powered transient \citep{Berger2013, Tanvir2013}, because its temporal delay, color,
and brightness are compatible with theoretical light curve calculations for an assumed ejecta mass
of some $10^{-2}\,M_\odot$, taking into account the fact that r-process nuclei, in particular the
abundant lanthanides, increase the opacity by roughly a factor of 100 compared to iron
\citep{Barnes2013, Kasen2013, Hotokezaka2013a, Tanaka2013, Tanaka2014, Grossman2014}.

Nucleosynthesis calculations agree \citep[e.g.,][]{Freiburghaus1999, Goriely2005, Roberts2011,
  Goriely2011, Korobkin2012, Bauswein2013, Rosswog2014, Wanajo2014a} that the dynamical ejecta of
the merging phase provide sufficiently neutron rich conditions for a robust r-processing up to the
third abundance peak and the actinides, although the theoretical predictions show considerable
differences in important details, depending on the exact conditions in the ejecta (electron
fraction, $Y_e$, expansion velocity, and specific entropy) and the employed nuclear physics, in
particular the theoretical assumptions about the fission-fragment distribution
\citep[cf.][]{Goriely2013}. Relativistic NS-NS mergers \citep[e.g.,][]{Bauswein2013}, for example,
tend to produce faster ejecta than Newtonian simulations \citep[e.g.,][]{Roberts2011, Korobkin2012,
  Rosswog2014, Grossman2014}, and ---compared to models that disregard neutrino effects
\citep[e.g.,][]{Goriely2011, Bauswein2013}---, the $Y_e$ in the ejecta is slightly higher when
neutrino emission is included \citep[e.g.,][]{Korobkin2012} and could even be significantly higher
when neutrino emission and absorption are both taken into account \citep{Wanajo2014a}. Slower mass
ejection and a wider spread of the neutron excess enable the creation of r-nuclei with mass numbers
above $A\sim$80--110, whereas fast, very neutron-rich ejecta lead only to species with
$A\ga$130--140. The exact structure of the second and third peaks as well as the quality of
reproducing the solar r-element distribution in the rare-earth region depends sensitively on the
nuclear reaction rates and the fission fragment yields, if fission recycling plays an important
role. In any case, however, a robust and strong production of lanthanides and third-peak material
can be expected.

The hydrodynamics and nucleosynthetic output of merger remnants are less well studied than those of
the dynamical merging phase or immediate post-merging phase, mainly because simulations of the
evolution of relic massive NSs or BH-torus systems need to be carried out over much longer time
intervals. Magnetic fields, turbulence, and viscous effects play important roles and add
complications, and a reasonably good treatment of neutrino production, absorption, and transport is
indispensable to obtain reliable predictions of the nucleosynthetic conditions in outflows. First
interesting modeling attempts of rapidly rotating post-merger HMNSs or SMNSs including a
  neutrino treatment in two- and three dimensions (2D, 3D) exist \citep{Dessart2009, Metzger2014,
  Perego2014a}, but the enormous complexity of the problem still enforces considerable
simplifications, among them short evolution phases, purely Newtonian gravity, the disregard of
magnetic-field effects, simple neutrino treatments, or a parametric, not fully self-consistent
description of the neutron-star core as a neutrino emitting boundary irradiating a surrounding disk,
or a combination of several of these assumptions.

Hydrodynamical exploration of NS-BH mergers and the early development (over periods of several ten
milliseconds) of their relic BH-torus systems has only started to be done in full 3D general
relativity with a neutrino-leakage treatment \citep{Deaton2013, Foucart2014}, but magnetic and
viscous effects, which determine the mass-loss of the accretion tori on secular timescales, are
still not included in such models. \citet{Fernandez2013} have recently carried out the first (2D)
hydrodynamical simulations of post-merger BH-tori, which were able to follow the long-time evolution
over the whole period of intense neutrino emission and a large fraction of the mass-loss history
(for many seconds) including an approximative description for neutrino emission and
self-irradiation, viscosity terms, and a pseudo-Newtonian treatment of gravity for (nonrotating) BHs
following \citet{Paczy'nsky1980}. This spearheading work went clearly beyond simple parametric or
semi-analytic modeling \citep[e.g.,][]{Surman2008, Metzger2008c, Metzger2009b,
  Wanajo2012} and earlier \hbox{(magneto-)}hydrodynamical simulations in 2D \citep{Lee2005,
  Shibata2007, Shibata2012, Janiuk2013} and 3D
\citep{Ruffert1999a, Setiawan2004, Setiawan2006}. In nucleosynthesis and light-curve calculations of
the expanding ejecta of compact binary mergers, however, the contribution from the merger-remnant
phase has so far been included only highly schematically, either by applying simple multi-parameter
prescriptions for neutrino-driven wind conditions \citep{Rosswog2014, Grossman2014} or by making ad
hoc assumptions about the composition of the disk-wind ejecta \citep{Barnes2013}.

In this work we perform the first comprehensive study of the nucleosynthetic output associated with
the ejecta from the merging phase of NS-NS and NS-BH binaries {\em and} with the neutrino and
viscously driven outflows of the subsequent long-time evolution of relic BH-torus systems. The
compact binary mergers are simulated with a relativistic smooth-particle-hydrodynamics code
\citep{Oechslin2007, Bauswein2010e} in a conformally flat, dynamical spacetime including
  temperature-dependent, microphysical EOSs, while the BH-torus modeling is conducted with a
Eulerian finite-volume Godunov-type scheme, supplemented by a shear-viscosity treatment with a
Shakura-Sunyaev $\alpha$-prescription for the dynamic viscosity \citep{Shakura1973}. For the
long-time evolution we employ a pseudo-Newtonian approximation of the gravity potential for the
rotating relic BHs \citep{Artemova1996}. In the time-dependent BH-torus modeling we apply, for the
first time, detailed energy-dependent and velocity-dependent 2D neutrino transport based on a new
two-moment closure scheme (Just, Obergaulinger \& Janka, in preparation), which allows us to
determine the neutrino-driven wind and the neutron-to-proton ratio in the disk outflows with higher
accuracy than in previous simulations. All simulations include microphysical treatments of the gas
equation of state.

Our nucleosynthesis calculations are carried out in a post-processing step of the ejecta produced by
the hydrodynamical models, using a full r-process network including all relevant nuclear reactions
\citep{Goriely2008,Goriely2010,Xu2013}. For the combined analysis we pick cases from larger sets of
NS-NS merger models, which lead to prompt or slightly delayed BH formation, and NS-BH merger models
on the one side and BH-torus models on the other side such that the macroscopic system parameters
(BH and torus masses and BH spins of the merger remnants) match roughly on both sides. This also
yields good consistency of the mean values of the total specific energies of the torus gas in both
modeling approaches.

We find that the torus outflows complement the dynamical merger ejecta by contributing the
lower-mass r-process nuclei ($A \la 140$) that are massively underproduced by the strong r-process
taking place in the very neutron-rich material expelled during the binary coalescence. The combined
abundance distribution can reproduce the solar pattern amazingly closely, but in contrast to the
robustness of the high-mass number component ($A \ga 140$) the low-mass number component must be
expected to exhibit considerable variability, depending on the variations of the intrinsic outflow
properties. On the other hand, direction-dependent differences result from the combination of highly
asymmetric dynamical NS-BH merger ejecta with much more isotropically distributed torus
outflows. Cases with suppressed dynamical ejecta component resemble abundance patterns observed in
heavy r-element deficient metal-poor stars.

Our paper is structured as follows. In Sect.~\ref{sec:numerical-methods} we briefly summarize the
basic methodical aspects of our hydrodynamical modeling of the different considered scenarios and
evolution stages and of our nucleosynthesis analysis. In Sect.~\ref{sec:results} we describe the
conceptual aspects of the coherent modeling approach and present our results for the hydrodynamical
simulations of binary mergers and relic BH-torus systems and for the corresponding r-process
nucleosynthesis. Section~\ref{sec:summ-disc-concl} concludes with a summary and discussion of our
results. In Appendix A we present a comparison test of our neutrino-transport treament and briefly
discuss the accuracy of the latter.

\section{Numerical methods}\label{sec:numerical-methods}

\subsection{Compact binary mergers}\label{sec_numerics_merger}

The merging phase of our models is simulated with a smooth-particle-hydrodynamics (SPH) code with an
approximate treatment of relativistic gravity. The code computes the hydrodynamical properties of
fluid elements comoving with the matter flow
(see~\citealp{Oechslin2002,Oechslin2007,Bauswein2010,Bauswein2010e} for details). The Lagrangian
nature of this scheme is well suited to follow the unbound material during the coalescence. The
relativistic Lagrangian formulation of hydrodynamics evolves so-called ``conserved variables'',
which yields ordinary differential equations for the conserved rest-mass density $\rho^*$, the
conserved specific momenta $\tilde{u}_i$ and the conserved energy density
$\tau$~\citep{Siegler2000}. The quantities depend on the rest-mass density $\rho$, the coordinate
velocity $v_i$, the specific internal energy density $\epsilon$, the pressure $P$, and the metric
functions encoding the gravitational field.

The Einstein equations are solved by imposing a conformal flatness condition (CFC) for the spatial
metric
$\mathrm{d}s^2=-\alpha^2\mathrm{d}t^2+\psi^4\delta_{ij}(\mathrm{d}x^i+\beta^i\mathrm{d}t)(\mathrm{d}x^j+\beta^j\mathrm{d}t)$,
which results in nonlinear elliptical equations for the lapse function $\alpha$, the shift vector
$\beta_i$ and the conformal factor $\psi$~\citep{Isenberg1980,Wilson1996}. In the case of NS-NS
mergers this CFC approximation has been shown to yield quantitatively accurate results for the
oscillation modes of NS-merger remnants and for the properties of ejecta when compared to
calculations solving the full Einstein
equations~\citep{Bauswein2012,Bauswein2013,Hotokezaka2013,Takami2014}.

For the simulation of NS-BH mergers we use the extended CFC
formulation~\citep{Bonazzola2004,Cordero-Carrion2009} to avoid divergence problems in the source
terms in the proximity of the BH. In this formulation a BH can be treated by means of the puncture
approach~\citep{Brandt1997}, which factors out divergent terms in the metric function and which
calculates the deviations of the spacetime from a static BH solution, the latter of which is
analytically known. In a similar manner the BH momentum and spin are taken into account by adding
analytically known solutions to the extrinsic curvature~\citep{Bowen1979,Bowen1980,Bauswein2010e},
which appears in the source terms of the elliptical equations. This procedure is fully equivalent to
the construction of BH-BH or NS-BH binary initial data necessary for fully relativistic calculations
(see e.g.~\citealp{Kyutoku2009,Bauswein2010e,Baumgarte2010}, where the corresponding formulae can be
found; see also~\citealp{Faber2006} for a similar approach within CFC). In our implementation,
however, the method for computing initial data is employed in every time
step~\citep{Bauswein2010e}. In fully relativistic formulations of the Einstein evolution equations
the BH position is tracked passively by $\mathrm{d}x^i_{\mathrm{BH}} /
\mathrm{d}t=-\beta^i(x^i_{\mathrm{BH}})$ with $x^i_{\mathrm{BH}}$ being the center of the BH
(position of the puncture). For our calculations we promote this equation to an active equation of
motion for the BH to be solved in parallel with the hydrodynamical equations. For details of the
implementation, see~\citet{Bauswein2010}. More information will be provided in a forthcoming paper.

The BH treatment is validated by computing orbits of NS-BH binaries and comparing the orbital
angular velocity with stationary calculations of quasi-equilibrium initial data~\citep{Etienne2009},
which can be reproduced within a few per cent~\citep{Bauswein2010e}. More importantly for the
purposes of this paper, we also compare the resulting tori of NS-BH mergers with those computed by
grid-based fully relativistic simulations~\citep{Kyutoku2011,Kyutoku2013}. Torus masses agree with
an accuracy of about 20 per cent for binaries with initially non-rotating as well as rotating
BHs. Also the amount of dynamically ejected matter is consistent within a factor of two. We confirm
by convergence tests that the torus and ejecta masses are insensitive to the chosen grid setup for
the metric solver and the SPH particle resolution within a range of 20 per cent and 30 per cent,
respectively.

For NS-BH as well as NS-NS mergers the hydrodynamical equations are supplemented by an additional
equation $\mathrm{d}Y_e / \mathrm{d}t = 0$, which describes the advection of the initial electron
fraction $Y_e$. The electron fraction, $Y_e$, rest-mass density, $\rho$, and specific internal
energy density, $\epsilon$, are employed to call the equation of state (EOS) $P(\rho,\epsilon,Y_e)$,
which is required to close the system of hydrodynamical equations. Throughout the paper we use fully
temperature-dependent, microphysical high-density equations of state (see
Sect.~\ref{sec:choice-models-coher}), which are indispensable for an accurate description of the
merger dynamics and a reliable determination of torus and ejecta properties~\citep{Bauswein2010,
  Bauswein2013}. Moreover, the temperature and composition of the ejected material are needed as
input to the nucleosynthesis calculations. For NS-NS and NS-BH mergers the NSs are assumed to be
initially cold and in neutrinoless $\beta$-equilibrium, which sets the initial electron
fraction. The stars are placed on circular quasi-equilibrium orbits such that the binaries merge
after about two to three revolutions. For both types of systems we impose an initially non-rotating
velocity profile of the NSs, because NS spin periods are long compared to the orbital periods and
because tidally locked systems are not expected to form during the inspiral~\citep{Kochanek1992,
  Bildsten1992}. The initial rotation of the BH, $A_{\mathrm{BH},0}$, is a free parameter of our
setups (see Sect.~\ref{sec:choice-models-coher}). The NSs are modeled with about 150,000 SPH
particles per star.

\subsection{Merger remnants}\label{sec_numerics_remnant}
In this study we exclusively consider BH-torus systems as possible merger remnants, i.e. we do not
explore the case of hyper-/supermassive NSs resulting from binary NS mergers. Since during the
remnant evolution neutrino-transport effects in the accretion torus become non-negligible, we use a
different simulation code for the remnants than for the actual mergers (see
Sect.~\ref{sec_numerics_merger}), namely a recently developed finite-volume neutrino-hydrodynamics
code described in \cite{Obergaulinger2008} and Just et al. (in preparation).

The BH-torus simulations are performed in axisymmetry and employing Newtonian hydrodynamics. We
ignore the self-gravity of the torus, whose mass is much smaller than the BH mass, and for the
gravitational potential of the BH we use the pseudo-Newtonian Artemova-potential \citep[][as given
by their Eq.~13]{Artemova1996}. 

The Artemova--potential is an extension of the widely employed Paczy\'nsky-Wiita--potential
\citep{Paczy'nsky1980} that additionally takes into account essential effects of BH
rotation. Specifically, compared to the Kerr metric \citep{Bardeen1972} the potential by
\citet{Artemova1996} accurately reproduces the radius of the innermost stable circular orbit (ISCO)
as function of the BH mass, $M_{\mathrm{BH}}$, and BH spin parameter, $A_{\mathrm{BH}}$, and it
approximately (up to a few per cent) reproduces the gravitational plus kinetic specific binding
energy of a test particle orbiting the central BH at the ISCO. These features make the
Artemova--potential the preferred choice compared to the Paczy\'nsky-Wiita-potential since
post-merger BHs are in general fast rotators and therefore the reduced ISCO radius and enhanced
(absolute) binding energy can have a sizable impact on the disk evolution and outflow
generation. However, the gas kinematics are still treated nonrelativistically and improvements are
desirable \citep[see, e.g.,][]{Tejeda2013}. Therefore, although we consider the qualitative features
of BH rotation that are most relevant to our study to be captured by the Artemova--potential, the
quantitative accuracy of our treatment is of course limited and has to be compared against future
general relativistic studies.

The BH mass and spin are kept fix during the simulations. Given that the considered torus masses are
only a fraction $\xi \leq 0.1$ of the BH masses (see Sec.~\ref{sec:choice-models-coher}), we expect
no qualitative and only a small quantitative impact of this approximation on our main results. Note
that the maximum possible change $\Delta A_{\mathrm{BH}}$ of the BH spin parameter due to disk
accretion can be estimated by assuming that the entire torus is instantly (i.e. for a fixed BH spin)
accreted from the ISCO as
$\Delta A_{\mathrm{BH}} \la \xi [\tilde{l} - A_{\mathrm{BH}}(2+\xi)]/(1+\xi)^2 < \xi$, where
$\tilde{l}$ is the dimensionless Keplerian specific angular momentum at the ISCO (in units of
$G M_{\mathrm{BH}}/c$, where $G$ and $c$ are the gravitational constant and the speed of light,
respectively) and the last inequality holds for the case $A_{\mathrm{BH}}=0.8$ (with
$\tilde{l}\approx 2.4$) which is exclusively considered in this paper.

For the EOS we assume the fluid to be composed of an ideal Fermi gas of electrons and
positrons, of a thermal photon gas, and of a 4-species Boltzmann gas of neutrons, protons,
$\alpha$-particles and a representative, heavy nucleus in the form of $^{54}$Mn (whose particular
choice does not have any significant relevance for our simulations). The baryonic species are taken
to be in nuclear statistical equilibrium (NSE).

The neutrino transport is described by a truncated two-moment scheme, which solves the evolution
equations of the energy density and flux density of neutrinos augmented with a closure relation that
expresses the Eddington tensor as a function of the evolved quantities. For a detailed description
of the formalism and its implementation we refer to a dedicated paper (Just et al., in preparation)
and for other implementations of similar schemes the reader may consult, e.g., \cite{Audit2002,
  Shibata2011a, OConnor2013, Skinner2013, Scadowski2013}. A test problem in which the radiation
field of a representative torus configuration is compared against ray-tracing results, together with
a brief discussion on the accuracy of our transport treatment can be found in Appendix A. We employ
the closure prescription by \cite{Minerbo1978} for the calculations in this study. The transport
implementation basically includes all velocity-dependent terms up to $\mathcal{O}(v/c)$. However,
since the rotational velocities close to the BH as well as the polar velocities in the axial funnel
can reach values comparable to the speed of light but incompatible with our $\mathcal{O}(v/c)$
transport treatment, we limit velocities in the transport equations to $0.2\,c$. We also omit terms
depending on the azimuthal velocity component $v^\phi$, thus ignoring the azimuthal components of
the neutrino propagation and advection as well as $r\phi$ and $\theta\phi$ components of the
neutrino viscosity. This procedure may appear too radical to cope with high rotation velocities
occurring just in the close vicinity of the BH and would numerically not be necessary in the whole
torus, but it still seems reasonably compatible with our modeling constraint to axisymmetry. Because
the torus is axisymmetric the time evolution of the neutrino radiation quantities is mainly driven
by radial and lateral transport while the rotational aberration is a secondary effect, at least for
rotational velocities not too close to the speed of light.

In our energy-dependent transport scheme we evolve the radiation quantities in discrete energy
groups for each species, electron neutrinos and electron antineutrinos. We ignore the effects of
heavy-lepton neutrinos, which are of minor relevance during the torus evolution
\citep[e.g.][]{Janka1999,Ruffert1999a,Deaton2013,Fernandez2013, Foucart2014}. For the neutrino
interaction channels we take into account the $\beta$-processes with free nucleons and scattering of
neutrinos off free nucleons (as formulated in \citealp{Bruenn1985}), as well as annihilation of
neutrino-antineutrino pairs (based on \citealp{Dicus1972, Schinder1987, Cooperstein1987}). Since at
later stages of the torus evolution neutrino absorptions become negligible, we switch from the full
neutrino-transport scheme to a computationally cheaper treatment assuming neutrinos to be only
emitted but not reabsorbed or scattered by the medium. We typically switch to this simplification at
some time $t_{\mathrm{switch}}\ge t_{\eta}$, where $t_{\eta}$ is the time when the neutrino emission
efficiency $\eta_{\nu}$ (see Sect.~\ref{sect3_remnant} for the definition) drops below $0.005$. We
have validated by tests that $t_{\mathrm{switch}}$ is chosen sufficiently late to ensure that
effects due to non-vanishing optical-depths are negligible. Depending on the model, the values of
$t_{\mathrm{switch}}$ lie typically between $0.3\,$s and $1.5\,$s.

The initial disk models are constructed as rotational equilibrium configurations with constant
specific angular momentum \citep[see, e.g.,][where similar disk models have been used as initial
models]{Igumenshchev1996, Stone1999, Fernandez2013}. For given BH properties $M_{\mathrm{BH}}$ and
$A_{\mathrm{BH}}$ and given torus mass $M_{\mathrm{torus}}$, the fluid configuration is defined by
setting the initial electron fraction to $Y_{e,0}=0.1$ everywhere in the disk (roughly guided by
post-merging results), by imposing a polytropic relation $P_{\mathrm{0}} \propto \rho_0^{4/3}$ to
hold between the initial density $\rho_0$ and pressure $P_0$, by fixing the maximum density of the
torus to
$\rho_{\mathrm{max}} \approx 2.2\times 10^{12}\times (M_{\mathrm{torus}}/M_\odot) \,$g\,cm$^{-3}$,
and by placing the inner torus edge to a radius of $r_{\mathrm{min}}= 3\,r_{\mathrm{S}}$, where
$r_{\mathrm{S}}$ is the Schwarzschild radius corresponding to the chosen BH mass $M_{\mathrm{BH}}$.

We employ the $\alpha$-viscosity approach by \cite{Shakura1973} to include the effects of turbulent
angular momentum transport in an approximate, parametrized fashion. Since a number of different
formulations for the (shear) viscosity tensor $T_{ij}\equiv \eta_{\mathrm{vis}}(\nabla_iv_j +
\nabla_jv_i-(2/3)\delta_{ij}\nabla_kv^k)$ and the dynamic viscosity coefficient
$\eta_{\mathrm{vis}}$ exist in the literature (with $v_i$ being the velocity components), we
exploratively consider two different formulations to test the dependence of our nucleosynthesis
results on the explicit choice of the viscosity prescription: In the ``type 1'' case, which
is used for most BH-torus models, we include all components of $T_{ij}$ and the dynamic viscosity
coefficient is defined as
\begin{equation}
  \label{eq:alpha1}
  \eta_{\mathrm{vis, type 1}}\equiv \alpha_{\mathrm{vis}}\,\rho \,c_s^2 \,\Omega_{\mathrm{K}}^{-1} \, ,
\end{equation}
where $c_s\equiv \sqrt{\gamma P/\rho}$ with the adiabatic index $\gamma$ and $\Omega_{\mathrm{K}}$
is the Keplerian angular velocity. This prescription was employed, for instance, by
\cite{Igumenshchev1996, Lee2005, Setiawan2006}. In the ``type 2'' case, in contrast, only the
components $T_{r\phi}$ and $T_{\theta\phi}$ are included (i.e. the remaining components are set to
0) and the dynamic viscosity coefficient $\eta_{\mathrm{vis, type 2}}$ is computed like
$\eta_{\mathrm{vis, type 1}}$, cf. Eq.~(\ref{eq:alpha1}), but with the adiabatic sound speed $c_s$
replaced by the isothermal sound speed $c_i\equiv \sqrt{P / \rho}$. This prescription was used, for
instance, by \cite{Fernandez2013}. Note that for the type 2 viscosity, in contrast to the type 1
case, shear motions in meridional planes remain unaffected by viscosity. As a consequence, the
development of small-scale vorticity is less suppressed and the fluid pattern appears less laminar
than for the viscosity treatment of type 1. Furthermore, for a given value of
$\alpha_{\mathrm{vis}}$ the dynamic viscosity coefficients (which are a measure of the ``strength''
of viscosity) for both prescriptions are related by
\begin{equation}
  \label{eq:dynvis}
  \eta_{\mathrm{vis,type\,2}}=
  \frac{1}{\gamma}\eta_{\mathrm{vis,type\,1}} \approx(0.6\ldots0.8)\times \eta_{\mathrm{vis,type\,1}} \, .
\end{equation}

For all BH-torus simulations we take $416\times 160$ grid cells to cover the domain $(r,\theta) \in
[10\,\mathrm{km},2.5\times 10^4\,\mathrm{km}] \times [0,\pi]$. We use non-uniform grids in both
coordinate directions with the radial cell size increasing roughly proportionally to the radius and
the angular grid ensuring a resolution around the equatorial plane slightly better than at the
poles. We employ 10 logarithmically spaced energy groups spanning the range of $0-80\,$MeV to cover
the neutrino-energy space. For numerical reasons we need to define a (local) minimum gas density
value to which the evolved density is reset whenever it drops below this value. We choose this floor
density, $\rho_{\mathrm{floor}}$, to have a radially decreasing and time-dependent profile:
Initially, $\rho_{\mathrm{floor}}$ monotonically decreases with $\rho_{\mathrm{floor}}\simeq 10^5\,,
10^2\, ,10^{-3}\,$g\,cm$^{-3}$ at radii $r=10\, ,10^3\, ,2.5\times 10^4\,$km, respectively. During
the first $\sim 200\,$ms of evolution we continuously decrement the floor density in the region
$r\in [10\,\mathrm{km},1000\,\mathrm{km}]$ down to a final, constant value of
$10^2\,$g\,cm$^{-3}$. At the start of each simulation, we fill the volume surrounding the torus with
an ambient medium that has 1.5 times the floor density, resulting in a total mass of $\sim 4\times
10^{-6}\,M_\odot$ for the initial ambient medium.

\subsection{Nucleosynthesis}\label{sec_numerics_nuc}
The r-process nucleosynthesis is calculated by post-processing a representative set of trajectories
for the ejected matter, taking into account the dynamics as determined by the hydrodynamics
simulation.  The number of post-processed trajectories per model typically lies between a few
hundred for the merger models and a few thousand for the BH-torus models. The temperature evolution
is estimated on the basis of the laws of thermodynamics, allowing for possible nuclear heating
through beta-decays, fission, and alpha-decays, as described in \cite{Meyer1989}. Also, for the
late-time evolution the ejecta dynamics includes the pressure feedback by nuclear heating through
the approximative model described in \cite{Goriely2011a}. The nucleosynthesis is followed with a
reaction network including all 5000 species from protons up to Z=110 that lie between the valley of
$\beta$-stability and the neutron-drip line. All charged-particle fusion reactions on light and
medium-mass elements that play a role when the NSE freezes out are included in addition to radiative
neutron captures and photodisintegrations.  The reaction rates on light species are taken from the
NETGEN library, which includes all the latest compilations of experimentally determined reaction
rates \citep{Xu2013}. Experimentally unknown reactions are estimated with the TALYS code
\citep{Koning2005,Goriely2008} on the basis of the Skyrme Hartree-Fock-Bogolyubov (HFB) nuclear mass
model, HFB-21 \citep{Goriely2010}. On top of these reactions, $\beta$-decays as well as
$\beta$-delayed neutron emission probabilities are also included, the corresponding rates being
taken from the updated version of the Gross Theory \citep{Tachibana1990} based on the same HFB-21
$Q$-values.

All fission rates, i.e. the neutron-induced, photo-induced, $\beta$-delayed and spontaneous fission
rates, are estimated on the basis of the HFB-14 fission paths \citep{Goriely2007} and the nuclear level
densities within the combinatorial approach \citep{Goriely2008} obtained with the same single-particle
scheme and pairing strength. The neutron- and photo-induced fission rates are estimated on the basis
of the TALYS code for all nuclei with $90 \le Z \le 110$ \citep{Goriely2009}. Similarly, the
$\beta$-delayed and spontaneous fission rates are estimated with the same TALYS fission barrier
penetration calculation. The $\beta$-delayed fission rate takes into account the full competition
between the fission, neutron and photon channels, weighted by the population probability given by
the $\beta$-decay strength function \citep{Kodama1975}. The fission fragment yield distribution is
estimated with the renewed statistical scission-point model based on microscopic ingredients, the
so-called SPY model, and described in \citet{Panebianco2012,Goriely2013}.

\begin{figure}
  \includegraphics[width=84mm]{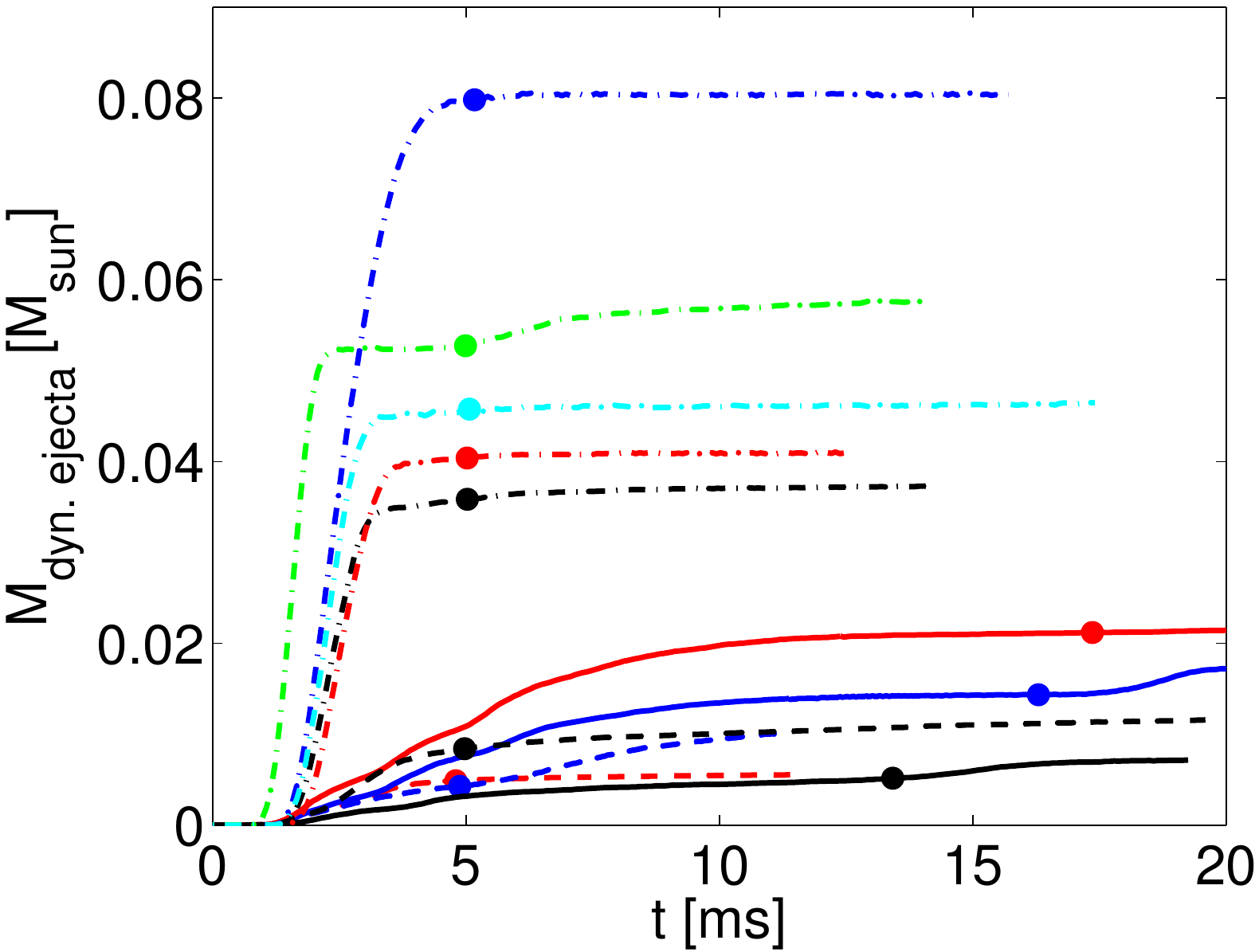}
  \caption{Prompt (``dynamical'') mass ejection in our NS-NS and NS-BH merger simulations as
    function of time.  The curves show the cumulative mass beyond a radius of $150\,$km that
    fulfills the criterion to be gravitationally unbound \citep[see][]{Bauswein2013}. The solid
  lines correspond to NS-NS cases with delayed collapse of the merger remnant to a BH (red:
  SFHX\_1515, blue: SFHO\_145145, black: TMA\_1616), the dashed lines to NS-NS cases with prompt BH
  formation (red: SFHO\_1218, blue: SFHO\_13518, black: TM1\_175175), and the dash-dotted lines
  belong to the NS-BH mergers (blue: TM1\_1123, green: TM1\_14051, cyan: TM1\_1430, red: SFHO\_1123,
  black: DD2\_14529). The bullets indicate the times when the ejecta masses for the merger phase are
  determined, assuming that the subsequent remnant evolution is covered by our BH-torus
  simulations.}
  \label{fig:binaries_massejection}
\end{figure}

\setlength{\tabcolsep}{3.3pt}
\begin{table*}
  \caption{Parameters and outflow properties for the merger models. The columns contain from left
    to right: Model name, masses of the binary components (with $M_2$ being the initial BH mass in the
    NS-BH cases), spin parameter of the initial BH (only NS-BH; in all cases prograde with the 
    orbital motion), equation of state, prompt (pc) or
    delayed (dc) collapse (only NS-NS), mass and spin parameter of the BH remnant, mass of the
    remnant torus, mass of the dynamical ejecta, asymmetry parameter (see
    Sect.~\ref{sec:comp-binary-merg} for the definition), mean electron fraction $\overline{Y}_e$, mean
    entropy $\bar{s}$ and mean outflow velocity $\bar{v}$ of the dynamical ejecta, and the name of
    the corresponding BH-torus model. The quantities $\overline{Y}_e, \bar{s}$ and $\bar{v}$ are 
    measured 5\,ms after the merger.}
  \label{table_mergers}
  \centering
  \begin{tabularx}{\textwidth}{lccccccccccccccc}

    \hline \hline
                 &             &             &                     &      &       &                   &                   &                      &                      &                    &                  &                          &                     &  & \vspace{-2.5mm}      \\
    Merger       & $M_{1}$     & $M_{2}$     & $A_{\mathrm{BH},0}$ & EOS  & pc/dc & $M_{\mathrm{BH}}$ & $A_{\mathrm{BH}}$ & $M_{\mathrm{torus}}$ & $M_{\mathrm{dyn}}$   & $B_{\mathrm{asy}}$ & $\overline{Y}_e$ & $\bar{s}/k_{\mathrm{B}}$ & $\bar{v}$           & Remnant  \vspace{0.5mm} \\
    model        & $[M_\odot]$ & $[M_\odot]$ &                     &      &       & $[M_\odot]$       &                   & $[M_\odot]$          & $[10^{-3}\,M_\odot]$ &                    &                  &                          & $[10^{10}\,$cm/s$]$ & model                   \\
    \hline
    SFHO\_1218   & 1.2         & 1.8         &                     & SFHO & pc    & 2.78              & 0.76              & 0.137                & 4.9                  & 0.28               & 0.036            & 9.9                      & 1.19                & M3A8m1\ldots            \\
    SFHO\_13518  & 1.35        & 1.8         &                     & SFHO & pc    & 2.97              & 0.78              & 0.099                & 4.3                  & 0.16               & 0.036            & 6.7                      & 1.28                & M3A8m1\ldots            \\
    SFHX\_1515   & 1.5         & 1.5         &                     & SFHX & dc    & 2.77              & 0.78              & 0.106                & 21.2                 & 0.01               & 0.032            & 8.2                      & 0.67                & M3A8m1\ldots            \\
    SFHO\_145145 & 1.45        & 1.45        &                     & SFHO & dc    & 2.68              & 0.79              & 0.091                & 14.3                 & 0.02               & 0.033            & 7.9                      & 0.64                & M3A8m1\ldots            \\
    TM1\_175175  & 1.75        & 1.75        &                     & TM1  & pc    & 3.37              & 0.85              & 0.027                & 8.4                  & 0.07               & 0.027            & 10.0                     & 1.12                & M3A8m03\ldots           \\
    TMA\_1616    & 1.6         & 1.6         &                     & TMA  & dc    & 3.04              & 0.83              & 0.037                & 5.2                  & 0.07               & 0.012            & 5.4                      & 0.62                & M3A8m03\ldots           \\
    \hline                                                                                                                                                                                                                                  
    TM1\_1123    & 1.1         & 2.29        & 0.54                & TM1  &       & 3.04              & 0.81              & 0.30                 & 79.8                 & 0.93               & 0.056            & 0.64                     & 0.66                & M3A8m3\ldots \\
    SFHO\_1123   & 1.1         & 2.3         & 0.53                & SFHO &       & 3.09              & 0.82              & 0.26                 & 40.4                 & 0.96               & 0.042            & 0.73                     & 0.60                & M3A8m3\ldots \\
    DD2\_14529   & 1.45        & 2.91        & 0.53                & DD2  &       & 4.00              & 0.83              & 0.27                 & 35.9                 & 0.96               & 0.056            & 0.62                     & 0.67                & M4A8m3\ldots \\
    TM1\_1430    & 1.4         & 3.0         & 0.52                & TM1  &       & 4.03              & 0.81              & 0.30                 & 45.8                 & 0.97               & 0.054            & 0.50                     & 0.67                & M4A8m3\ldots \\
    TM1\_14051   & 1.4         & 5.08        & 0.70                & TM1  &       & 6.08              & 0.83              & 0.32                 & 55.8                 & 0.98               & 0.050            & 0.41                     & 0.75                & M6A8m3\ldots \\
    \hline \hline

  \end{tabularx}
\end{table*}

\setlength{\tabcolsep}{3.5pt}
\begin{table*}
  \caption{Parameters and outflow properties for the relic BH-torus models. The
    columns from left to right contain: Model name, mass and spin parameter of the BH, torus mass,
    value of viscosity parameter, settings for viscosity type, radioactive heating and
    neutrino heating, total outflow mass, neutrino-driven outflow mass, and the mean
    electron fraction $\overline{Y}_e$, mean entropy $\bar{s}$ and mean velocity $\bar{v}$ of the outflow.
    The two quantities $\overline{Y}_e$ and $\bar{s}$ are measured when the outflow
    temperatures drop below 5\,GK , and $\bar{v}$ is evaluated at a fixed radius of $r=10^4\,$km.}
  \label{table_remnants}
  \centering
    \begin{tabular}{lcccccccccccc}
      \hline
      \hline
      &                   &                   &                      &                         &        &        &        &                                           &                                           &             & \vspace{-2.5mm}                                           \\
      Remnant      & $M_{\mathrm{BH}}$ & $A_{\mathrm{BH}}$ & $M_{\mathrm{torus}}$ & $\alpha_{\mathrm{vis}}$ & visc.  & radio. & neutr. & $M_{\mathrm{out}}$                        & $M_{\mathrm{out,\nu}}$                    & $\overline{Y}_e$ & $\bar{s}/k_{\mathrm{B}}$ & $\bar{v}$      \vspace{0.5mm} \\ 
      model        & $[M_\odot]$       &                   & $[M_\odot]$          &                         & type   & corr.  & heat.  & $[10^{-3}\,M_\odot (M_{\mathrm{torus}})]$ & $[10^{-3}\,M_\odot (M_{\mathrm{torus}})]$ &             &                          & $[10^9\,$cm/s$]$ \vspace{0.5mm} \\
      \hline                                                                                                                                                                                                      
      M3A8m3a2     & 3                 & 0.8               & 0.3                  & 0.02                    & type 1 & No     & Yes    & 66.8 (22.3$\,\%$)                         & 3.00   (1.00$\,\%$)                       & 0.28        & 20.9                     & 1.16 \\
      M3A8m3a5     & 3                 & 0.8               & 0.3                  & 0.05                    & type 1 & No     & Yes    & 78.3 (26.1$\,\%$)                         & 3.51   (1.17$\,\%$)                       & 0.25        & 23.0                     & 1.55 \\
      M3A8m1a2     & 3                 & 0.8               & 0.1                  & 0.02                    & type 1 & No     & Yes    & 22.7 (22.7$\,\%$)                         & 0.09   (0.09$\,\%$)                       & 0.28        & 24.8                     & 1.03 \\
      M3A8m1a5     & 3                 & 0.8               & 0.1                  & 0.05                    & type 1 & No     & Yes    & 24.7 (24.7$\,\%$)                         & 0.35   (0.35$\,\%$)                       & 0.24        & 28.0                     & 1.56 \\
      M3A8m03a2    & 3                 & 0.8               & 0.03                 & 0.02                    & type 1 & No     & Yes    & $\phantom{0}$7.0 (23.4$\,\%$)             & 0.0005 (0.002$\,\%$)                      & 0.27        & 29.5                     & 0.96 \\
      M3A8m03a5    & 3                 & 0.8               & 0.03                 & 0.05                    & type 1 & No     & Yes    & $\phantom{0}$7.3 (24.3$\,\%$)             & 0.002  (0.007$\,\%$)                      & 0.25        & 32.7                     & 1.70 \\
      M4A8m3a5     & 4                 & 0.8               & 0.3                  & 0.05                    & type 1 & No     & Yes    & 66.2 (22.1$\,\%$)                         & 1.47   (0.49$\,\%$)                       & 0.26        & 28.1                     & 1.66 \\
      M6A8m3a5     & 6                 & 0.8               & 0.3                  & 0.05                    & type 1 & No     & Yes    & 56.3 (18.8$\,\%$)                         & 0.07   (0.02$\,\%$)                       & 0.27        & 29.4                     & 1.45 \\
      M3A8m3a2-v2  & 3                 & 0.8               & 0.3                  & 0.02                    & type 2 & No     & Yes    & 64.2 (21.4$\,\%$)                         & 2.58   (0.86$\,\%$)                       & 0.29        & 19.3                     & 0.97 \\
      M3A8m3a5-v2  & 3                 & 0.8               & 0.3                  & 0.05                    & type 2 & No     & Yes    & 70.1 (23.4$\,\%$)                         & 2.63   (0.88$\,\%$)                       & 0.26        & 19.7                     & 1.39 \\
      M4A8m3a5-rh  & 4                 & 0.8               & 0.3                  & 0.05                    & type 1 & Yes    & Yes    & 67.3 (22.4$\,\%$)                         & 1.51   (0.50$\,\%$)                       & 0.26        & 26.4                     & 1.62 \\
      M3A8m1a2-rh  & 3                 & 0.8               & 0.1                  & 0.02                    & type 1 & Yes    & Yes    & 22.8 (22.8$\,\%$)                         & 0.09   (0.09$\,\%$)                       & 0.28        & 25.1                     & 1.05 \\
      M3A8m3a2-noh & 3                 & 0.8               & 0.3                  & 0.02                    & type 1 & No     & No     & 56.6 (18.7$\,\%$)                         & --                                        & 0.24        & 21.7                     & 0.90 \\
      \hline
      \hline
  \end{tabular}
\end{table*}

\section{Results}\label{sec:results}

\subsection{Choice of global parameters and coherence of modeling
  approach}\label{sec:choice-models-coher}

In Table~\ref{table_mergers} we summarize the investigated merger models and their properties. The
employed microphysical EOSs are SFHO~\citep{Steiner2013}, SFHX~\citep{Steiner2013},
TM1~\citep{Sugahara1994,Hempel2012}, TMA~\citep{Toki1995,Hempel2012}, and
DD2 \: \citep{Typel2010,Hempel2010}, which are all derived within the relativistic mean-field
approach. These EOSs are compatible with the current limit on the maximum NS mass of about
$2\,M_{\odot}$ \citep{Demorest2010, Antoniadis2013}. For the NS-NS mergers we consider three models
which result in the direct formation of a BH immediately after merging (within $\la 1-2\,$ms,
``prompt collapse''). In addition, we simulate three setups which lead to a ``delayed collapse''
after sufficient angular momentum has been redistributed and radiated away by gravitational
waves. The lifetimes of the relic NSs in these cases are roughly 10\,ms. In our study we do not
consider cases where a metastable, long-lived ($\ga 20\,$ms) or even stable massive NS is formed. We
therefore do not investigate here the potentially relevant outflow component which may arise during
the lifetime of such a massive NS \citep{Dessart2009,Metzger2014,Perego2014a}. For the NS-BH merger
models we examine cases in which relic BHs with masses $M_{\mathrm{BH}}\simeq 3,4$ and $6\,M_\odot$
are formed, varying the EOS between TM1, SFHO and DD2. In all simulated cases the initial BH spin
(with dimensionless spin parameter $A_{\mathrm{BH},0}$ between 0.5 and 0.7) is assumed to be
prograde with the orbital motion.

The merger models chosen here are not the most typical cases concerning what is expected
from observations of NS-NS binaries \citep[e.g.][]{Lattimer2012} and binary population models of
NS-NS and NS-BH systems \citep[e.g.][]{Dominik2012}. For the NS-NS case the most likely
configuration with $M_1\simeq M_2 \simeq 1.4 \,M_\odot$ leads to a long-lived massive NS for most
choices of the EOS \citep{Bauswein2013}. Instead, we focus here on NS-NS models which
collapse within the simulation time and produce remnant configurations that match 
the initial conditions for our long-term evolution studies of
BH-torus systems. Note, however, that the relic BH-torus models considered in our work may
also represent remnants formed in cases where the NS collapses to a BH with some longer delay.

The transition from the merger models to the relic BH-torus models is realized in an approximate
manner by setting up BH-torus models in rotational equilibrium as described in
Sect.~\ref{sec_numerics_remnant}, specifying values for the BH mass $M_{\mathrm{BH}}$, BH spin
$A_{\mathrm{BH}}$ and torus mass $M_{\mathrm{torus}}$ motivated by and approximately equal to the
corresponding values in the merger models. Therefore, each merger model can be associated with one
or several BH-torus models (see Table~\ref{table_mergers}).  Given the limited compatibility of the
codes used for simulating the merger and merger remnant evolutions, this is a pragmatic approach
which should be replaced by a fully consistent treatment of both evolutionary phases in future
models. However, our approach of examining both phases separately is advantageous insofar as
different progenitors may lead to very similar BH-torus systems and therefore the BH-torus models
are representative of a larger variety of binary merger configurations than investigated in the
present study.

In Table~\ref{table_remnants} the properties of the investigated BH-torus models are
summarized. Given the selection of merger models, the set of BH-torus models allows us to explore
the dependence of the nucleosynthesis production on variations of the torus mass, the BH mass, and
the viscosity parameter $\alpha_{\mathrm{vis}}$. The two additional models with suffix `v2' are set
up to study the impact of the specific choice for the viscosity prescription
(cf. Sect.~\ref{sec_numerics_remnant}). Moreover, since the heating feedback due to radioactive
decays of newly formed r-process nuclei is not consistently included in our standard hydrodynamical
simulations (but only in the post-processing calculations for the nucleosynthesis), we examine two
more models (with suffix `rh'), in which radioactive heating is approximated in the following way:
First, the radioactive heating rates, $Q(t_{\mathrm{nuc}})$, and temperatures,
$T(t_{\mathrm{nuc}})$, for all outflow trajectories of the standard models are averaged (where
$t_{\mathrm{nuc}}$ is the time during nucleosynthesis, normalized such that $t_{\mathrm{nuc}}=0$ at
$T=10$\,GK). Then the resulting average heating rate, $\langle Q\rangle(t_{\mathrm{nuc}})$, is
interpreted as a function of the average temperature, $\langle Q\rangle(T) \equiv \langle
Q\rangle(t(\langle T\rangle))$, and applied in the `rh'-simulations as additional energy source term
in the energy equation for all cells with $T \le 10\,$GK and positive radial velocities. We also
examine the effects of neutrino heating by considering model M3A8m3a2-noh, which is set up in
analogy to model M3A8m3a2 but with all neutrino source terms in the gas-energy and momentum
equations being switched off in regions where net heating by neutrinos applies, i.e., where energy
is transferred from neutrinos to the stellar medium.

\begin{figure*}
\begin{center}
\minipage{0.5\textwidth}
\includegraphics[width=0.95\textwidth]{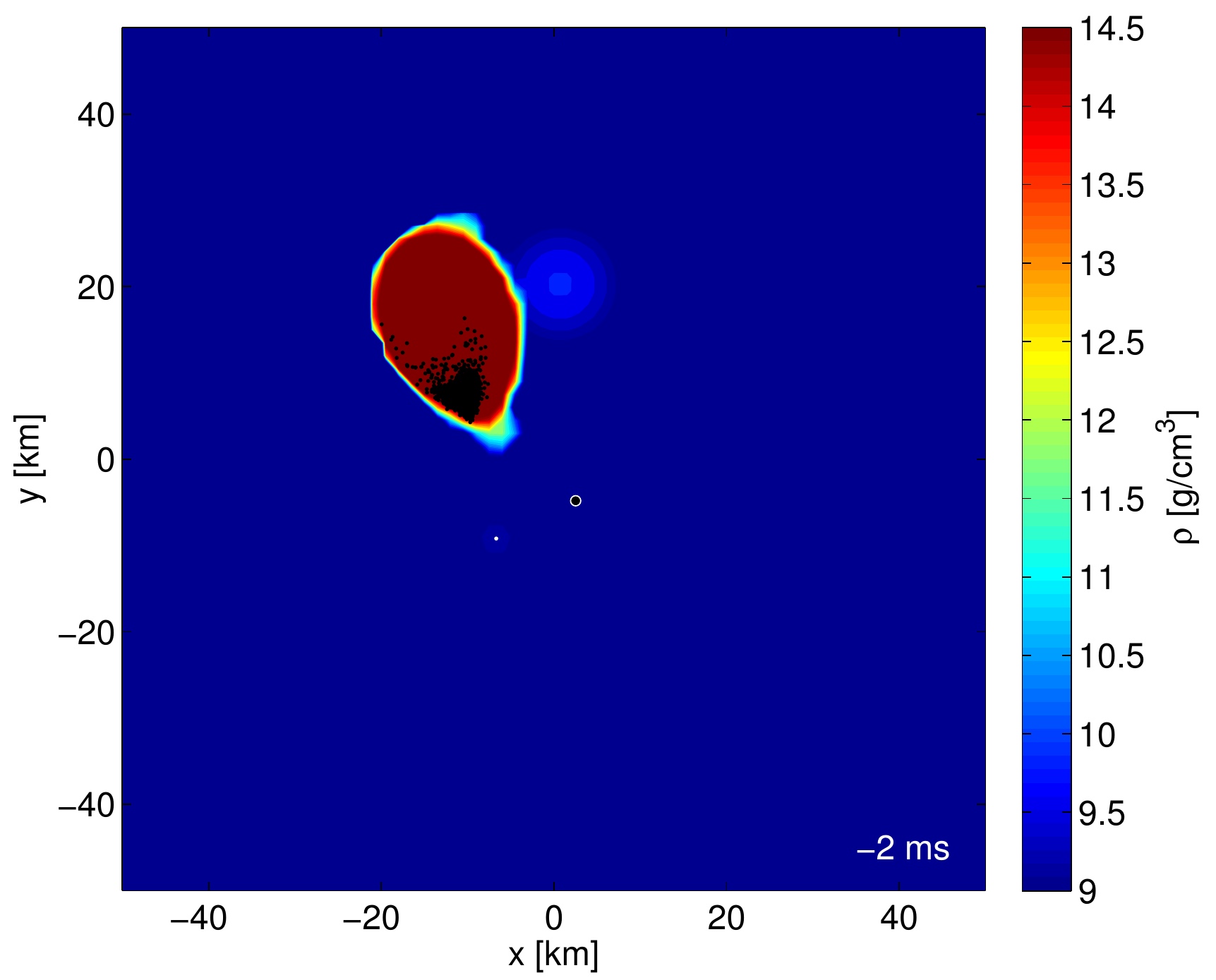}
\endminipage\hfill
\minipage{0.5\textwidth}
\includegraphics[width=0.95\textwidth]{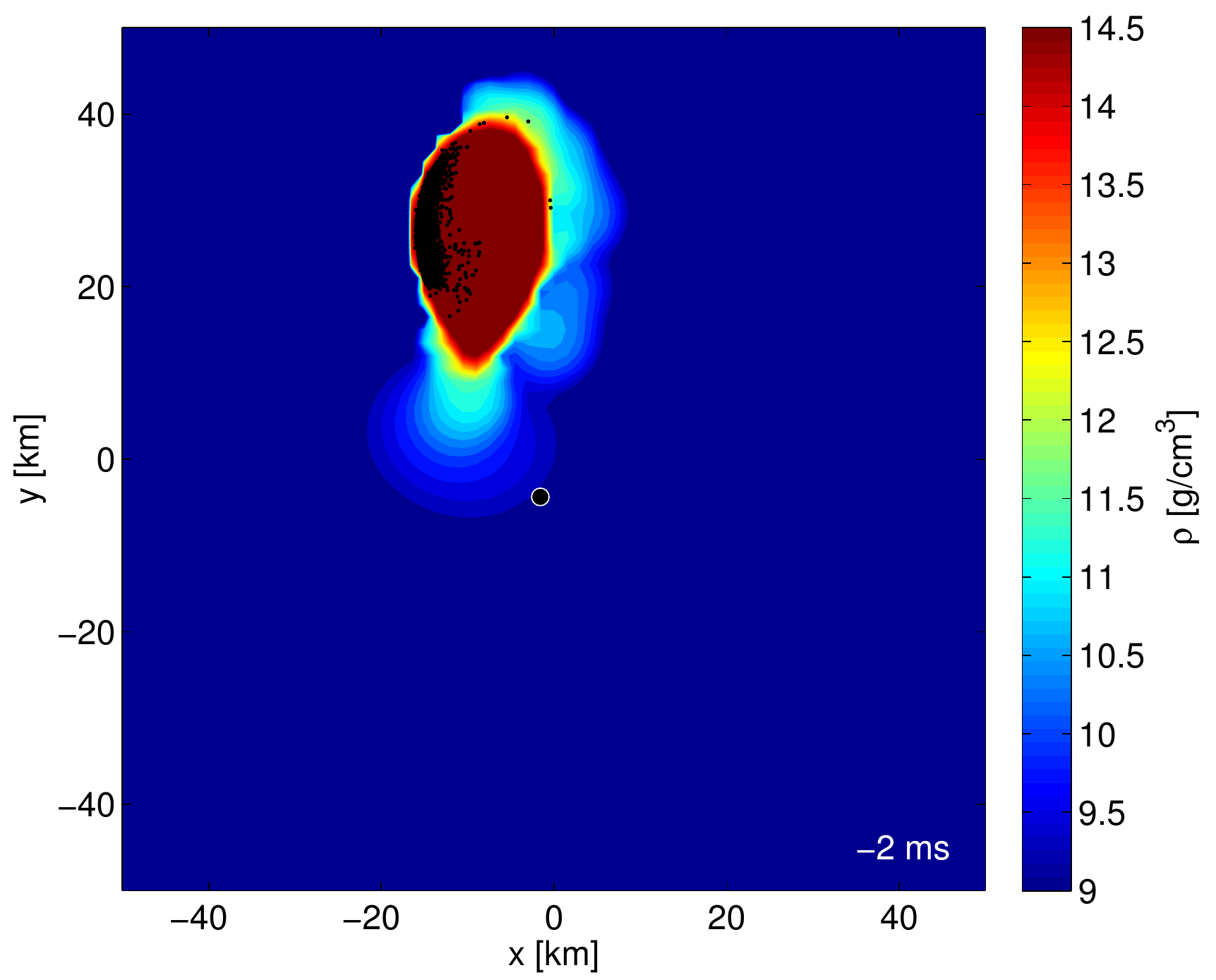}
\endminipage\hfill
\\
\minipage{0.5\textwidth}
\includegraphics[width=0.95\textwidth]{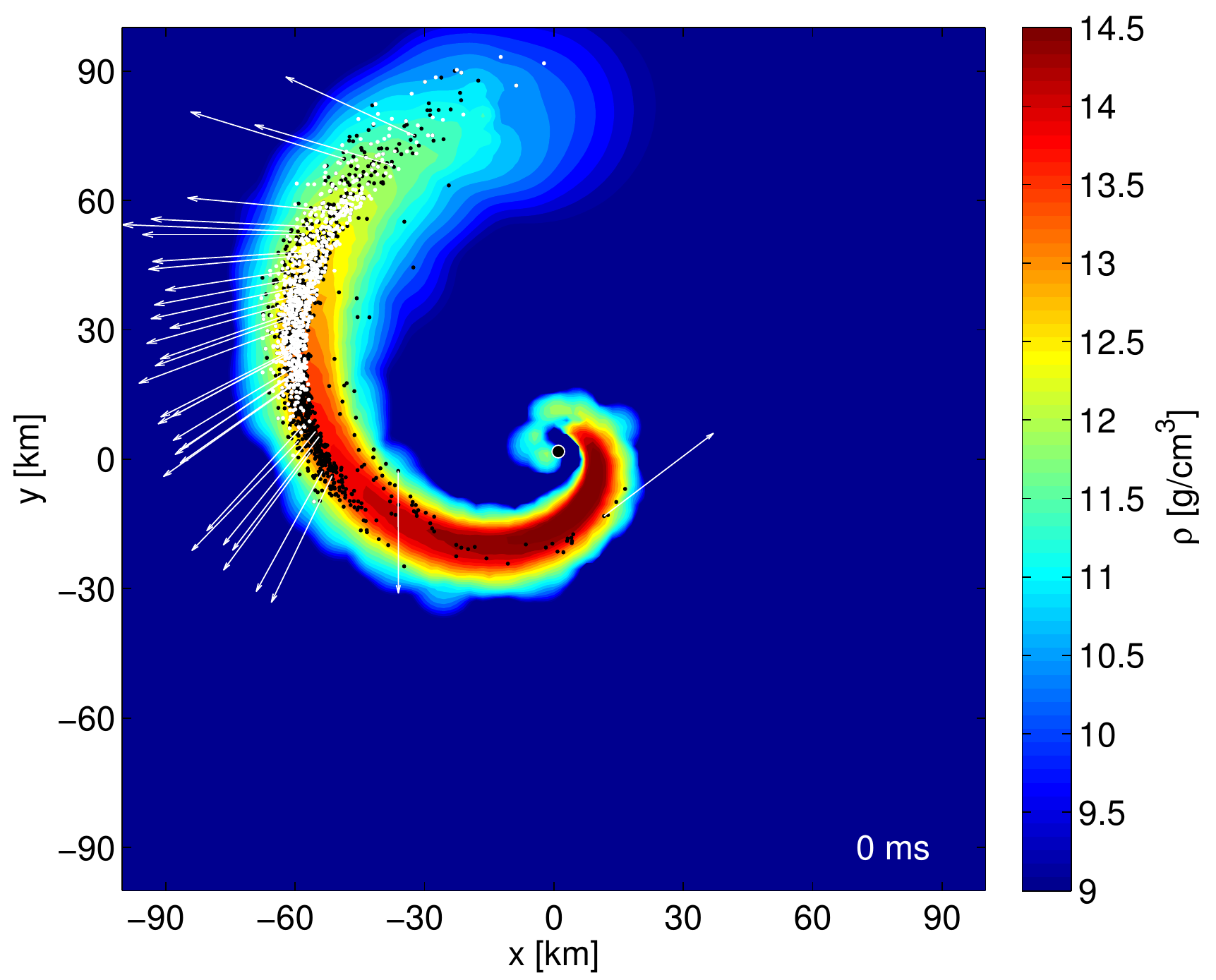}
\endminipage\hfill
\minipage{0.5\textwidth}
\includegraphics[width=0.95\textwidth]{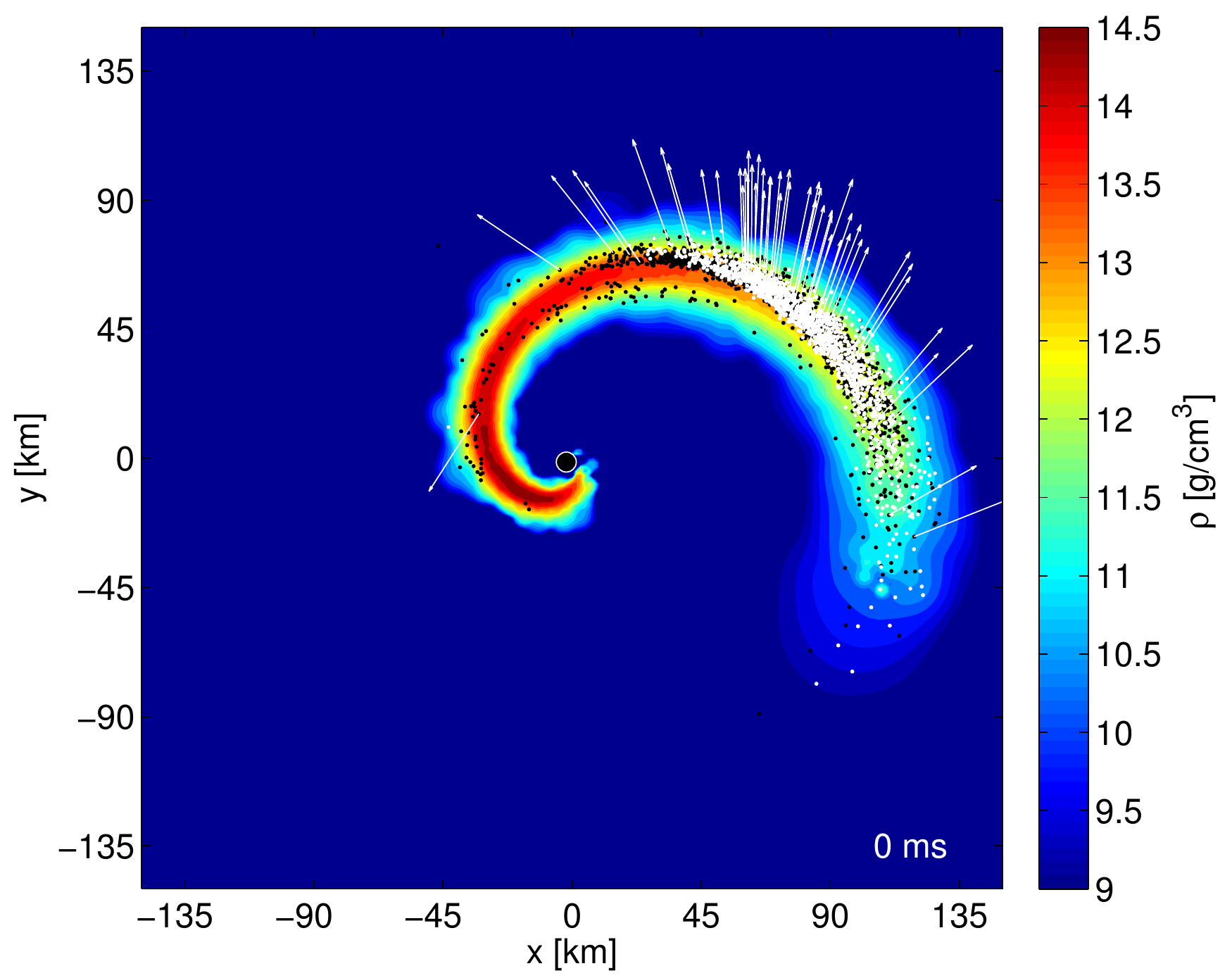}
\endminipage\hfill
\\
\minipage{0.5\textwidth}
\includegraphics[width=0.95\textwidth]{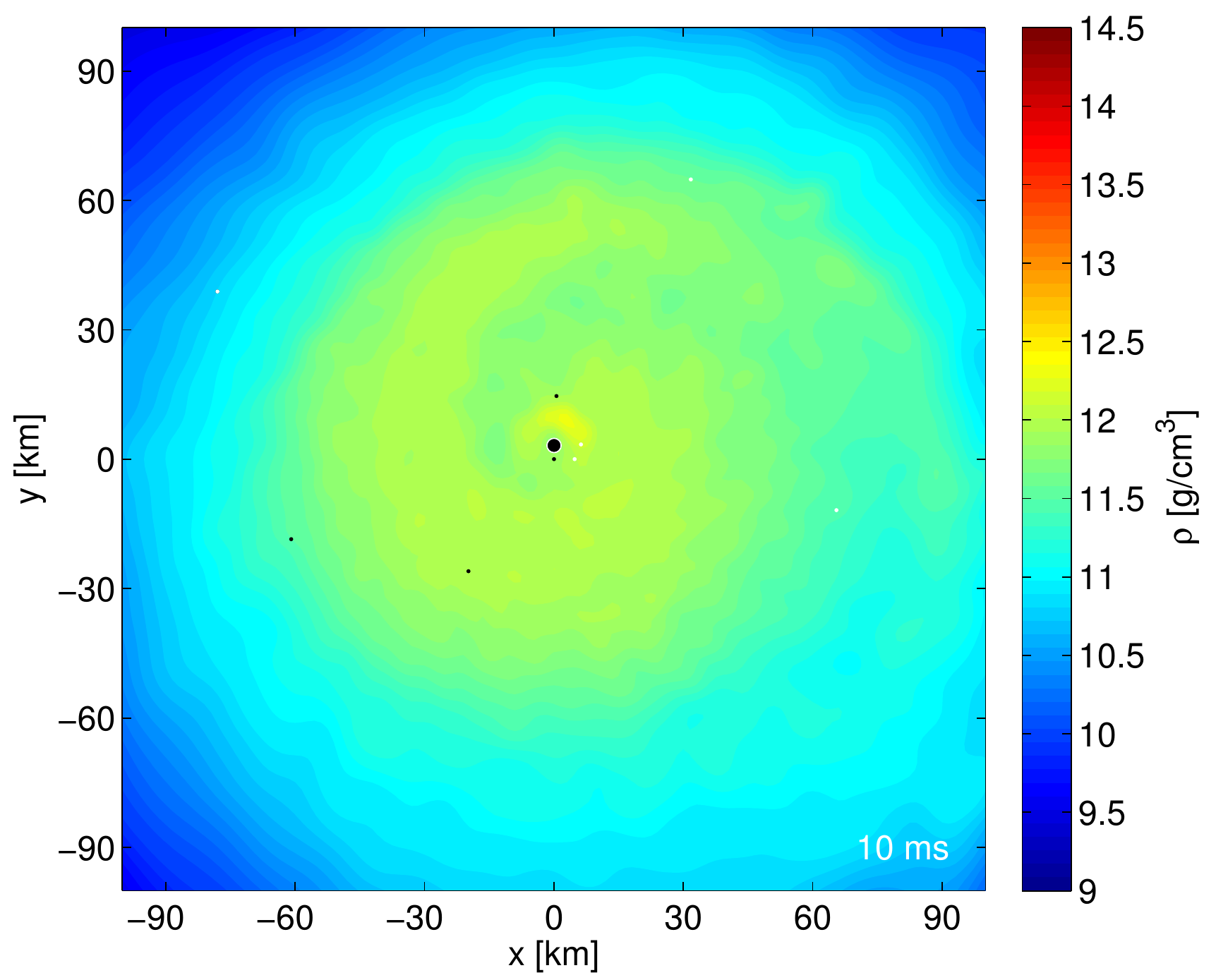}
\endminipage\hfill
\minipage{0.5\textwidth}
\includegraphics[width=0.95\textwidth]{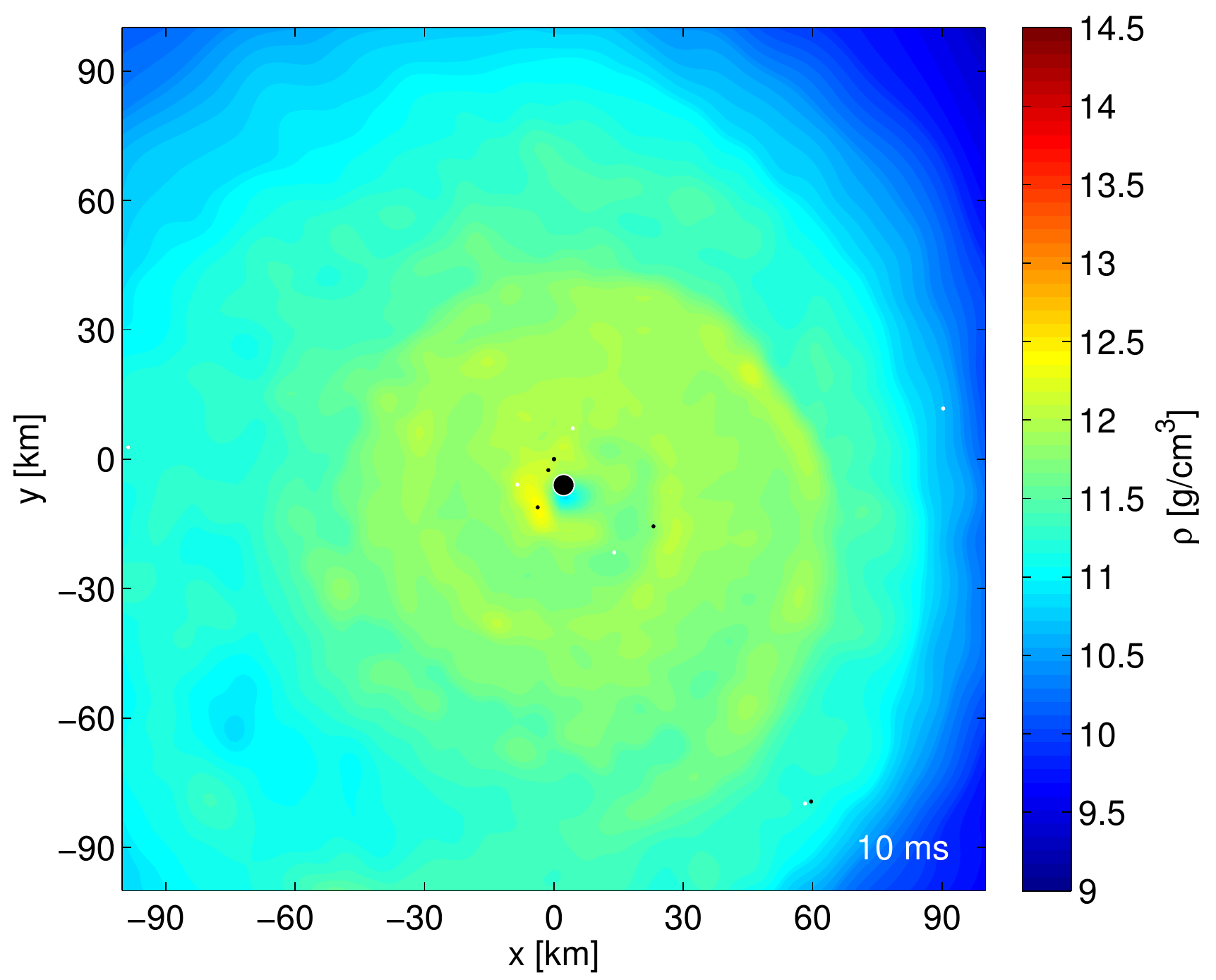}
\endminipage\hfill
\end{center}
\caption{Evolution of the rest-mass density in the equatorial plane for two NS-BH merger models.
  The left panels correspond to model DD2\_14529, which describes the merging of a 1.45~$M_{\odot}$
  NS and a 2.9~$M_{\odot}$ BH with a spin parameter of 0.53 for the DD2 EOS, the right panels
  display model TM1\_14051, which is the coalescence of a 1.4~$M_{\odot}$ NS and a 5.1~$M_{\odot}$
  BH with a spin parameter of 0.7 for the TM1 EOS.  The time is given in the lower right corner and
  is normalized to the moment when the BH has accreted half of the NS matter. The position of the
  BH is indicated by a filled black circle, whose radius in isotropic coordinates is 
  defined by the gravitational mass of an isolated, nonrotating BH. The dots trace SPH
  particles (projected into the equatorial plane) which eventually become unbound during the
  simulations. White dots denote particles which fulfill the ejecta criterion at the current time,
  while black dots indicate particles that will only become unbound later on. For a subset of SPH
  particles the arrows visualize the coordinate velocities of the corresponding fluid elements with
  the length of the arrows being proportional to the velocity. An arrow length of 10\,km corresponds
  to 0.2 times the speed of light.  The visualization tool SPLASH was used to convert SPH data to
  grid data \citep{Price2007}.}
  \label{fig:merger_snapshots}
\end{figure*}

\subsection{Compact binary mergers}\label{sec:comp-binary-merg}

For any type of compact-object binary the merging occurs after an inspiral phase, in which the
gravitational-wave emission and the corresponding angular momentum loss lead to a shrinking of the
orbital separation. The secular inspiral of the binary components proceeds increasingly faster
until the system enters the dynamical merging phase. 
The ejecta masses (Fig.~\ref{fig:binaries_massejection}) as well as the 
exact conditions in the ejecta and the
properties of the remaining tori depend on the dynamics of the compact object mergers. We briefly
describe the dynamics of three typical models which represent different formation channels for
relic BH-torus systems and for which the merging process possibly leaves an imprint on the properties
of the ejecta. All simulated binary mergers belong to one of these three categories of events.

{\it Delayed collapse (TMA\_1616)}. The coalescence of two NSs of 1.6\,$M_{\odot}$ with the TMA
EOS leads to the formation of a single, massive, differentially rotating NS, which is supported
against gravitational collapse by centrifugal forces and thermal pressure. The remnant consists
of a rotating double-core structure with the two dense cores bouncing against each other. The
associated redistribution of angular momentum and the emission of gravitational waves generated by
the oscillations of the merger remnant cause the collapse of the central object to a BH about 9~ms after
merging. Matter with high angular momentum forms a torus of about 0.037\,$M_{\odot}$ surrounding the
central BH, which has a spin parameter of $\sim$0.83.

During the final collision of two NSs, matter is scraped off from the NS surfaces and squeezed out
from their joint contact interface to become unbound in waves created by violent pulsations of the
rapidly rotating, massive NS remnant. Most of the dynamical ejecta are produced within less than
10\,ms after the merging (Fig.~\ref{fig:binaries_massejection}).  The ejecta mass still increases at
later times and even after the BH formation. However, we do not take into account such late ejecta
produced in our merger models because of their low resolution after the BH formation and because
this evolutionary phase is intended to be partially covered by our BH-torus models, where neutrino
effects are included, in contrast to the merger simulations. In this work we conservatively
determine the dynamical ejecta (cf.\ Table~\ref{table_mergers} and
Fig.~\ref{fig:binaries_massejection}) by requiring that matter has moved at least 150~km away from
the merger site in addition to applying the formal ejecta criterion of positive total specific
energy. This condition ensures that the matter truly becomes unbound, but it may slightly
underestimate the total ejecta mass by about 10 per cent.  As documented in
Table~\ref{table_mergers}, the ejecta are very neutron rich, because they originate from the inner
crust of the original NSs. The outflow has an average velocity of about $0.6\times
10^{10}$\,cm\,s$^{-1}$ and is fairly isotropic with a moderate preference towards the orbital plane
and corresponding suppression in the polar regions \citep[see, e.g.,][]{Bauswein2013}. A detailed
description of the formation and dynamics of such ejecta from NS mergers with delayed BH formation
was provided by~\cite{Bauswein2013}.  Recently, \cite{Wanajo2014a} pointed out a possible importance
of neutrino emission and absorption for reducing the neutron excess in the escaping material. We
will discuss this aspect in Sect.~\ref{sec:nucdynej}.

We note in passing that the stability or instability of the NS-NS merger remnant, the timescale
until it collapses to a BH, and the ejecta (and torus) masses depend sensitively on the total binary
mass and the NS EOS.  Less massive systems and/or stiff nuclear EOSs tend to yield remnants that do
not collapse within typical simulation times of $\sim$20\,ms after the merging.  More compact NSs
(corresponding to softer nuclear EOSs) lead to higher ejecta masses, more massive binaries produce
more dynamical ejecta for a given EOS, and unequal-mass binary mergers also result in more unbound
matter than mergers of symmetric systems \citep{Bauswein2013}. The TMA EOS is relatively stiff in
the sense that its NS radii are large. The ejecta and torus masses are therefore small. In contrast,
the SFHX and SFHO EOS are on the softer side with smaller NS radii and correspondingly larger ejecta
and torus masses.

{\it Prompt collapse (SFHO\_13518)}. For the SFHO EOS the merging of a 1.35\,$M_{\odot}$ NS and a
1.8\,$M_{\odot}$ NS results in a direct collapse, i.e., in BH formation only about one millisecond 
after the stars touch. The less massive NS is tidally strongly stretched before and during the
merging and forms a massive tidal tail. This tidal tail winds around the BH, collides with itself
after one orbit, and most of it ends up in a torus. The torus comprises a mass of about 
0.1\,$M_{\odot}$, and the BH obtains angular momentum which corresponds to a spin parameter of 
roughly 0.8.

Prompt mass ejection happens during a short period of less than 5\,ms
(Fig.~\ref{fig:binaries_massejection}). 
A substantial fraction of several ten per cent of these ejecta originate from the 
collision interface, from where they are squeezed out and pushed away by the expanding torus.
The rest of the ejecta is originally located in the tidal tail and becomes unbound by
tidal forces. The mass loss can slowly continue even later than 10\,ms after the merging
due to angular momentum transfer within the torus, but
for better comparison with the other models we do not include these late ejecta obtained
in the merger simulations, because we consider the remnant evolution to be followed by our
BH-torus models. 

Also in the prompt collapse scenario the spatial distribution of unbound material is
nearly isotropic with somewhat less matter being expelled towards the poles than close to 
the orbital plane (in the two polar cones with half-opening angle of 45$^\circ$ the ejecta mass
is roughly a factor of two smaller than in the other directions). Again the
ejecta originate from very neutron-rich layers of the initial NSs, which implies an average 
electron fraction of 0.04. Typical expansion velocities of about 0.4 times the speed of light are
somewhat higher than in the models resulting in a delayed collapse, which have average outflow
velocities of $\sim$20 per cent of the speed of light. The other merger models with prompt
collapse show outflow properties similar to those of model SFHO\_13518.

Prompt BH formation is unlikely to be the generic outcome of NS mergers, because NS binary
observations (see compilation in \citealt{Lattimer2012}) and theoretical population synthesis
studies of binaries (e.g., \citealt{Dominik2012}) suggest that most binary systems possess a total
mass of about 2.7\,$M_{\odot}$. For such systems a prompt BH-collapse of the merger remnant is not
expected for any EOS that is compatible with the current observational lower limit of the maximum NS
mass (cf.~\citealt{Bauswein2013a} for a determination of the threshold mass for direct BH formation
in dependence of the binary parameters and EOS).

\begin{figure*}
  \includegraphics[scale=1.1]{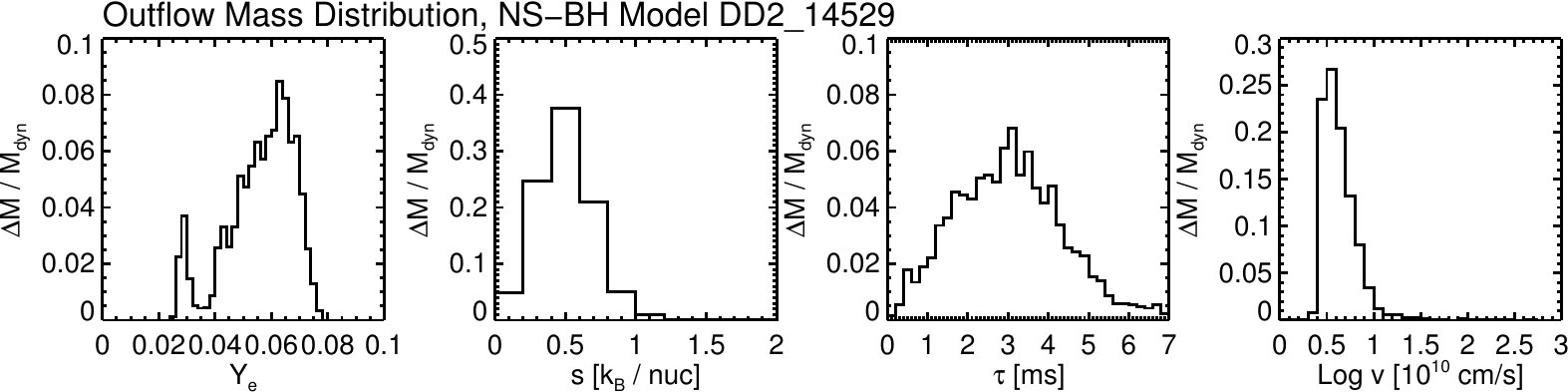}
  \caption{Mass-distribution histograms for the dynamical ejecta of the NS-BH merger model
    DD2\_14529. All distributions are normalized to the total mass of dynamical ejecta,
    $M_{\mathrm{dyn}}=3.59\times 10^{-2}\,M_\odot$. From left to right the panels show the mass
    distributions for the electron fraction $Y_e$, entropy per baryon $s$ (both measured about 5\,ms
    after the merging), expansion timescale $\tau$ (measured here as the time within which the
    density decreases from $10^{12}$\,g\,cm$^{-3}$ to $10^{10}$\,g\,cm$^{-3}$), and outflow velocity
    $v$ (measured about 10\,ms after the merging). The small peak at low $Y_e$ and the adjacent local 
    minimum are artifacts caused by SPH-particle clustering in the initial model of the NS. This
    distortion of the $Y_e$ mass distribution does not affect the nucleosynthesis results.}
  \label{fig:merger_histogram}
\end{figure*}

{\it NS-BH merger (DD2\_14529)}. Figure~\ref{fig:merger_snapshots} (left panels) exemplifies the
merging of a NS-BH binary with a mass ratio of two and an initial BH spin parameter of roughly
0.5. For the same model, we show in Fig.~\ref{fig:merger_histogram} the properties of the dynamical
ejecta in terms of mass distribution histograms for the electron fraction, entropy per baryon,
expansion timescale and outflow velocity. During the late inspiral phase the NS is tidally deformed
and develops a cusp pointing towards the BH. As the star approaches the BH, mass transfer sets in
from this cusp and the NS is stretched into an extended, spiral-arm-like tidal tail.  While mass is
shed off the far end of the tidal arm to expand outwards, the matter at the near end wraps the BH
and, if it has sufficiently high angular momentum, performs a full orbit around the BH to collide
with the spiral arm. Finally, the matter remaining outside the BH assembles into a nearly
axisymmetric torus of about 0.27\,$M_{\odot}$. Accretion from this torus continues at a lower rate
and the configuration reaches a quasi-steady state. At this time the BH mass has grown to
4~$M_{\odot}$ and the BH spin parameter has increased to approximately 0.8.

When about half of the NS mass has been accreted, matter located at the outer edge of the spiral arm
becomes gravitationally unbound by tidal effects within about one millisecond (visualized in
Fig.~\ref{fig:merger_snapshots}, middle panels for $t = 0$).  A smaller fraction (approximately 10
per cent of the total ejecta mass) is expelled during the subsequent evolution, in particular when
the tip of the spiral arm collides with the main body of the arm. The unbound matter originates
mostly from the inner crust of the NS and therefore is very neutron-rich ($Y_e$ ranges between 0.02
and 0.08, see Fig.~\ref{fig:merger_histogram}, left panel). The dominant ejection of decompressed,
unshocked matter from the tidally disrupted NS is reflected by the pronounced peak at entropies
below 1\,$k_\mathrm{B}$ per nucleon\footnote{Note that in Table~\ref{table_mergers} we report 
entropies obtained from a post-processing procedure that reduces the effects caused by heating
due to numerical viscosity in the SPH code. These entropies are also applied 
for our nucleosynthesis calculations \citep[see][]{Goriely2011}.}
in the second panel of Fig.~\ref{fig:merger_histogram}. Typical ejecta velocities (at
10\,ms after the NS-BH merging) are 15--25\% of the speed of light (right panel of
Fig.~\ref{fig:merger_histogram}), similar to those obtained for the delayed collapse cases of NS-NS
mergers.
 
As a consequence of the ejection mechanism, the outflow is highly asymmetric. To quantify the
asymmetry we identify the hemisphere where the main ejecta mass, $M_+$, is produced and the opposite
hemisphere (ejecta mass $M_-$) and define the asymmetry parameter $B_{\mathrm{asy}}$ by
$B_{\mathrm{asy}} \equiv (M_+ - M_-)/(M_+ + M_-)$ with $(M_+ + M_-)$ being the total ejecta mass.
Symmetric mass ejection corresponds to $B_{\mathrm{asy}}\sim 0$, highly asymmetric to
$B_{\mathrm{asy}}\sim 1$.  For model DD2\_14529 we determine an asymmetry of 0.96 and observe
similar values for the other NS-BH merger cases investigated in our study. These findings are in
line with the anisotropy reported in~\cite{Kyutoku2013}. In contrast, NS mergers show much
smaller asymmetries of the order of only a few per cent for symmetric and nearly symmetric systems
and some 10\% for highly asymmetric binaries (see Table~\ref{table_mergers}).

The right panels of Fig.~\ref{fig:merger_snapshots} display the merging of a NS with a more massive
BH of about 5\,$M_{\odot}$ and a spin parameter of 0.7 (model TM1\_14051), which according to
population synthesis studies~\citep{Dominik2012} may represent an astrophysically more typical
case. The merger dynamics and ejecta production proceeds qualitatively similarly to the case of
model DD2\_14529. Comparing the rest-mass density in the equatorial plane the more massive model
evolves in close analogy to the simulation with a mass ratio of 3 in~\citep[][see their
Fig.~2]{Kyutoku2011}. Note that we employ a stiff EOS (TM1), which tends to compensate for the
somewhat lower mass ratio and slightly higher spin used
in~\cite{Kyutoku2011}. Quantitatively, our model compares well with the calculation
2H-Q3M135a75 of this paper (stiff EOS, mass ratio 3, initial BH spin parameter 0.75), which results
in a torus of 0.35\,$M_{\odot}$ and a final value of the spin parameter of 0.86. Our model
TM1\_14051 is also compatible with the $Q=3$ simulation with the MS1 EOS and a BH spin parameter of
0.75 in~\cite{Kyutoku2013}. The authors of the latter work report an ejecta mass of
0.07~$M_{\odot}$, which is in satisfactory agreement with 0.055~$M_{\odot}$ of unbound matter found
in our simulation.

\subsection{Merger Remnants}\label{sect3_remnant} 

\begin{figure*}
  \begin{center}
    \includegraphics[scale=1.1]{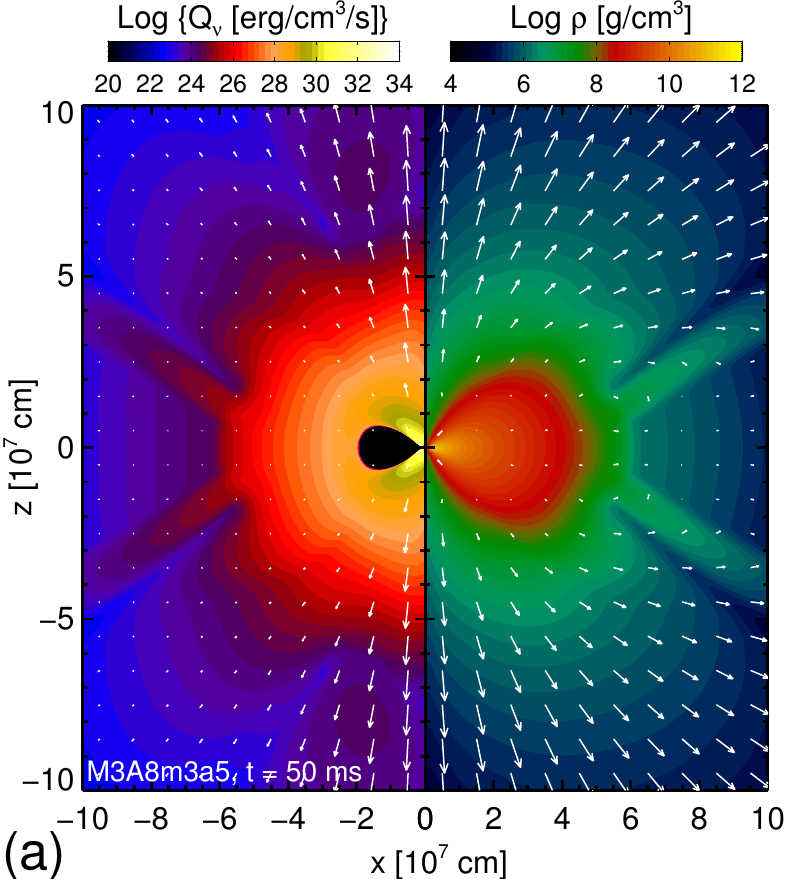}
    \includegraphics[scale=1.1]{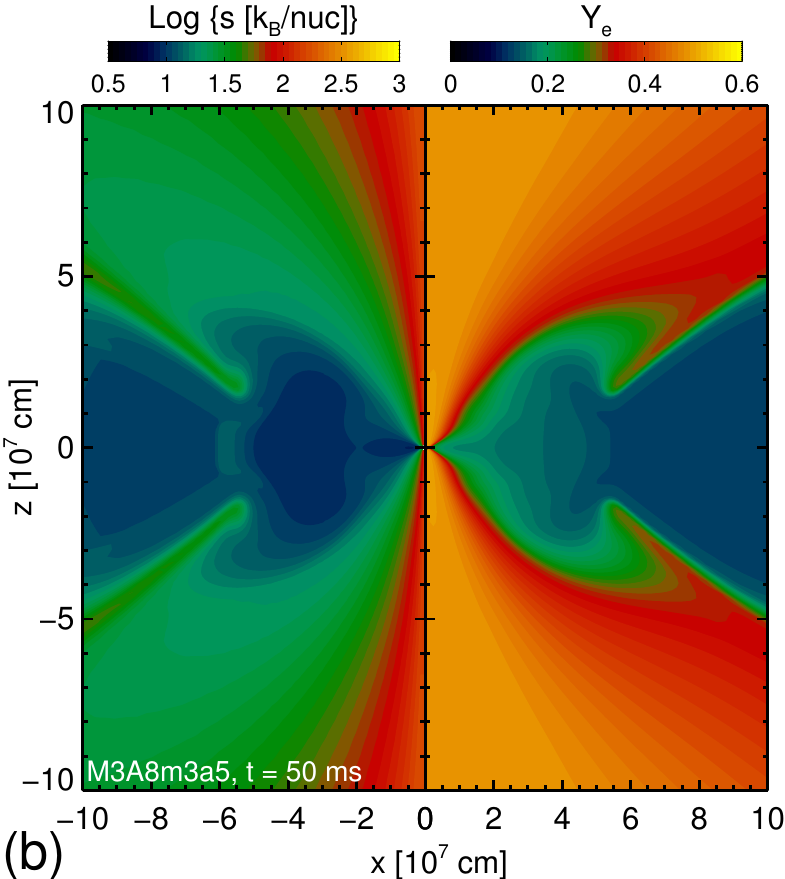}
    \includegraphics[scale=1.1]{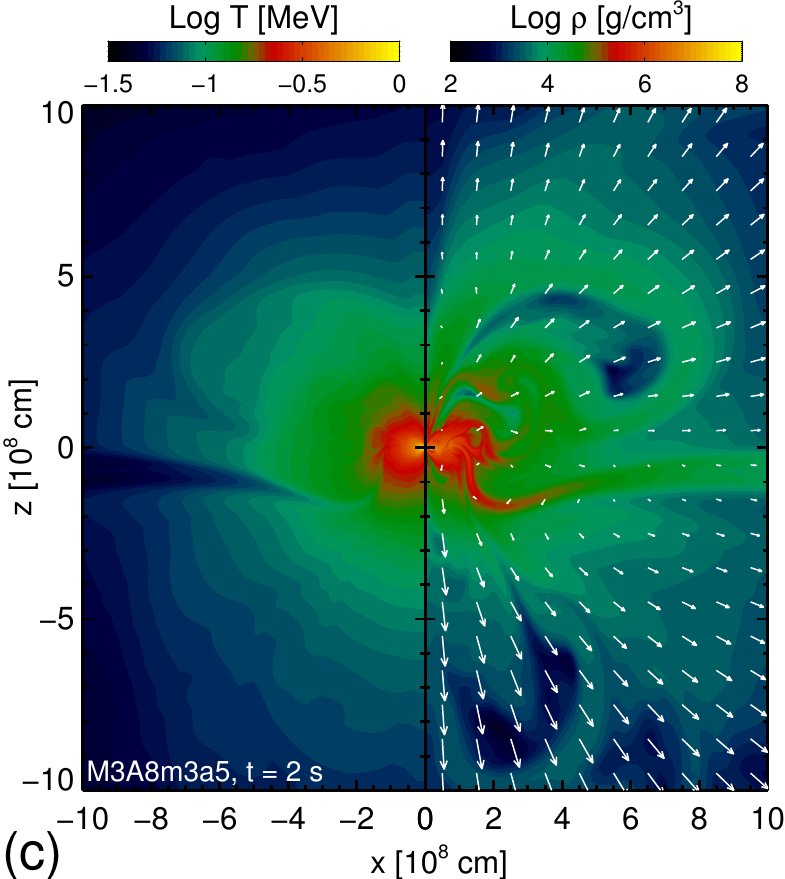}
    \includegraphics[scale=1.1]{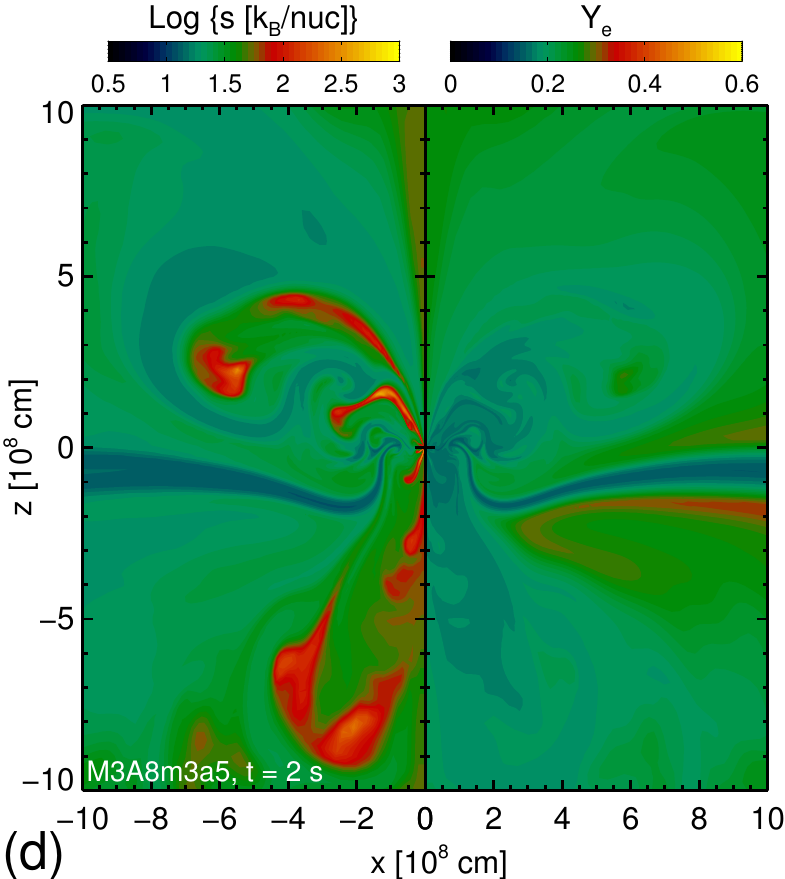}
  \end{center}
  \caption{Snapshots for the BH-torus model M3A8m3a5 at $t=50\,$ms ({\em top row}) and $t=2\,$s
    ({\em bottom row}). At early times neutrino-driven ejecta dominate the outflow, whereas at late
    times the viscous component is more important. Plot~(a) shows the total net neutrino-heating
    rate (left panel), overlaid with arrows indicating the vectors of the energy-integrated energy
    flux for electron neutrinos multiplied by $4\pi r^2$, and the density (right panel), overlaid
    with arrows for the velocity vectors. The neutrino-heating rate includes neutrino absorption on
    nucleons as well as neutrino-pair annihilation; in the black region neutrino cooling dominates
    heating. The maximum arrow lengths correspond to values of $5\times 10^{53}$\,erg\,s$^{-1}$ and
    $7\times 10^9$\,cm\,s$^{-1}$ in the left and right panels, respectively. Plot~(b) shows the
    entropy (left panel) and the electron fraction (right panel). Plots~(c) and~(d) display the
    same quantities as plots~(a) and~(b), respectively, but with the neutrino information replaced
    by temperature. The longest arrow in plot~(c) defines a velocity of $2\times
    10^9$\,cm\,s$^{-1}$. Note that the radial scales as well as the density ranges for the plots at
    $t=50\,$ms and $2\,$s are different.}
  \label{fig:torus_contour}
\end{figure*}

In Sect.~\ref{sec:origin-ejecta} we describe basic features of the evolution of BH-tori as NS-NS or
NS-BH merger remnants and the corresponding phases and components of mass loss.
Sects.~\ref{sec:ejecta-masses} and~\ref{sec:ejecta-properties} will present an overview of the
corresponding ejecta masses and properties, respectively. In Sect.~\ref{sec:comp-with-prev} our
results will be briefly compared with recent models published by \cite{Fernandez2013}.  More details
of our BH-torus evolution simulations will be provided in a forthcoming paper by Just, Janka \&
Obergaulinger (in preparation).

\subsubsection{Torus evolution, neutrino emission, and ejecta production}\label{sec:origin-ejecta}

Depending on their initial mass, accretion tori as merger remnants evolve on timescales of
milliseconds to seconds by losing mass to the accreting BH and in thermally, magnetically and
viscously driven outflows. In our hydrodynamic models the effects of magnetic fields are simplified
by the standard treatment through Shakura-Sunyaev $\alpha$-viscosity terms \citep{Shakura1973}. For
the high densities ($10^9$\,g\,cm$^{-3}$ and higher) and mass accretion rates
($\sim$10$^{-2}$\,$M_\odot$\,s$^{-1}$ to several $M_\odot$\,s$^{-1}$) characteristic of the
post-merger configurations, the tori cool predominantly by neutrino emission while photons are
trapped and contribute to the pressure as a plasma component in thermal equilibrium with electrons,
positrons, nucleons and nuclei \citep[see, e.g.,][]{Popham1999, Kohri2002, DiMatteo2002}.

Sufficiently massive and initially dense tori are known to evolve through three stages of radiation
efficiency and correspondingly different mass-loss properties \citep[cf., for
example,][]{Metzger2008c, Metzger2009b, Beloborodov2008, Fernandez2013}: (1) High initial densities
produce optically thick conditions for neutrinos so that neutrinos are trapped and therefore
neutrino cooling is inefficient \citep[see, e.g.][]{DiMatteo2002}.  (2) As the torus mass decreases
and densities drop, the so-called NDAF \citep[Neutrino-Dominated Accretion Flow,][]{Popham1999}
state is reached, in which gravitational energy that is converted to internal energy by viscous
heating is essentially completely radiated away by neutrinos.  (3) Eventually, when mass, density,
and temperature of the torus decrease further, the neutrino production rate becomes so low that
neutrino cooling is inefficient again. In this phase of an advection-dominated accretion flow
\citep[ADAF,][]{Narayan1994} viscous heating leads to large-scale convective motions and powers a
strong expansion of the torus \citep[for simulations of viscous ADAFs, see,
e.g.,][]{Igumenshchev1996,Igumenshchev1999,Stone1999}.

During stages (1) and (2) the high neutrino luminosities emitted by the dense, inner parts of the
torus irradiate and heat the outer, more dilute layers of the torus. Doing so they can launch a mass
outflow from this neutrino-heated (or ``gain'') region, which has similarities to the
neutrino-driven wind blown off the surface of new-born, hot neutron stars in supernova cores
\citep[e.g.,][]{Qian1996}.  The mass-loss rate as well as the thermodynamic conditions and the
neutron-to-proton ratio of these ejecta are sensitive functions of the neutrino emission properties
of the torus, but different from nascent neutron stars the BH-torus systems are extremely
nonspherical.  During the ADAF phase viscous angular momentum transport as well as dissipative
heating by shear viscosity lead to an inflation of the torus, driving outward mass flows near the
equatorial plane but also launching mass loss away from the equator. In contrast to the
neutrino-driven mass ejection, viscously driven expansion proceeds gradually and continuously, and
the matter becomes gravitationally unbound only at large radii and with lower velocities than the
supersonic neutrino-driven wind.

\begin{figure}
  \includegraphics[width=84mm]{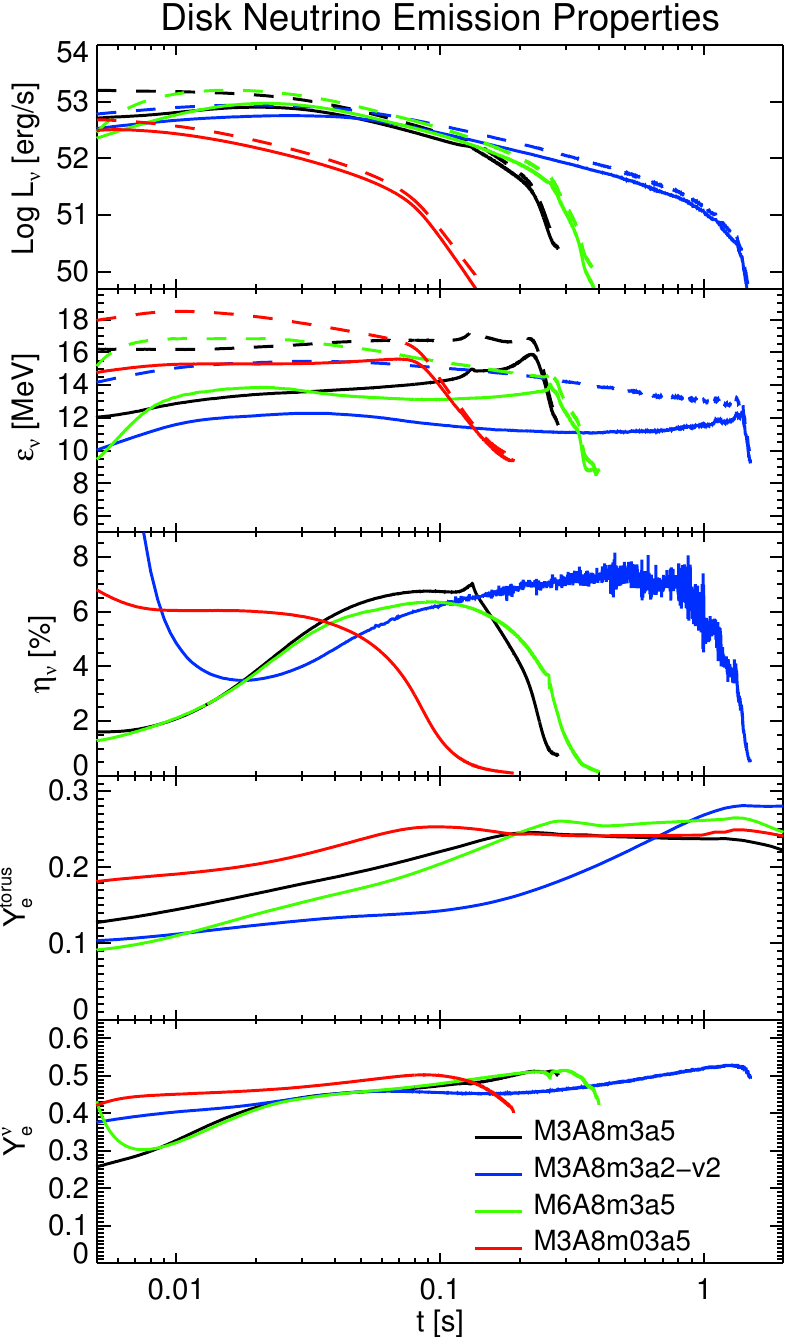}
  \caption{Time dependent neutrino emission properties for four BH-torus models.  The upper two
    panels show the energy-integrated luminosities, $L_\nu$, of $\nu_e$ (solid lines) and
    $\bar\nu_e$ (dashed lines) and the corresponding mean energies, $\varepsilon_\nu$,
    respectively. The latter are defined as $\varepsilon_\nu \equiv L_\nu/L_{N,\nu}$ with
    $L_{N,\nu}$ being the total neutrino number emission rate for neutrino $\nu$. The values are
    obtained by integrating the radial components of the neutrino flux densities in the laboratory
    frame over a spherical surface at radius $r=500$\,km. The third panel from the top shows the
    neutrino-emission efficiency parameter $\eta_\nu$ according to Eq.~(\ref{eq:eta}). The fourth
    panel from the top gives the mean electron fractions of the tori, $Y_e^{\mathrm{torus}}\equiv
    (\int\mathrm{d}V\,\rho\,Y_e) / (\int\mathrm{d}V\,\rho)$. The bottom panel displays estimates of
    the asymptotic electron fraction, $Y_e^{\nu}$, of the neutrino-driven wind, using the results of
    the two top panels in Eq.~(\ref{eq:yeequi}).}
  \label{fig:torus_lum}
\end{figure}

The first, optically thick period lasts at most $\sim$10--30\,ms for initial torus masses considered
in this work. Some smaller contribution to the mass ejection by neutrino heating happens already in
this phase. However, since our initial BH-tori are artificially constructed as equilibrium models
rather than being direct results from merger simulations, one should apply great caution in
interpreting the ejecta of this intermediate stage between the post-merger remnant and the relaxed,
neutrino radiating, quasi-steady-state torus.  Fortunately, the mass loss during this transition
phase is far subdominant in the total ejecta budget. In Fig.~\ref{fig:torus_contour} we provide
representative snapshots for the two following evolutionary phases for model M3A8m3a5, the NDAF
phase (top row of plots, 0.05\,s after the start of the simulation) and the ADAF phase
(bottom row of plots, for $t = 2.0$\,s). For this particular model the NDAF phase lasts about
250\,ms before a steep decrease of the neutrino luminosities\footnote{We stress that the term
  ``luminosity'' is used here for denoting the total energy loss rate of the torus. Of course,
  because of the pronounced direction dependence of the neutrino flux with a clear preference for
  small angles around the polar axis (cf. Fig.~\ref{fig:torus_contour}a, left panel)
  this quantity is not measurable for observers in any direction.} indicates the transition to the
ADAF stage (Fig.~\ref{fig:torus_lum}, top panel), but depending on the torus mass and viscosity
parameter ---the BH mass has less influence--- the duration of the NDAF can be between less than
$\sim$0.1\,s and more than $\sim$1\,s.

Because of the initially very neutron rich conditions ($Y_{e,0} = 0.1$), the torus protonizes (see
Fig.~\ref{fig:torus_lum}, where the time evolution of the mean electron fraction of the torus,
$Y_e^{\mathrm{torus}}$, is plotted) by radiating more electron antineutrinos, $\bar\nu_e$, produced
in positron captures on neutrons than electron neutrinos, $\nu_e$, emitted by electron captures on
protons. Since the mean energies are ordered in the same hierarchy, $\varepsilon_{\bar\nu_e} >
\varepsilon_{\nu_e}$, the $\bar\nu_e$ luminosity is also higher than the $\nu_e$ luminosity in all
cases (Fig.~\ref{fig:torus_lum}).  These findings are fully compatible with previous simulations of
post-merger accretion tori \citep[e.g.,][]{Ruffert1999a, Setiawan2006, Deaton2013, Foucart2014}.

Also the neutrino-emission efficiency $\eta_\nu$, defined by
\begin{equation}
L_{\nu_e} + L_{\bar\nu_e} = \eta_\nu \dot M_\mathrm{acc}c^2 
\label{eq:eta}
\end{equation}
as the ratio of the total neutrino luminosity to the rate of rest-mass energy accreted by the BH, is
a clear indicator of the three evolutionary stages of the accretion tori. During phase (1) the
efficiency is lower than in the NDAF phase (2)\footnote{The initial overshooting of $\eta_\nu$ is a
  consequence of the rotational equilibrium setup for the initial torus models, which causes some
  time delay until the mass accretion rate increases in response to viscous effects.} and drops at
the transition to the ADAF phase (Fig.~\ref{fig:torus_lum}, third panel). In the NDAF phase the
values of $\eta_\nu$ are approximately constant and range between a few per cent and nearly 10\%,
close to the energy-loss efficiency calculated for test masses accreted through the innermost stable
circular orbit.
 
In Fig.~\ref{fig:torus_contour}a, left panel, the black area indicates the dense, inner parts of the
torus, where neutrino cooling exceeds heating. Outside of this region the net heating (i.e., heating
minus cooling) rate is color coded. Heating is strongest above the torus and exhibits steep decrease
with growing vertical and horizontal distance from the BH, in whose vicinity the plasma temperature
is highest. The neutrino-driven wind develops at about 150\,km and fills a wide cone with
half-opening angle of about 65$^\circ$ around the polar axis (at a radius of 1000\,km).  Within this
cone the outflowing matter has electron fractions $Y_e$ larger than $\sim$0.3 and entropies, $s$, of
more than $\sim$10\,$k_\mathrm{B}$ per nucleon (Fig.~\ref{fig:torus_contour}b). The viscously driven
outflow during the ADAF phase, which is visible in plots~(c) and~(d) of
Fig.~\ref{fig:torus_contour}, is distinctly different not only by being driven outwards also close
to the equator, but also by considerably wider distributions of $Y_e$ (including values as low as
0.1) and entropy (see also Fig.~\ref{fig:torus_hist} and Sect.~\ref{sec:ejecta-properties}).

In our nucleosynthesis study we neglect all expelled material with entropies per baryon in excess of
1000\,$k_{\mathrm{B}}$. Ejecta with such high entropies are driven by the energy deposition through
neutrino-antineutrino pair annihilation in cases and phases where a low-density funnel around the
polar axis is formed. This jet-like, baryon-poor outflow can reach Lorentz factors beyond 100 and is
considered as possible source of short gamma-ray bursts \citep[e.g.][]{Paczynski1986, Eichler1989,
  Jaroszynski1993, Ruffert1999a, Aloy2005}. The physical and
numerical limitations of our present simulations (with a nonrelativistic code, which requires to set
a lower density limit in the axial funnel) do not allow us to treat these high-entropy outflows
accurately. We therefore do not investigate the associated nucleosynthesis.  Since the mass of these
ejecta is several orders of magnitude lower than that of the dominant merger ejecta, their
contribution to the overall abundance yields can safely be ignored.

\begin{figure}
  \includegraphics[width=84mm]{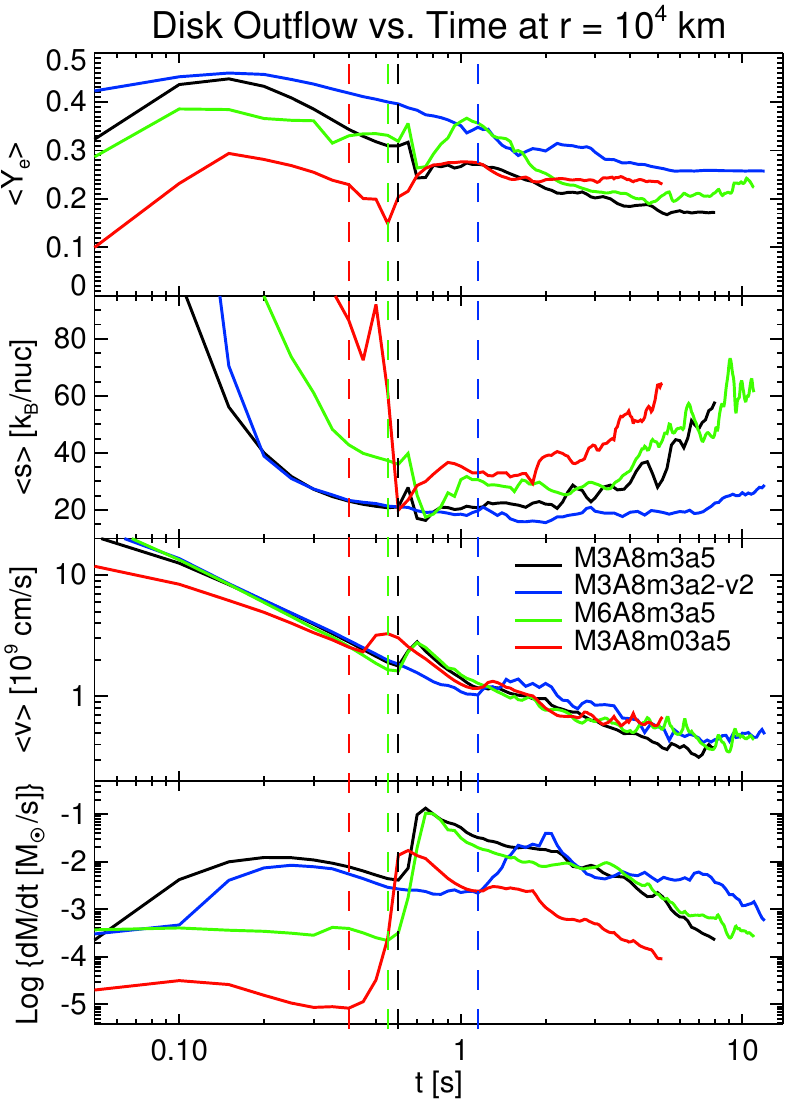}
  \caption{Outflow properties as functions of time for four BH-torus models. The upper three panels
    show the mass-flux weighted averages of the electron fraction, entropy, and velocity with mean
    value of quantity $A$ defined as $\langle A\rangle \equiv (2\pi\int\mathrm{d}\cos\theta\,
    A\,\rho v r^2) / \dot{M}_{\mathrm{out}}$, where $\dot{M}_{\mathrm{out}}=
    2\pi\int\mathrm{d}\cos\theta\,\rho v r^2$ is the total mass-outflow rate displayed in the bottom
    panel. All quantities are measured at a radius of $r=10^4$\,km, which implies a time retardation
    of roughly $0.1-1$\,s compared to the evolution shown in Fig.~\ref{fig:torus_lum}. For each model a
    vertical dashed line indicates the time $t_{\mathrm{\nu}}$ which approximately marks the
    transition from a mainly neutrino-driven outflow to a predominantly viscously-driven outflow
    (see Sect.~\ref{sec:ejecta-masses}).}
  \label{fig:torus_flux}
\end{figure}
\begin{figure}
  \includegraphics[width=84mm]{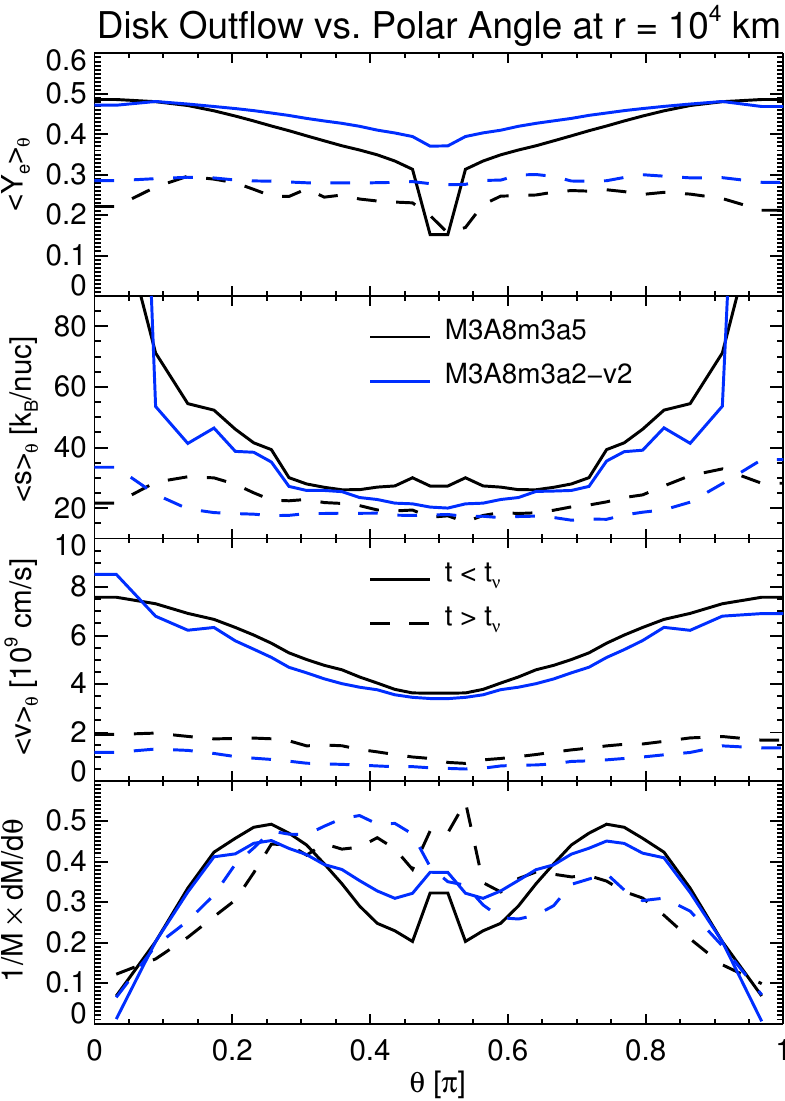}
  \caption{Polar-angle dependence of the mass ejection and ejecta properties for two representative
    BH-torus systems. Time-averaged quantities are shown as functions of polar angle, namely the
    electron fraction (top panel), entropy per baryon (second panel), and outflow velocity (third
    panel), all measured at a radius of 10$^4$\,km. The bottom panel displays the differential mass
    loss ${\mathrm{d}}M/{\mathrm{d}}\theta$, normalized by the angle integral, $M$, of this
    quantity. (Note that an isotropic mass ejection corresponds to
    $M^{-1}{\mathrm{d}}M/{\mathrm{d}}\theta = 0.5\sin\theta$.) The solid lines represent all ejecta
    (neutrino-driven plus viscously driven) for $t < t_\nu$ (for the meaning of $t_\nu$, see
    Fig.~\ref{fig:torus_flux}), the dashed lines correspond to all ejecta for the ADAF phase, i.e.,
    for $t \ge t_\nu$. The early ejecta include mostly neutrino-driven outflow (plus a viscously
    driven component near the equatorial plane at $\theta = \pi/2$) with dominant mass loss at
    intermediate angles.  Their properties exhibit a strong dependence on the polar angle. The late
    ejecta contain almost exclusively viscously driven matter and are expelled much more
    isotropically with little variations of the outflow properties in different directions.}
  \label{fig:ejecta_angledep}
\end{figure}

\subsubsection{Ejecta masses}\label{sec:ejecta-masses}

Figure~\ref{fig:torus_flux} shows the time evolution of the mass-outflow rates 
(integrated over all directions; bottom panel) for four representative torus models 
and the corresponding average properties of the ejecta (mean electron fraction, entropy, 
and velocity, from the top to the third panel), measured at a distance of 10$^4$\,km
from the BH. The approximate transition times $t_\nu$ between the NDAF state with its
strong neutrino-driven wind and the ADAF phase, where the outflows are mainly driven
by viscous effects, are indicated by vertical dashed lines for all models. Since the
viscously ejected material clearly dominates in mass (cf.\ Table~\ref{table_remnants})
and escapes with lower velocities than the fast, early neutrino wind, the transition 
typically coincides with a rapid increase of the outflow rate (Fig.~\ref{fig:torus_flux},
bottom panel). We stress that at times $t < t_\nu$ outflow with the 
characteristic properties of material accelerated by neutrino heating 
(see Sect.~\ref{sec:ejecta-properties}) is found only at latitudes somewhat away
from the equatorial plane (usually at more than 15$^\circ$), while in the plane of
the disk viscous transport drives the gas expansion. However, a distinction of the
two ejecta components is not always unambiguously possible, because the outflows are
affected both by neutrino and dissipative heating and therefore parts of them possess
mixed properties.

The angular distribution of the early ($t < t_\nu$) and late ($t \ge t_\nu$) mass
loss exhibits characteristic differences because of the different mass ejection 
dynamics of neutrino-driven and viscously driven outflows 
(Fig.~\ref{fig:ejecta_angledep}). Different from nascent neutron
stars, the torus is extremely nonspherical and therefore the neutrino emission
and wind ejection as well as the wind properties depend strongly on
the distance from the rotation axis and on the direction. Neutrino emission
and mass ejection are strongest in the vicinity of the BH. While the neutrino
flux is directed mainly parallel to the rotation axis (see the arrows in the
left panel of Fig.~\ref{fig:torus_contour}a),
the neutrino-driven wind carries away most mass at intermediate inclination
angles relative to the equatorial plane (Fig.~\ref{fig:ejecta_angledep}).
The characteristic properties (electron fraction, specific entropy, expansion
velocity) of this ejecta component vary strongly with the polar angle.
In contrast, the viscously driven outflows, which dominate the ejecta by far
in the ADAF phase at $t\ge t_\nu$, are much more spherical and their time-averaged
properties show much less angular variation (dashed lines in the panels of 
Fig.~\ref{fig:ejecta_angledep}). While outward angular momentum transport
perpendicular to the rotation axis drives equatorial mass loss, a considerable
fraction of the viscously heated disk matter is also blown away
with large angles relative to the torus equator, lifted out of the gravitational
potential by pressure forces and convective motions in the vertically inflated
configuration.

The integrated mass of the neutrino-driven outflow, $M_{\mathrm{out,\nu}}$, and the total ejecta
mass (neutrino plus viscosity driven), $M_{\mathrm{out}}$, are listed for all simulated BH-torus
models in Table~\ref{table_remnants}. The total outflow mass $M_{\mathrm{out}}$ takes into account
all material that has crossed the radius of $10^4\,$km until the end of the simulation\footnote{All
  BH-torus models are evolved up to a time at which the sphere of radius 10$^4$\,km contains less
  than one per cent of the original torus mass $M_{\mathrm{torus}}$.}, while $M_{\mathrm{out,\nu}}$
includes only material that has crossed $r=10^4\,$km at times $t<t_\nu$ and at angles $>15^\circ$
away from the equator. While the viscous outflows are a robust feature, carrying away $\sim$19--26\%
of the original torus mass, the neutrino-driven wind masses are considerably smaller, with
$M_{\mathrm{out,\nu}}$ reaching at most $\sim$1\% of the initial torus mass.  Since the
neutrino-wind properties are steep functions of the neutrino luminosity \citep[cf., e.g.,][for the
case of neutrino-driven winds of proto-neutron stars]{Qian1996}, the amount of such ejecta depends
much more sensitively on the torus mass and, particularly for less neutrino-opaque tori of low mass,
also on the viscosity. More massive tori and higher viscosities (which produce more heating) lead to
larger neutrino luminosities.

In contrast, the fraction of the torus that is expelled in viscously driven outflows varies only
weakly for different torus masses. For increasing BH mass (and all other system parameters fixed)
the masses of both kinds of ejecta decrease. Two effects contribute to this decline.  First, the
tori around higher-mass BHs in our models are more strongly gravitationally bound, i.e., a greater
part of their material possesses high specific binding energies, making it harder to lift this
material out of the gravitational potential well\footnote{Note that the qualitative tendency of
  higher mean specific binding energies for more massive BHs is consistent with the results of our
  NS-BH simulations and the corresponding energy values are even in good quantitative
  agreement.}. Second, for bigger BHs the neutrino-heating efficiency, $\eta_{\mathrm{\nu,heat}}
\equiv Q_{\mathrm{\nu,tot}}/(L_{\nu_e}+L_{\bar\nu_e})$ (with $Q_{\mathrm{\nu,tot}}$ being the total
net neutrino-heating rate), drops because the torus is geometrically smaller relative to the size of
the BH. This reduces the ``self-irradiation effect'' around the axial funnel above the poles of the
BH, by which neutrinos emitted from the neutrinosphere close to the BH, where the temperatures are
highest, can be reabsorbed in near-surface matter at other places of the torus ``walls'' around the
axial funnel. A torus that is big in comparison to the BH is supportive for this effect, a
vertically less extended torus (or, in the extreme case, a disk) diminishes it.

A stronger neutrino-energy transfer to the medium in the outer torus layers does not only enhance
the neutrino-driven wind but also the outflows caused by viscous effects. This can be seen by the
additional BH-torus model M3A8m3a2-noh that we computed without net neutrino heating. In this
simulation the total ejecta account for 18.7\% of the initial torus mass compared to 22.3\% in the
corresponding model M3A8m3a2, which includes all neutrino effects. This difference is considerably
larger than $M_{\mathrm{out},\nu}$ in model M3A8m3a2, which equals only about one per cent of the
torus mass. This can be understood by the fact that neutrino heating does not only drive mass
ejection but also causes a mass flow from torus regions closer to the BH, where the heating is
strongest, to more distant locations. This matter does not become unbound by neutrino-energy
transfer but settles into the torus again at larger radii, partly feeding the slower viscous-outflow
component from these regions. When net neutrino heating is switched off, the neutrino-aided
relocation of inner torus material does not happen but this matter is accreted by the BH during the
NDAF phase instead.

As mentioned above, higher values for the dynamic viscosity coefficient and the larger viscosity
effects connected with the type-1 prescription increase the neutrino-driven outflows by raising the
neutrino luminosities during the NDAF phase and also by improving the conditions for the neutrino
self-irradiation effect. The latter is a consequence of the torus inflation in reaction to viscous
heating.  Enhanced viscous dissipation of kinetic energy, which instigates more violent convective
mass motions in a bigger volume, as well as viscous angular momentum transport also have a direct,
positive impact on the mass loss from the outer regions of the torus.

Finally, radioactive decay heating that is approximately included in two test calculations (suffix
``rh'' in the model names) turns out to produce negligible differences in the ejecta mass of both
outflow components.

\begin{figure*}
  \includegraphics[scale=1.1]{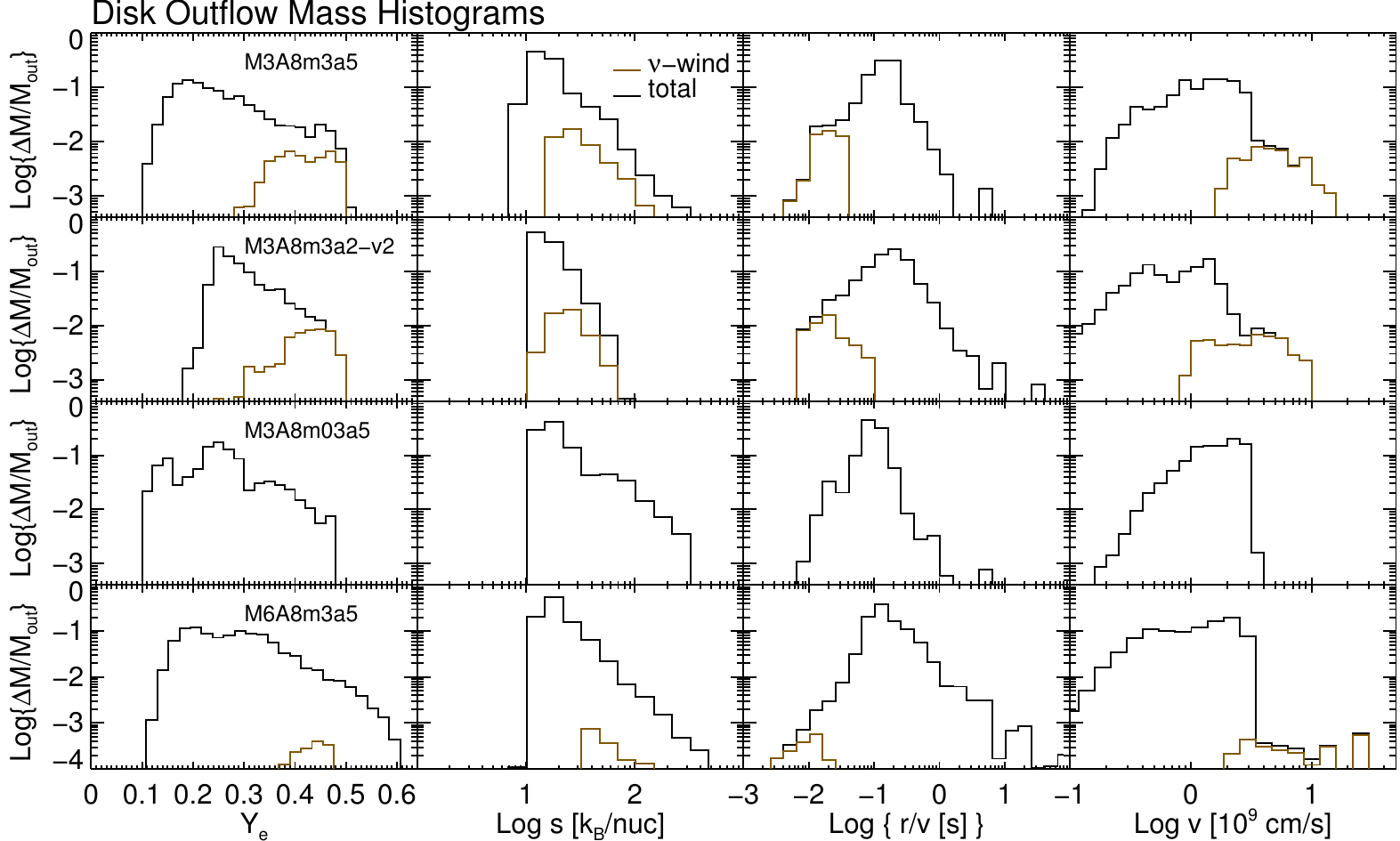}
  \caption{Mass-distribution histograms of the neutrino-driven outflow
    (brown lines) and of the whole (neutrino-driven plus viscously driven) ejecta 
    (black lines) for the four representative BH-torus models named in the left panels.
    All distributions are normalized by the total mass, $M_{\mathrm{out}}$, expelled
    in each case. From left to right the panels show the distributions for
    electron fraction, entropy per nucleon, dynamical expansion timescale, and velocity. 
    The first three quantities are measured at the time when the temperature of an
    escaping fluid element has dropped to 5\,GK, the outflow velocities
    are measured at a fixed radius of $r=10^4$\,km.}
  \label{fig:torus_hist}
\end{figure*}

\subsubsection{Ejecta properties}\label{sec:ejecta-properties}

The nucleosynthesis-relevant ejecta properties of our four representative
merger remnant models are plotted in Figs.~\ref{fig:torus_flux} and \ref{fig:torus_hist}.
Figure~\ref{fig:torus_flux} displays the time evolution of the mass-flux averaged electron
fraction, $\left\langle Y_e\right\rangle$, 
entropy per nucleon, $\left\langle s\right\rangle$, and 
outflow velocity, $\left\langle v\right\rangle$, for all ejecta leaving
a sphere of radius $r=10^4\,$km. In Fig.~\ref{fig:torus_hist} we show the corresponding
histograms of the ejecta mass distribution for $Y_e$, $s$, the dynamical expansion timescale
$\tau_{\mathrm{dyn}} \equiv r/v$ (all measured at $T=5$\,GK) and $v$ (measured at
$r=10^4$\,km), distinguishing between neutrino-driven outflow and the total 
(neutrino-driven plus viscously driven) mass loss.
Average values of $Y_e$, $s$ and $v$ of the whole expelled matter for each of our
simulated BH-torus configurations are listed in Table~\ref{table_remnants}.

For both neutrino-driven and viscously driven ejecta the
basic features of the time evolution of the average electron fraction, specific    
entropy, and expansion velocity (Fig.~\ref{fig:torus_flux}) as well as the overall
properties of the mass distributions of these quantities (Fig.~\ref{fig:torus_hist})
are similar in all cases and their dependence on the investigated global parameters
of the BH-torus systems is fairly weak.
Overall, the neutrino-driven wind in all models is less neutron rich, has
higher entropies, and expands faster than the viscous wind
(Fig.~\ref{fig:torus_hist}). The bulk of the neutrino-driven wind material
has $Y_e\sim 0.3$--0.5, $s/k_{\mathrm{B}}\sim 20$--50 per nucleon, $\tau_{\mathrm{dyn}}$
around 10\,ms, and asymptotic outflow velocities $v$ significantly above 
$10^9$\,cm\,s$^{-1}$ up to more than $10^{10}$\,cm\,s$^{-1}$.
In contrast, the viscous outflows exhibit a wider spread of conditions with the maxima
of the distributions being around $Y_e\sim 0.1$--0.35, 
$s/k_{\mathrm{B}}\sim 10$--30 per nucleon, $\tau_{\mathrm{dyn}}\sim 100$\,ms, and
expansion velocities between $0.5\times 10^9$\,cm\,s$^{-1}$ and about
$3\times 10^9$\,cm\,s$^{-1}$.

A lower value of the dynamic viscosity coefficient (or viscosity of type~2 instead of type~1)
notably increases the average and minimum electron fractions in the viscously driven ejecta. This
can be qualitatively understood from the fact that for a slower expansion $Y_e$ has more time to
increase from its low initial values towards higher values near weak $\beta$-equilibrium in the hot,
more dilute outflow. $Y_e$ thus rises until weak beta-processes freeze out when their reaction
timescale exceeds the expansion timescale, which happens earlier in faster outflows.  In the case of
more massive BHs, expelled matter not classified as neutrino driven develops a tail of the $Y_e$
distribution that extends far beyond 0.5. The strength of this tail increases with the BH mass (cf.\
the lower left panel for model M6A8m3a5 in Fig.~\ref{fig:torus_hist}). These special ejecta are
connected to a high-temperature, low-density bubble that forms in the inner torus (close to the BH)
at the time when the transition from the NDAF phase to the ADAF stage takes place. The bubble
accelerates outward due to buoyancy forces and for our models with BH masses $\ge$4\,$M_\odot$
triggers mass ejection of material whose $Y_e$ is close to the equilibrium value on the proton-rich
side.

The basic behavior of the neutron-to-proton ratio in the torus medium and in the neutrino-driven and
viscously driven outflows can be understood on grounds of simple considerations (which involve
physics in close analogy to what applies for matter in some regions and phases of supernova
cores). During the NDAF phase, where neutrino production by electron and positron captures on free
nucleons is fast but the created neutrinos are not trapped, the bulk of the torus matter resides
close to neutrino-less $\beta$-equilibrium corresponding to vanishing electron neutrino chemical
potential, $\mu_\nu = 0$. For typical torus densities and temperatures ($\rho\sim
10^{10}$--$10^{12}$\,g\,cm$^{-3}$ and $T\sim 10$--100\,GK\,$\sim$1--10\,MeV, respectively) such
conditions imply low values of the electron fraction, $Y_e^\beta \equiv Y_e(\rho,T,\mu_\nu = 0)\sim
0.1-0.2$, for which electrons are moderately degenerate \citep[see, e.g.,][]{Ruffert1997,
  Beloborodov2003, Chen2007, Arcones2010}.

In contrast, when the density and temperature in the expanding, neutrino-driven 
outflow drops, electron and positron captures, whose rates scale roughly with $T^5$,
freeze out quickly. Being exposed to the intense neutrino fluxes produced by the
torus in the NDAF phase, however, nucleons in the cooling ejecta
continue to absorb $\nu_e$ and $\bar\nu_e$. If these reactions 
achieve kinetic equilibrium, the corresponding value of the
electron fraction can be approximately expressed in terms of the neutrino
luminosities, $L_\nu$, and mean energies, $\varepsilon_\nu$, as \citep{Qian1996}:
\begin{equation}
  \label{eq:yeequi}
  Y_e^{\nu} \simeq \left( 1 +
     \frac{L_{\bar\nu_e}}{L_{\nu_e}}\frac{\varepsilon_{\bar\nu_e} - 2 Q_{\mathrm{np}}}
    {\varepsilon_{\nu_e} + 2 Q_{\mathrm{np}}} \right)^{\!-1} .
\end{equation}
Here $Q_{\mathrm{np}}\simeq 1.29$\,MeV is the rest-mass energy difference between neutrons and
protons. Note that $\varepsilon_\nu$ as defined in Fig.~\ref{fig:torus_lum} differs from the mean
energy that appears in the derivation of $Y_e^{\nu}$ by \citet{Qian1996}, but the numerical estimate
of $Y_e^{\nu}$ is very similar. The lower panel of Fig.~\ref{fig:torus_lum} shows our estimates of
$Y_e^{\nu}$ as functions of time for four representative BH-torus systems, whose neutrino
luminosities and mean energies are displayed in the top two panels of this figure. The values of
$Y_e^{\nu}\sim 0.3$--0.5 are considerably higher than in the torus itself and in the same ballpark
as those obtained for the neutrino-driven outflows in the hydrodynamical simulations (cf.\ top panel
of Fig.~\ref{fig:torus_flux}; note the retardation time of $\sim 0.1-1$\,s that needs to be taken
into account in comparing Figs.~\ref{fig:torus_flux} and \ref{fig:torus_lum}). Of course, more than
rough qualitative agreement cannot be expected because Eq.~(\ref{eq:yeequi}) provides only a crude
estimate.  On the one hand it assumes the neutrino luminosities to be radiated isotropically while
the neutrino fluxes emitted by the accretion tori depend strongly on the direction and position. On
the other hand the assumption of kinetic equilibrium of $\nu_e$ and $\bar\nu_e$ absorptions breaks
down when the wind expansion becomes too fast, in which case the electron fraction in the wind
freezes out at a value lower than $Y_e^{\nu}$. This is most obvious in the case of model M3A8m03a5
with its low torus mass and correspondingly low neutrino luminosities, where $\langle Y_e\rangle$ in
the neutrino-driven ejecta hardly climbs up to 0.3 (Fig.~\ref{fig:torus_flux}, top panel).

Despite similarities in the dynamics and thermodynamics of neutrino-driven and viscously driven
outflows the conditions that determine their nuclear composition are drastically different. At the
end of the NDAF phase the neutrino emission becomes so low that neutrino absorption does not play a
role any longer. Instead, during the ADAF stage $e^\pm$-captures drive the electron fraction in the
outflowing material towards the local equilibrium value given by $Y_e^\beta(\rho,T,\mu_\nu =
0)$. Because the electron degeneracy in the warm, dilute outer torus regions is low, the
neutron-proton mass difference suppresses electron captures on protons compared to positron captures
on neutrons, for which reason one expects $Y_e^\beta > 0.5$. Nevertheless, the true electron
fractions computed for the ejecta during the ADAF phase usually remain clearly on the neutron-rich
side (except in the matter ejected in hot, buoyant bubbles at the transition from the NDAF to the
ADAF stage, see above). This is explained by the fact that the torus expands and cools by viscous
effects during the ADAF evolution. The rates of $e^\pm$-captures therefore decrease continuously and
$Y_e$ in the torus and outflow freezes out when the viscous timescale becomes shorter than the
capture timescales of electrons and positrons \citep[e.g.,][]{Metzger2009b}. In agreement with
\cite{Metzger2009b} and \cite{Fernandez2013} we find that much of the viscously driven ejecta remain
very neutron rich with $Y_e\sim$0.15--0.25, but some fraction develops higher $Y_e$ (see
Fig.~\ref{fig:torus_hist}, left panels) so that the average values are more around $Y_e\sim$0.2--0.3
(cf.\ Fig.~\ref{fig:torus_flux}, top panel, and Table~\ref{table_remnants}).

\subsubsection{Comparison with
\citet{Fernandez2013}}\label{sec:comp-with-prev}

The basic evolutionary features of the models described here are similar to those found by
\cite{Fernandez2013}. However, the outflow properties we obtain exhibit some differences in
important details.  \cite{Fernandez2013} made use of a much simpler neutrino treatment than ours,
which essentially prevented them from modeling systems in which significant neutrino-driven winds
can be expected.

In a test simulation with a nonrotating BH corresponding to model S-def investigated by
\cite{Fernandez2013}, we find that roughly 8\% of the torus mass leave a sphere of $10^4$\,km almost
exclusively as viscously driven outflow, but only a fraction of this mass (4--8\% of the torus mass)
fulfills the criterion that the total (internal plus kinetic plus gravitational) specific energy is
positive when crossing the radius of $10^4\,$km. This criterion is employed to estimate the actual
amount of material that will finally become gravitationally unbound and be mixed into the
interstellar medium. However, since thermal effects are still dynamically relevant at $r=10^4$\,km,
the criterion depends on the thermal energy that ultimately contributes to unbind the outflow, i.e.,
on the normalization of the internal energy used to evaluate the criterion at $r=10^4$\,km. In our
test simulation we obtained the lower estimate of the unbound outflow mass ($\sim 4\%$ of the torus
mass) when taking for the internal energy the thermal energy instantly resulting from our 4-species
EOS at $r=10^4\,$km, while we determined almost all of the outflow as nominally unbound when
assuming that the thermal energy at $r=10^4\,$km corresponds to all nucleons being recombined with
an average nuclear binding energy of 6\,MeV per nucleon\footnote{In our models with $A_\mathrm{BH} =
  0.8$, typically $\ga 90\%$ of the material that reaches a radius of more than $10^4$\,km are
  nominally unbound independent of the precise energy normalization, and we therefore count all mass
  at $r > 10^4$\,km as ejecta.}. \cite{Fernandez2013} quoted a fraction of 10\% of the initial torus
mass to become unbound, which, however, was reassessed by \citep{Metzger2014} as being too large by
a factor of $\sim$few due to an erroneous overestimation of the recombination heating by $\alpha$
particles. Hence, without having more detailed published information available for now, we consider
our masses of unbound outflows to be approximately consistent with \cite{Fernandez2013}. An
interesting feature worth to record from the comparison with \cite{Fernandez2013} is that the BH
spin appears to have a sizable impact on the amount of ejecta produced by the disk.

While the ejecta masses seem to be in reasonable agreement, the bulk of the outflow in the model of
\cite{Fernandez2013} is slightly more neutron rich, $Y_e\sim 0.1$--0.21, than in our simulation,
$Y_e\sim 0.18$--0.30. The reason of this difference cannot be unambiguously identified. The lower
values of $Y_e$ could simply also be a consequence of the error in the $\alpha$-recombination
heating, leading to higher escape velocities and therefore faster freeze-out from weak-reaction
equilibrium. However, \cite{Metzger2014} mention that the changes of nucleosynthesis-relevant
properties associated with their mistake are irrelevant. Another possible reason for the $Y_e$
difference could be connected to the different treatments of neutrino reactions and transport.

\subsection{Nucleosynthesis}\label{sec:nucleosynthesis}

\subsubsection{Nucleosynthesis in the dynamical ejecta}
\label{sec:nucdynej}
The nucleosynthesis resulting from the dynamical ejecta has been studied previously either based on
parametrized ejecta trajectories \citep{Freiburghaus1999,Goriely2005} or more ``realistic''
hydrodynamical Newtonian \citep{Korobkin2012,Rosswog2014} and (CFC-) relativistic
\citep{Goriely2011,Bauswein2013,Goriely2013} simulations of NS-NS binary systems, recently also
including neutrino effects in relativistic merger models \citep{Wanajo2014a}. In all these
simulations, the number of free neutrons per seed nucleus reaches a few hundred. With such a neutron
richness, fission plays a fundamental role by recycling the matter during the neutron irradiation
and by shaping the final r-abundance distribution in the $110 \la A \la 170$ mass region at the end
of the neutron irradiation. Thanks to this property, the final composition of the ejecta is rather
insensitive to details of the initial abundances and the astrophysical conditions, in particular the
mass ratio of the two NSs, the quantity of matter ejected, and the
EOS~\citep{Goriely2011,Bauswein2013,Korobkin2012}. The mentioned calculations essentially correspond
to the delayed-collapse scenario and their typical yields are displayed in
Fig.~\ref{fig_delayed-collapse}. In the following we present our first nucleosynthesis results
obtained for cases with prompt collapse of the NS-NS binary merger remnant as well as for NS-BH
binary systems (cf.\ Table~\ref{table_mergers}).

\begin{figure}
\includegraphics[width=84mm]{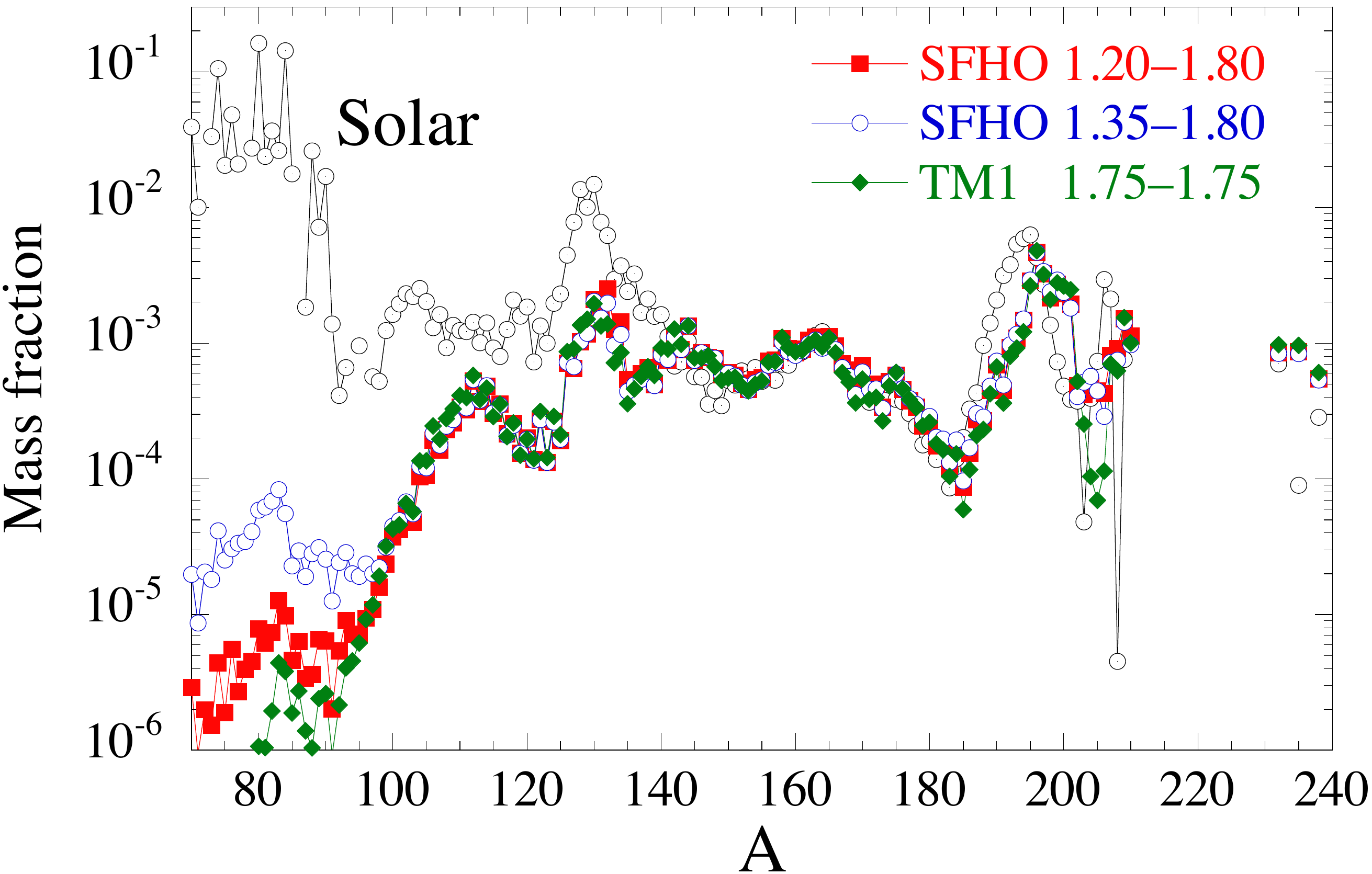}
\caption{Abundance distributions as functions of the atomic mass for the dynamical ejecta of three
  NS merger cases with prompt collapse of the remnants.  Each binary system is characterized in the
  legend by the EOS used in the simulation and the mass (in $M_{\odot}$) of the two NSs. All
  distributions are normalized to the same $A=196$ abundance. The dotted circles show the solar
  r-abundance distribution \citep{Goriely1999a}.}
 \label{fig_prompt-collapse}
\end{figure}

\vskip 0.2cm
\noindent {\it Prompt collapse cases}

Figure~\ref{fig_prompt-collapse} shows the abundance distributions resulting from the prompt
collapse of NS-NS binary systems. The three cases exhibit similar pattern characteristics of the
fission recycling nucleosynthesis, as described in \cite{Goriely2013,Goriely2014}.  In the
relativistic simulations, in contrast to the Newtonian approximation (where ejecta originate mostly
from cold extended spiral arms; \citealp{Korobkin2012,Rosswog2014}), not all the mass elements lead
to the same composition after ejection. Mass elements are ejected with considerably different
velocities so that the density evolution may vary significantly from one trajectory to the next
\citep[cf. ][]{Bauswein2013}. Fast expanding mass elements might not have time to capture all
available free neutrons leading to no or only one fission cycle while more slowly expanding mass
elements allow for all free neutrons to be captured and typically three fission cycles to take place
\citep{Goriely2014}.

\vskip 0.2cm
\noindent {\it Delayed collapse cases}

In the delayed collapse cases, the integrated mass associated with rapidly expanding material
compared to the total ejecta mass remains relatively smaller than in the prompt collapse cases. For
this reason, similar to Newtonian models of the NS-merger hydrodynamics, which tend to predict
rather slow expansion \citep{Korobkin2012}, the abundance distributions of relativistic models with
delayed remnant collapse exhibit a trough around $A\simeq 200$
(Fig.~\ref{fig_delayed-collapse}). This deep trough is absent in the prompt collapse cases
(Fig.~\ref{fig_prompt-collapse}), where the relative contributions of quickly expanding trajectories
to the final abundance distribution are found to be appreciable and to fill the trough in the
$A\simeq 200$ region.

\begin{figure}
\includegraphics[width=84mm]{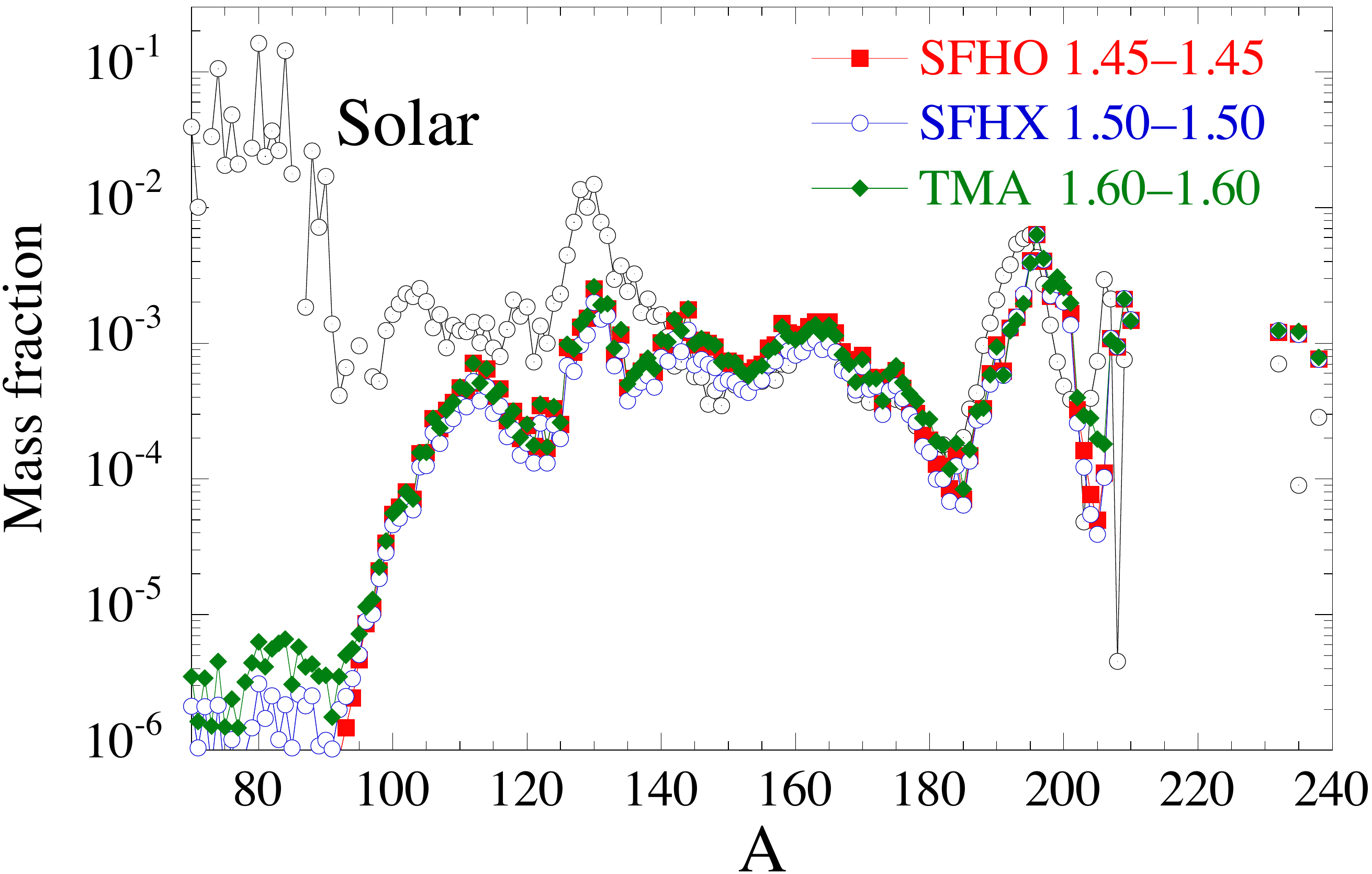}
\caption{Abundance distributions as functions of the atomic mass for the dynamical ejecta of three
  NS merger cases with delayed collapse of the remnants.  Each binary system is characterized in the
  legend by the EOS used in the simulation and the mass (in $M_{\odot}$) of the two NSs. All
  distributions are normalized to the same $A=196$ abundance. The dotted circles show the solar
    r-abundance distribution \citep{Goriely1999a}.}
 \label{fig_delayed-collapse}
\end{figure}

\vskip 0.2cm
\noindent {\it Comparison between NS-NS and NS-BH nucleosynthesis}

\begin{figure}
\includegraphics[width=84mm]{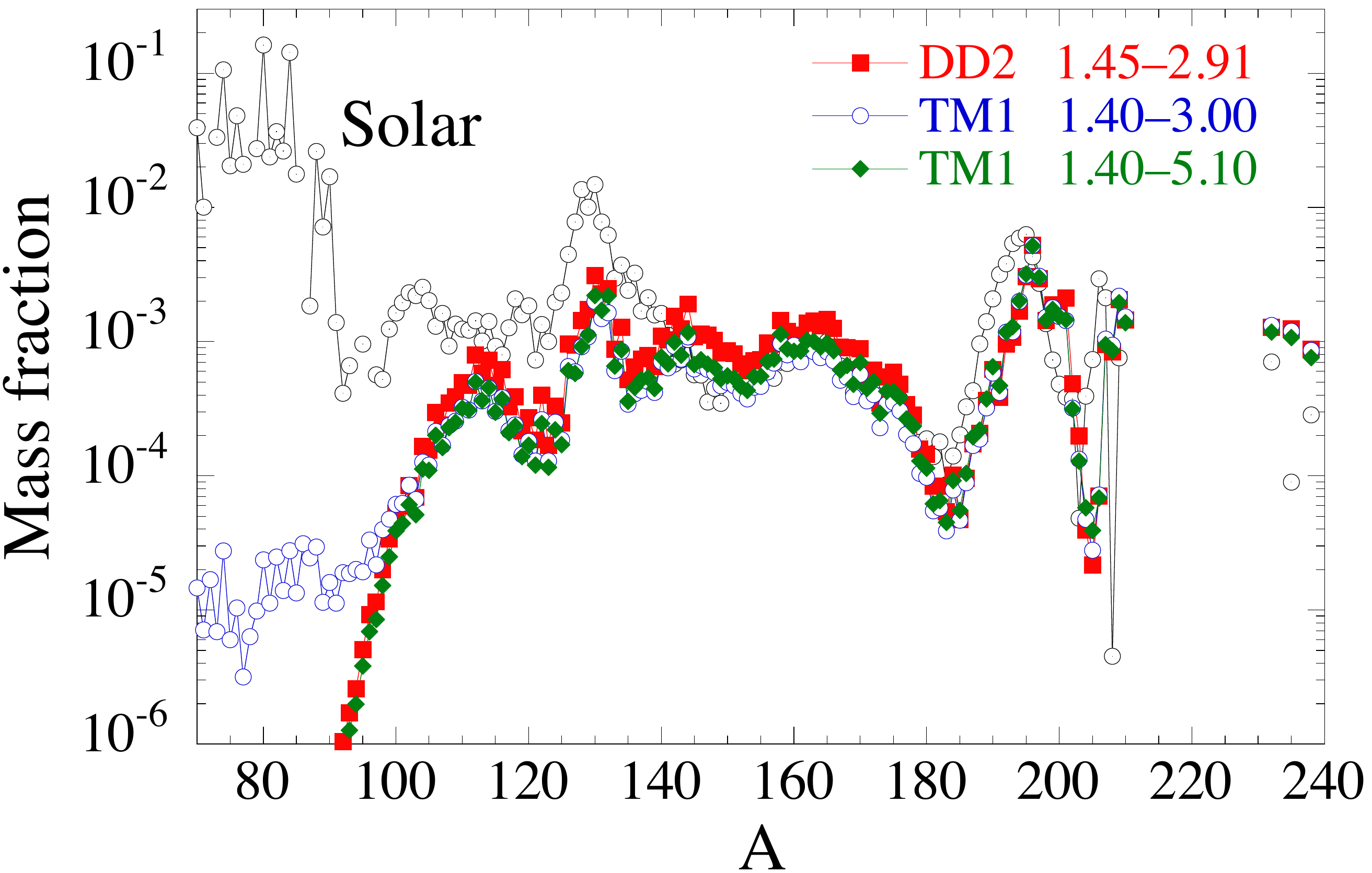}
\caption{Abundance distributions as functions of the atomic mass for the dynamical ejecta of three
  NS-BH merger cases. Each binary system is characterized in the legend by the EOS used in the
  simulation and the mass (in $M_{\odot}$) of the NS and BH, respectively. All distributions are
  normalized to the same $A=196$ abundance. The dotted circles show the solar r-abundance
    distribution \citep{Goriely1999a}.}
 \label{fig_NS-BH}
\end{figure}

Figure~\ref{fig_NS-BH} illustrates the abundance distributions representative of three NS-BH binary
systems. In the NS-BH cases, the expansion velocities are rather low (see
Table~\ref{table_mergers}), comparable to the delayed collapse cases of the NS-NS systems and lower than
those in the prompt collapse cases. 
The corresponding abundance distributions are therefore rather similar to
those found in the NS-NS delayed collapse with an underproduction of the $A\simeq 202$ nuclei.

\vskip 0.2cm
\noindent {\it Impact of neutrinos}

During the binary merging phase we disregard the effects of neutrinos in our hydrodynamic
simulations as well as in the nucleosynthesis studies. In a few recent simulations with
nucleosynthesis discussion neutrino effects were explored. While~\cite{Korobkin2012} did not find a
significant impact in their Newtonian studies, the relativistic NS-NS merger calculations
of~\cite{Wanajo2014a} for a ``soft'' nuclear EOS (i.e., neutron stars with small radii) point
towards a substantial impact of neutrino reactions on the electron fraction $Y_e$ in the dynamical
ejecta of cases with delayed collapse of the merger remnant. While positron captures on the abundant
neutrons in shock-heated matter stripped by the neutron stars must be expected to raise $Y_e$,
\cite{Wanajo2014a} also reported an additional $Y_e$ increase by electron neutrino absorption in
escaping ejecta. In essence, these weak interactions widen the distribution of $Y_e$ in the outflow
such that in addition to heavy r-process elements ($A>140$) also nuclei with lower mass numbers are
created. It is still unclear whether the observed effects apply similarly strongly to all
high-density EOSs and all binary systems that lead to a delayed collapse of the merger remnant.
Also the prompt collapse scenario and NS-BH mergers have not yet been investigated in all relevant
aspects in this context. The impact of neutrinos may be less important in the latter two cases,
because mass ejection there can be faster and more immediate so that strong shock heating
(associated with violent pulsations of the compact remnant or the collision of torus matter with
itself) affects only a smaller fraction of the ejecta, and because of the absence of a hot, massive
central object emitting copious neutrinos. For NS-BH mergers this expectation seems to receive
confirmation by recent results in \cite{Foucart2014}.

In any case, the simulations by \cite{Wanajo2014a} confirm that the dynamical ejecta are a robust
source of heavy r-process nuclei, but in addition may also produce a contribution to r-process
material with $A<140$. The relative amounts of nuclei above and below $A\sim 140$, however, are
massively affected by the long-lasting mass loss from the merger remnant that follows the immediate
ejection of material during the merging phase and early afterwards. As we will demonstrate below for
BH-torus systems, the final nucleosynthetic abundance distribution depends strongly on the detailed
properties of the BH-torus configuration and the spatial asymmetry of the ejecta expelled by the
binary merger and post-merger remnant.

\subsubsection{Nucleosynthesis in the disk ejecta}\label{sec:nucl-disk-ejecta}
In modeling the hydrodynamics and nucleosynthesis of the remnant ejecta, we consistently included
the electron and positron captures on free nucleons as well as the inverse captures of electron
neutrinos and antineutrinos. Note that the neutrino absorption rates (per nucleon) as obtained in
the hydrodynamic torus simulations were provided as functions of time along the tracer trajectories
for all nucleosynthetically processed mass elements of the ejecta. This procedure ensured that the
asymmetry of the neutrino emission was taken into account in its effects on the element formation.

\vskip 0.2cm
\noindent {\it Neutrino-driven and viscous components}

\begin{figure}
\includegraphics[width=84mm]{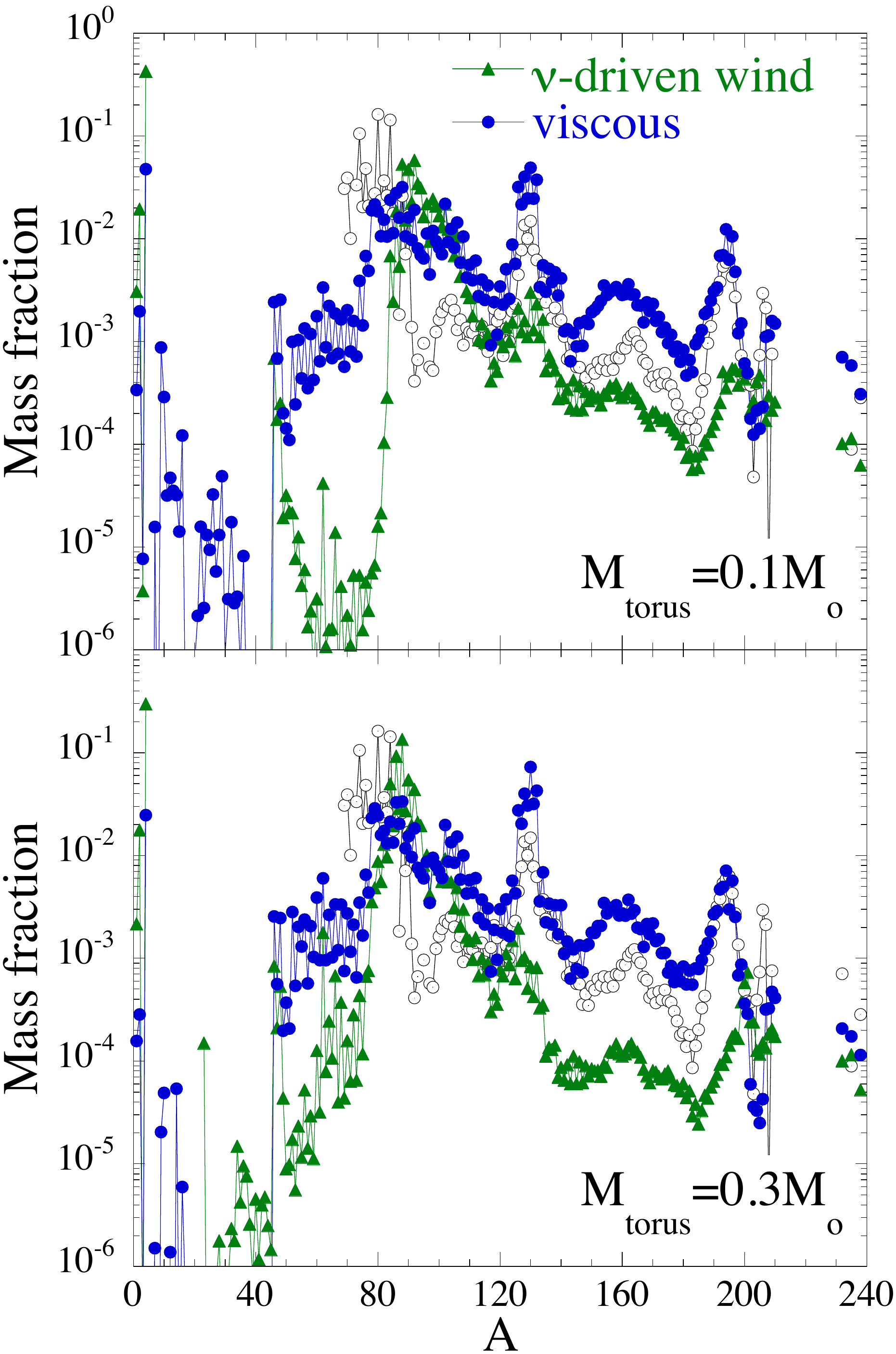}
\caption{Abundance distributions as functions of the atomic mass for the neutrino-driven and viscous
  components of the disk ejecta for tori of 0.1\,\Msun\ (top) and a 0.3\,\Msun\ (bottom). All
  distributions are normalized so that $\sum_X=1$. Calculations correspond to the M3A8m1a5 and
  M3A8m3a5 models. The dotted circles show the solar r-abundance distribution \citep{Goriely1999a}.}
 \label{fig_w+v} 
\end{figure}

The neutrino-driven and viscous outflow components identified in Sect.~\ref{sect3_remnant} are
characterized by different properties, i.e., the neutrino-driven component exhibits larger average
electron fractions and higher escape velocities (cf.\ Fig.~\ref{fig:torus_hist} and
Sect.~\ref{sec:ejecta-properties}). Consequently, the abundance patterns are different with less
strong r-processing in the neutrino-driven ejecta (Fig.~\ref{fig_w+v}). In the neutrino-driven wind,
the trajectories with electron fractions $Y_e\la 0.35$ and with the shortest expansion timescales
can still be responsible for the production of the heavy r-nuclei with $A>140$. In the viscous
ejecta, the average electron fraction is sufficiently lower such that the third abundance peak is
reached for a substantial amount of outflow trajectories. In all studied cases, the mass of the
neutrino-driven outflow remains small compared to the one associated with the viscous component
(cf. Table~\ref{table_remnants}). The final, ejected (combined) abundance distribution in the disk
outflows is therefore essentially identical to the viscous component. The ejected matter is roughly
composed of 80 to 94\% of r-process material, the remaining 6 to 20\% being made essentially of
$^4$He.

\vskip 0.2cm
\noindent {\it Sensitivity to global parameters}

\begin{figure}
\includegraphics[width=84mm]{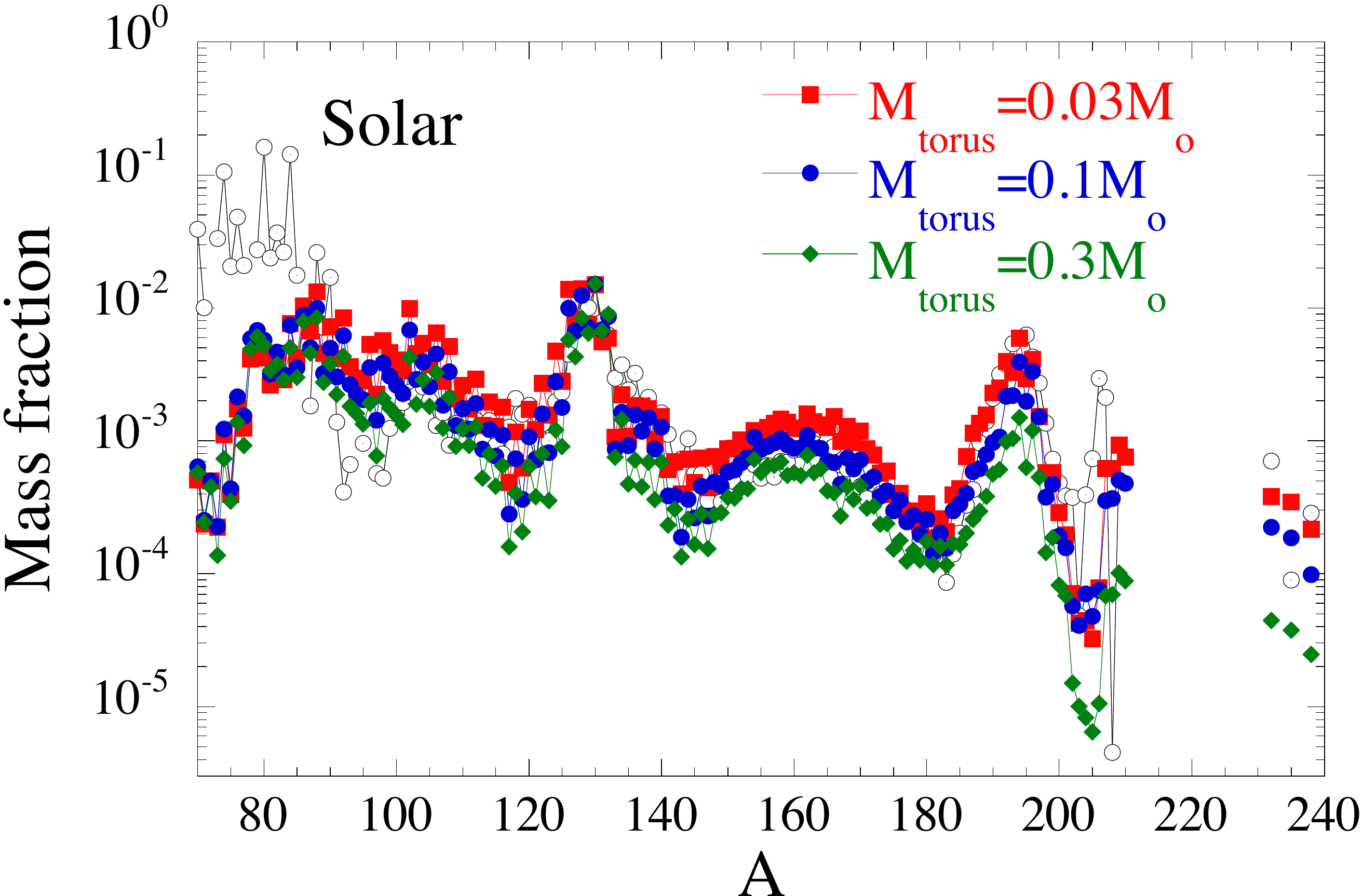}
\caption{Abundance distributions as functions of the atomic mass for three systems with torus masses
  of $M_{\mathrm{torus}}=0.03$, 0.1 and 0.3\,\Msun\ and a 3\,$M_\odot$ BH. All distributions are
  normalized to the same solar $A=130$ abundance. Calculations correspond to the M3A8m03a5, M3A8m1a5
  and M3A8m3a5 models. The dotted circles show the solar r-abundance distribution \citep{Goriely1999a}.}
 \label{fig_Mtorus}
\end{figure}

\begin{figure}
\includegraphics[width=84mm]{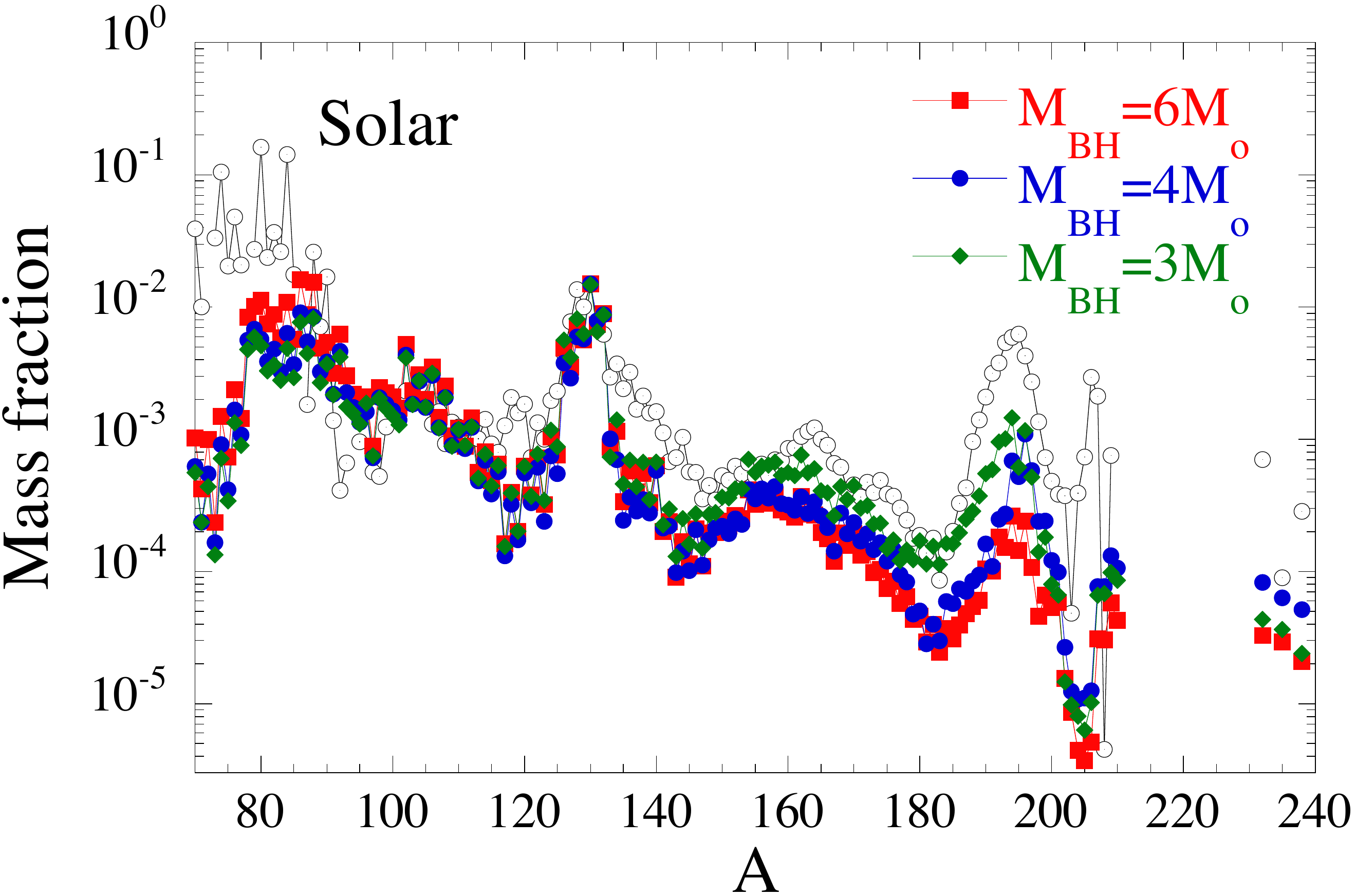}
\caption{Abundance distributions as functions of the atomic mass for three BH-torus systems with BH
  masses of $M_{\mathrm{BH}}= 3$, 4 and 6\,\Msun\ and the same 0.3\,\Msun\ tori. All distributions
  are normalized to the same solar $A=130$ abundance. Calculations correspond to the M3A8m3a5,
  M4A8m3a5 and M6A8m3a5 models. The dotted circles show the solar r-abundance distribution
    \citep{Goriely1999a}.}
 \label{fig_MBH} 
\end{figure}

\begin{figure}
\includegraphics[width=84mm]{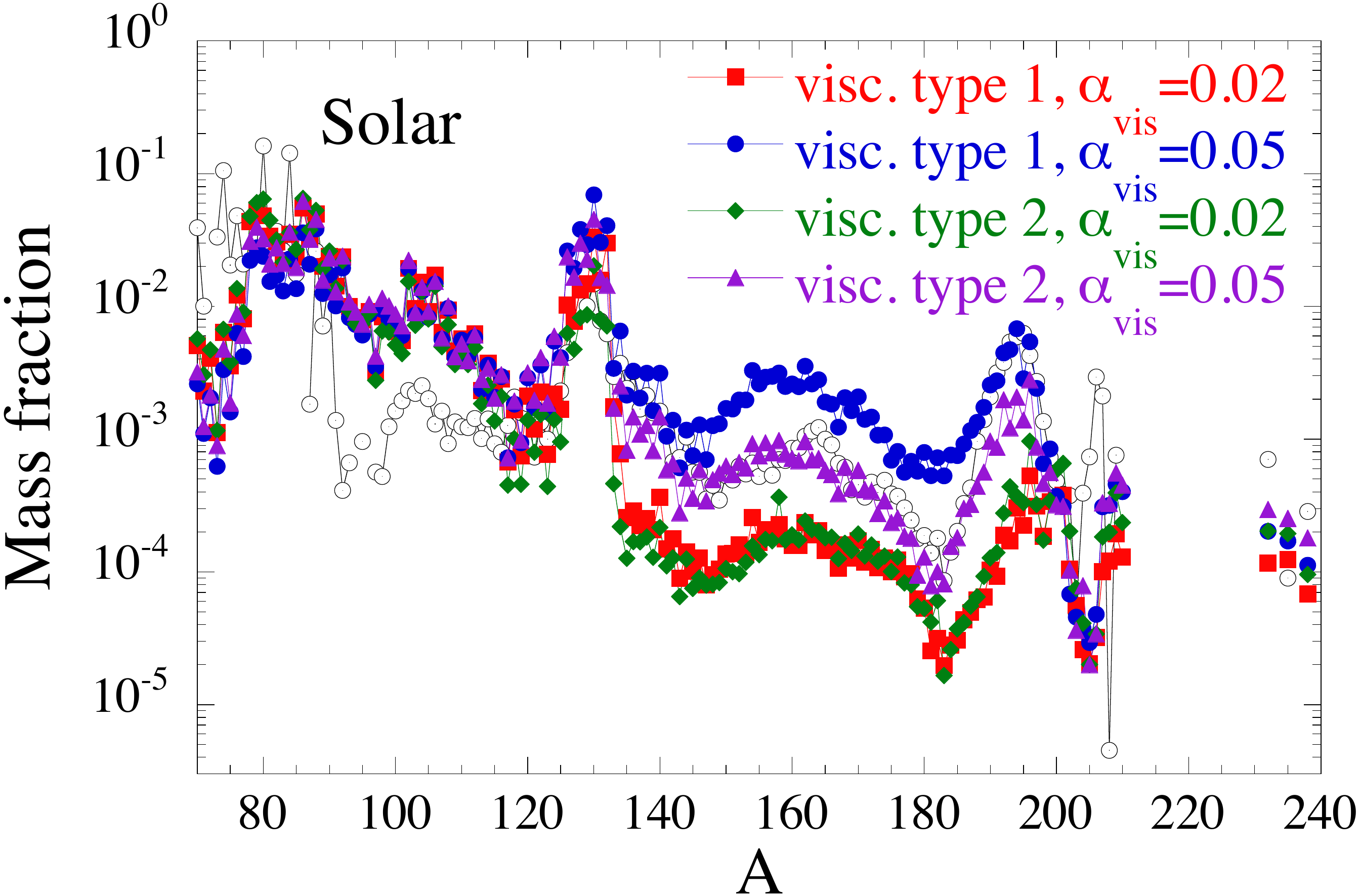}
\caption{Abundance distributions as functions of the atomic mass for four identical BH-torus systems
  ($M_{\mathrm{BH}}=3$\Msun; $A_{\mathrm{BH}}=0.8$; $M_{\mathrm{torus}}=0.3$\Msun) computed with two
  different values of the viscosity parameter $\alpha_{\mathrm{vis}}$, namely
  $\alpha_{\mathrm{vis}}=0.02$ and $\alpha_{\mathrm{vis}}=0.05$, and two different prescriptions of
  the viscosity tensor (cf. Sect.~\ref{sec:choice-models-coher}). All distributions are normalized
  so that $\sum_X=1$. Calculations correspond to the M3A8m3a2, M3A8m3a5, M3A8m3a2-v2 and M3A8m3a5-v2
  models. The dotted circles show the solar r-abundance distribution \citep{Goriely1999a}.}
 \label{fig_viscosity}
\end{figure}

The abundance distributions are found to be only weakly sensitive to the torus mass, as shown in
Fig.~\ref{fig_Mtorus} for cases with the same BH mass and spin and the same viscosity.  The
differences result from the two subtle trends that the fraction of material with $Y_e<0.2$ as well
as the mean entropy slightly increase for lower torus masses. As can be seen in Fig.~\ref{fig_MBH},
the abundance distribution is also found to be only moderately sensitive to the BH mass. The
observed slight trend towards relatively heavier elements for lower BH masses can be ascribed to the
lower mean electron fractions of the ejecta.
In Fig.~\ref{fig_viscosity}, we compare the abundance distributions obtained with two values of the
viscosity parameter $\alpha_{\mathrm{vis}}$, both for the type 1 and type 2 prescriptions, as
described in Sect.~\ref{sec_numerics_remnant}. In general terms, the abundance distribution is quite
robust with respect to the viscosity treatment for the intermediate-mass elements $80\le A \le 130$,
while it is rather sensitive to the viscosity for the $A>130$ elements. We observe that for a higher
dynamic viscosity coefficient $\eta_{\mathrm{vis}}$ a higher relative amount of $A>130$ elements is
obtained (remembering also that the $\eta_{\mathrm{vis}}$ for the two viscosity types are related by
Eq.~\ref{eq:dynvis}). This result can be explained by the fact that a higher dynamic viscosity leads
to a smaller mean electron fraction $\bar{Y}_e$ of the ejecta in our models (cf.\
Table~\ref{table_remnants}).


\vskip 0.2cm
\noindent {\it  Sensitivity to r-process heating feedback}

\begin{figure}
\includegraphics[width=84mm]{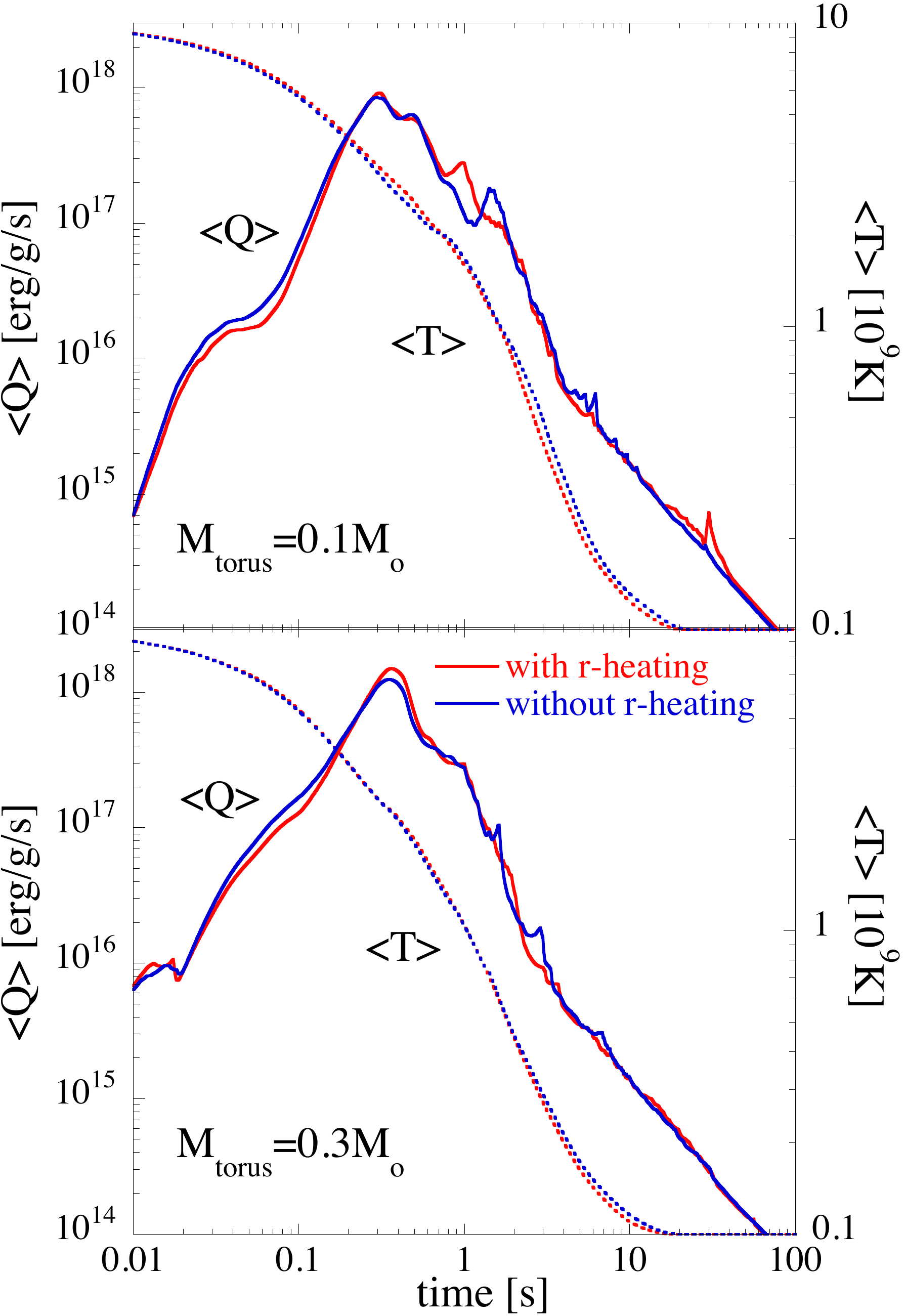}
\caption{Time evolution of the average radioactive heating rate per unit mass, $\langle Q \rangle$,
  for two BH-torus systems with $M_{\mathrm{BH}}=3\,M_\odot$, $A_{\mathrm{BH}}=0.8$,
  $M_{\mathrm{torus}}=0.1\,M_\odot$ (top) and $M_{\mathrm{BH}}=4\,M_\odot$, $A_{\mathrm{BH}}=0.8$,
  $M_{\mathrm{torus}}=0.3\,M_\odot$ (bottom), when the heating feedback due to the r-process
  $\beta$-decays and fission is included or not. Calculations correspond to models M3A8m1a2,
  M3A8m1a2-rh, M4A8m3a5 and M4A8m3a5-rh.}
 \label{fig_Qheat} 
\end{figure}

\begin{figure}
\includegraphics[width=84mm]{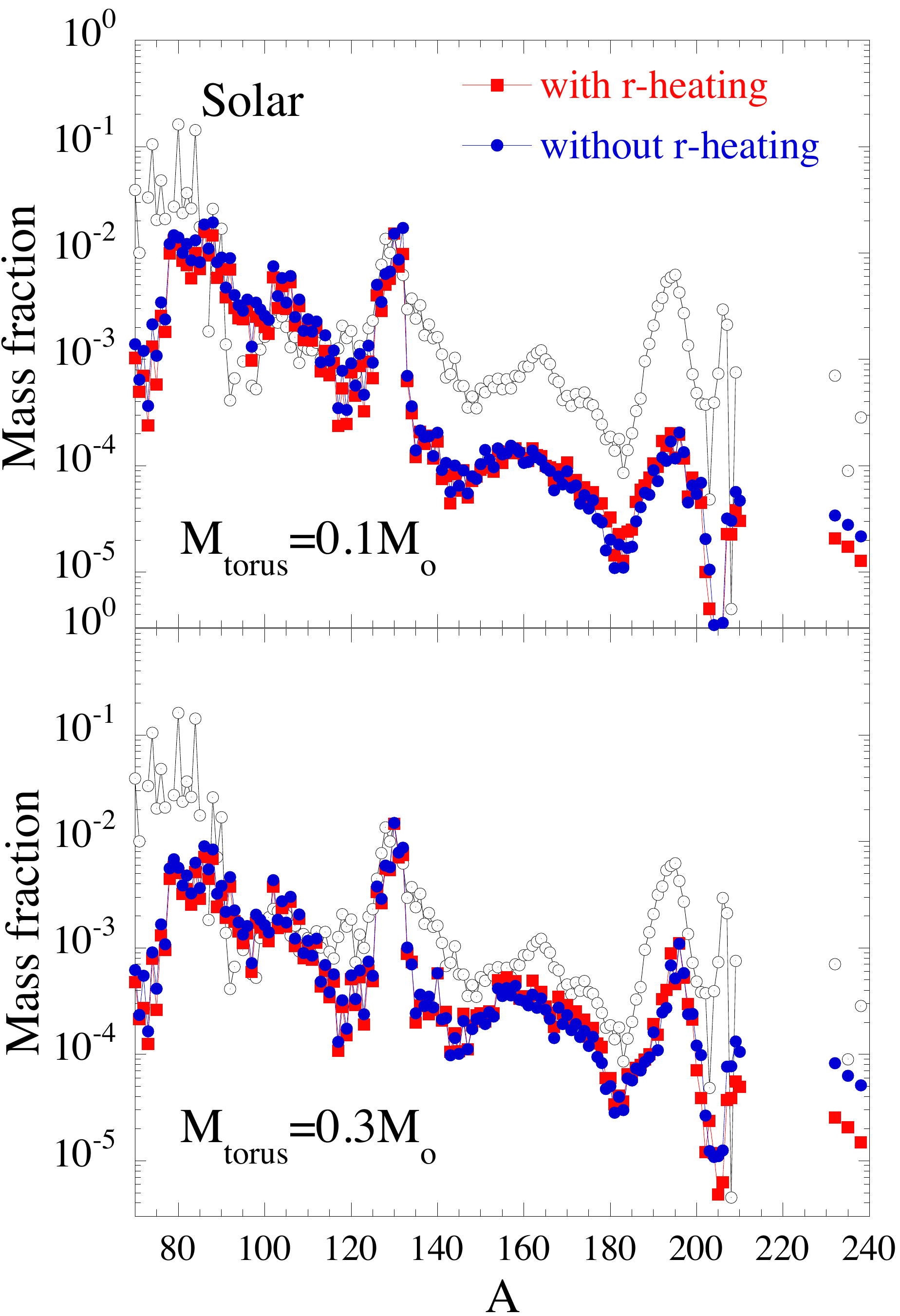}
\caption{Abundance distributions as functions of the atomic mass for two BH-torus systems with
  $M_{\mathrm{BH}}=3\,M_\odot$, $A_{\mathrm{BH}}=0.8$, $M_{\mathrm{torus}}=0.1\,M_\odot$ (top) and
  $M_{\mathrm{BH}}=4\,M_\odot$, $A_{\mathrm{BH}}=0.8$, $M_{\mathrm{torus}}=0.3\,M_\odot$ (bottom),
  when the heating feedback due to the r-process $\beta$-decays and fission are included or not.
  All distributions are normalized to the same solar $A=130$ abundance. Calculations correspond to
  models M3A8m1a2, M3A8m1a2-rh, M4A8m3a5 and M4A8m3a5-rh. The dotted circles show the solar
  r-abundance distribution \citep{Goriely1999a}.}
 \label{fig_reheating}
\end{figure}

In Fig.~\ref{fig_Qheat} we compare the average temperatures as well as the average heating rates for
the models with and without radioactive heating as implemented by the approximate method described
in Sect.~\ref{sec:choice-models-coher}. Correspondingly, in Fig.~\ref{fig_reheating} we compare the
abundance distributions for these models. The variations introduced by such a lowest-order
correction for radioactive heating are only marginal. Besides minor
differences in the abundance distributions, also the total ejecta masses are hardly affected. With
heating $M_{\mathrm{out}}$ increases from $22.1$ to $22.4$ per cent and from $22.7$ to $22.8$ per
cent of the original torus mass for models M4A8m3a5 and M3A8m1a2, respectively (see
Table~\ref{table_remnants}). This indicates that including the radioactive heating in a fully
consistent manner is not necessary or at least does not lead to any significant changes of our
modeling predictions \citep[see also][where a similar conclusion was drawn concerning the dynamical
ejecta]{Rosswog2014}.

\subsubsection{Combined nucleosynthesis in the dynamical and disk ejecta}
\begin{figure}
\includegraphics[width=84mm]{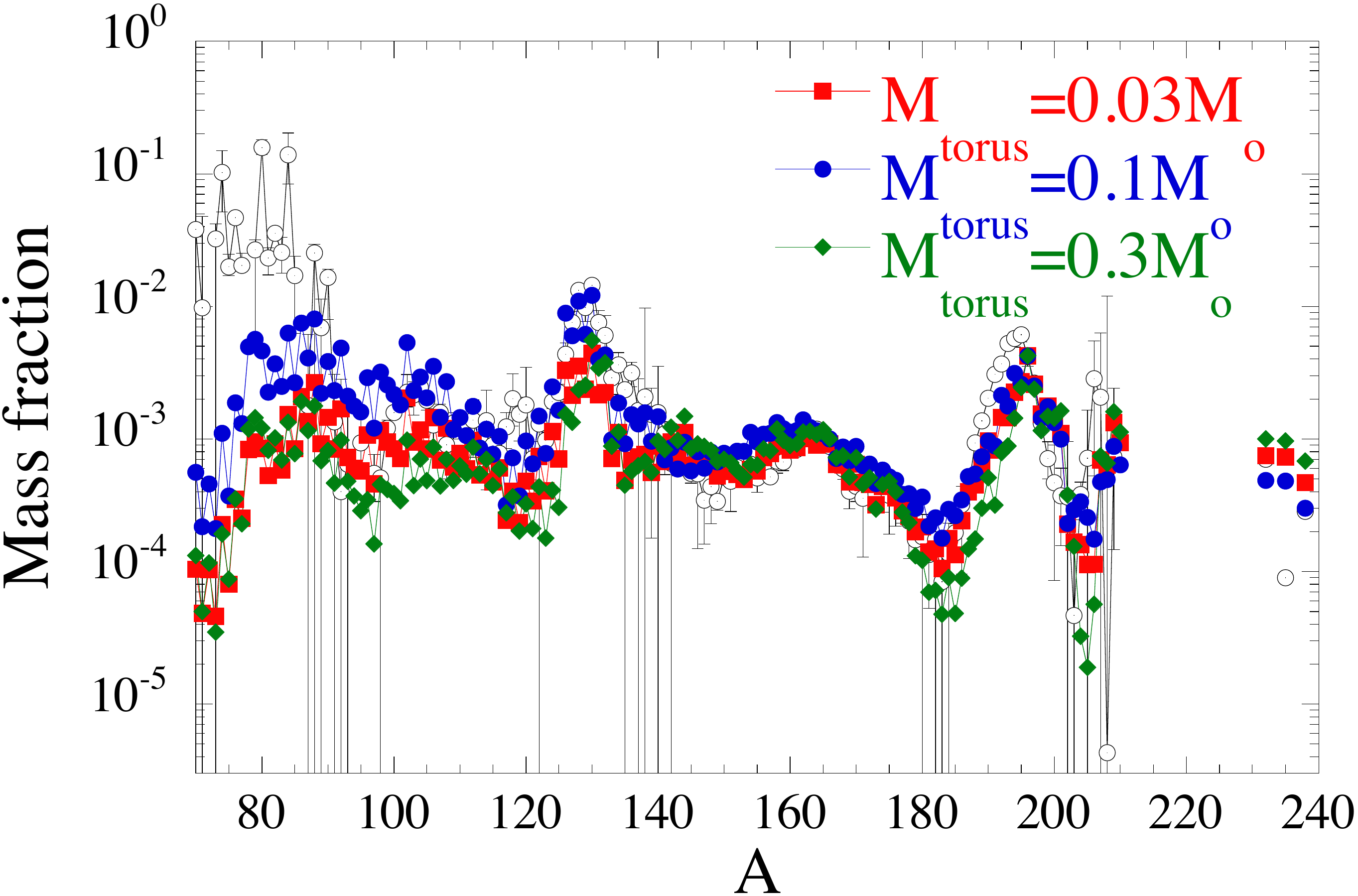}
\caption{Abundance distributions as functions of the atomic mass for three combined systems (merger
  model plus remnant model) corresponding to models with torus masses $M_{\mathrm{torus}}=0.03, 0.1$
  and 0.3\,\Msun. All distributions are normalized to the same solar $A=196$ abundance. Calculations
  correspond to the model combinations TMA\_1616--M3A8m03a5, SFHO\_13518--M3A8m1a5 and
  DD2\_14529--M4A8m3a5. The dotted circles show the solar r-abundance distribution
    \citep{Goriely1999a}.}
 \label{fig_dyn+disk}
\end{figure}

Assuming that the dynamical and disk nucleosynthesis components are both ejected isotropically, we
can combine the yields by summing up the mass fractions for both components, weighted by their
corresponding total ejected masses. We only combine systems corresponding to the same BH mass and
spin and the same torus mass in a roughly consistent manner. For models with 0.03, 0.1 and
0.3\,\Msun\ tori, the abundance distributions for such combined models are given in
Fig.~\ref{fig_dyn+disk}. The dynamical ejecta contribute mainly to the production of the $A>140$
nuclei, whereas the disk ejecta produce mostly the $90\le A \le 140$ nuclei.  The combined
distribution is in surprisingly good agreement with the solar system r-abundance distribution
(Fig.~\ref{fig_dyn+disk}). However, the relative strength of the second to the third r-process peak
depends sensitively on the ratio between the amount of mass ejected from the disk and the one
ejected dynamically (both components being composed of about 80$-$98\% of r-process material). For
the three cases shown in Fig.~\ref{fig_dyn+disk}, this ratio amounts to 1.4, 5.3 and 1.7 for the
0.03, 0.1 and 0.3\,\Msun\ torus models, respectively. For this reason, when normalizing to the
$A=196$ abundance, the lighter elements, and in particular the $A\simeq 130$ peak, are found to
typically vary within a factor of 3.

Note that it is well known that calculations of the r-process abundances are still affected by large
nuclear physics uncertainties \citep{Arnould2007}. Such uncertainties have been extensively studied in
the past, but each site provides its specific conditions and behaves in its own special manner so
that an assessment of the sensitivity to theoretical nuclear physics input requires careful and
dedicated exploration. While we already partly investigated the sensitivity of the nucleosynthesis
in the dynamical ejecta to masses, $\beta$-decay rates, and fission probabilities \citep{Goriely2013},
we defer such a sensitivity analysis for the composition of the disk ejecta to a future study.

\section{Comparison with observations}\label{sec:comp-with-observ}
\begin{figure}
\includegraphics[width=84mm]{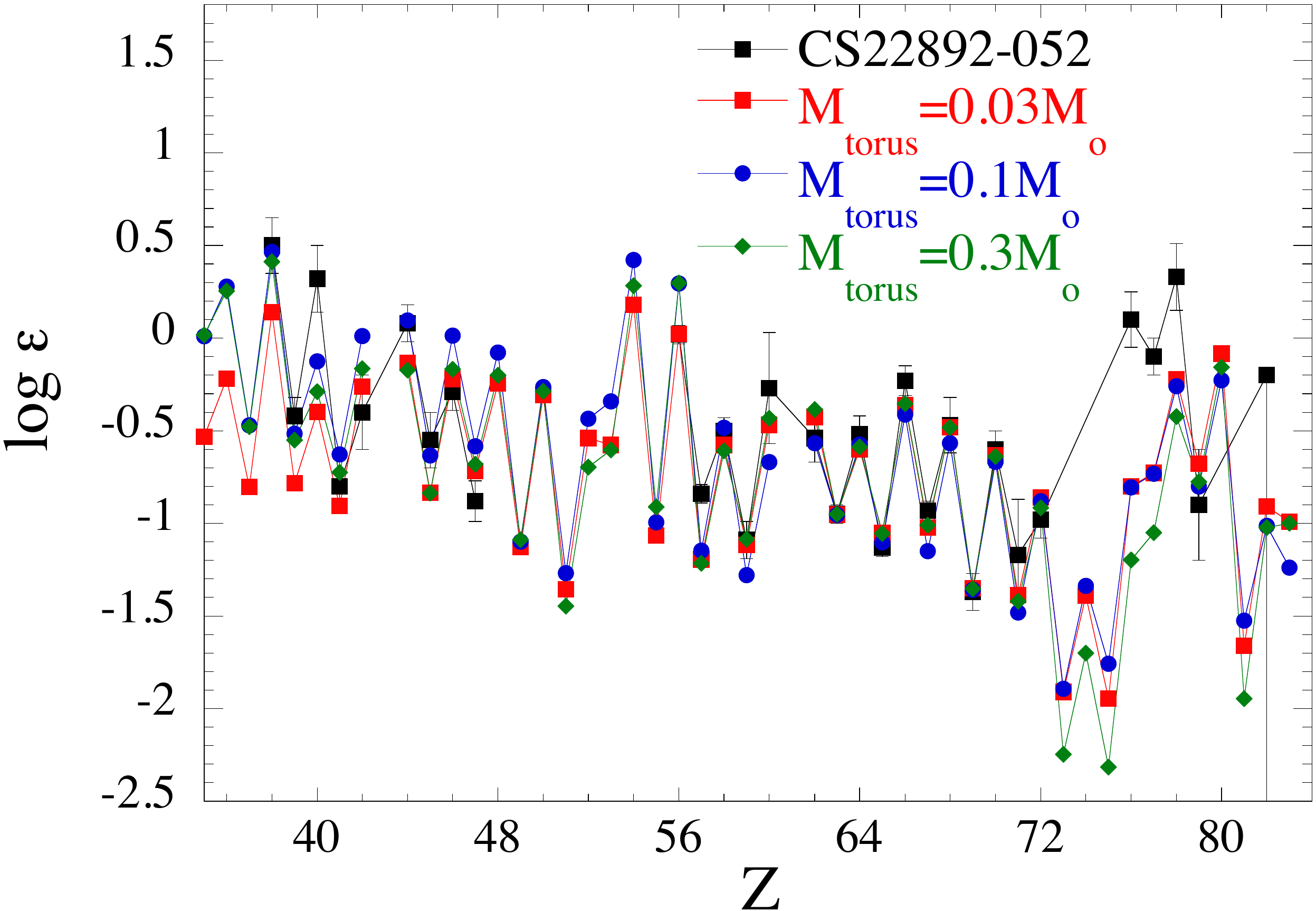}
\caption{Comparison between the elemental abundances of CS22892-052 (in the usual logarithmic scale
  relative to hydrogen, $\log \varepsilon = \log_{10}(N_{\mathrm{A}}/N_{\mathrm{H}}) + 12$ for
  element A and number density $N_{\mathrm{A}}$) and those obtained for the three combined systems
  corresponding to remnants with 0.03, 0.1 and 0.3\,\Msun\ tori as shown in Fig.~\ref{fig_dyn+disk}.
  All distributions are normalized to the observed Eu abundance. Calculations correspond to the
  model combinations TMA\_1616--M3A8m03a5, SFHO\_13518--M3A8m1a5 and DD2\_14529--M4A8m3a5.}
 \label{fig_CS22892}
\end{figure}

The striking similarity between the solar distribution of r-element abundances in the $56 \le Z \le
76$ range and the corresponding abundance pattern observed in ultra-metal-poor stars like CS
22892-052 \citep{Sneden2003,Sneden2008,Sneden2009} led to the conclusion that any astrophysical event
producing r-elements gives rise to a solar system r-abundance distribution, at least for elements
above Ba. In such r-process-enriched low-metallicity stars, some variation of about 0.5 dex,
however, is found for the elements lighter than Ba. The amazingly robust r-process for elements
above Ba could point to the possible creation of these elements by fission recycling in dynamical
NS-NS or NS-BH merger ejecta, as already argued in previous studies \citep[e.g.,][]{Goriely2011}. As
seen in Fig.~\ref{fig_dyn+disk}, variations of the abundances of the lighter elements with $40 \le Z
\le 56$ relative to those of the heavier elements by factors of a few could be accounted for when
the contributions of the disk ejecta in dependence on different torus masses are considered.
Figure~\ref{fig_CS22892} shows that the elemental distribution observed in the ultra-metal-poor star
CS22892-052 \citep{Sneden2003} can be fairly well reproduced by the nucleosynthesis from the combined
dynamical and disk ejecta. Discrepancies are found around the Os elements due to the shift of the
third r-process peak (see Fig.~\ref{fig_dyn+disk}), the exact position of which is affected by
nuclear uncertainties.

Recent observations also indicate that star to star variations in the r-process content of
metal-poor globular clusters may be a common, although not ubiquitous, phenomenon
\citep{Roederer2010,Roederer2011}. Stars such as HD 88609 or HD 122563 have been found to be
significantly deficient in their heavy elements \citep{Honda2007}. CS 22892-052 and HD 122563 are now
interpreted as two extreme cases representative of a continuous range of r-process nucleosynthesis
patterns \citep{Roederer2010}.

\begin{figure}
\includegraphics[width=84mm]{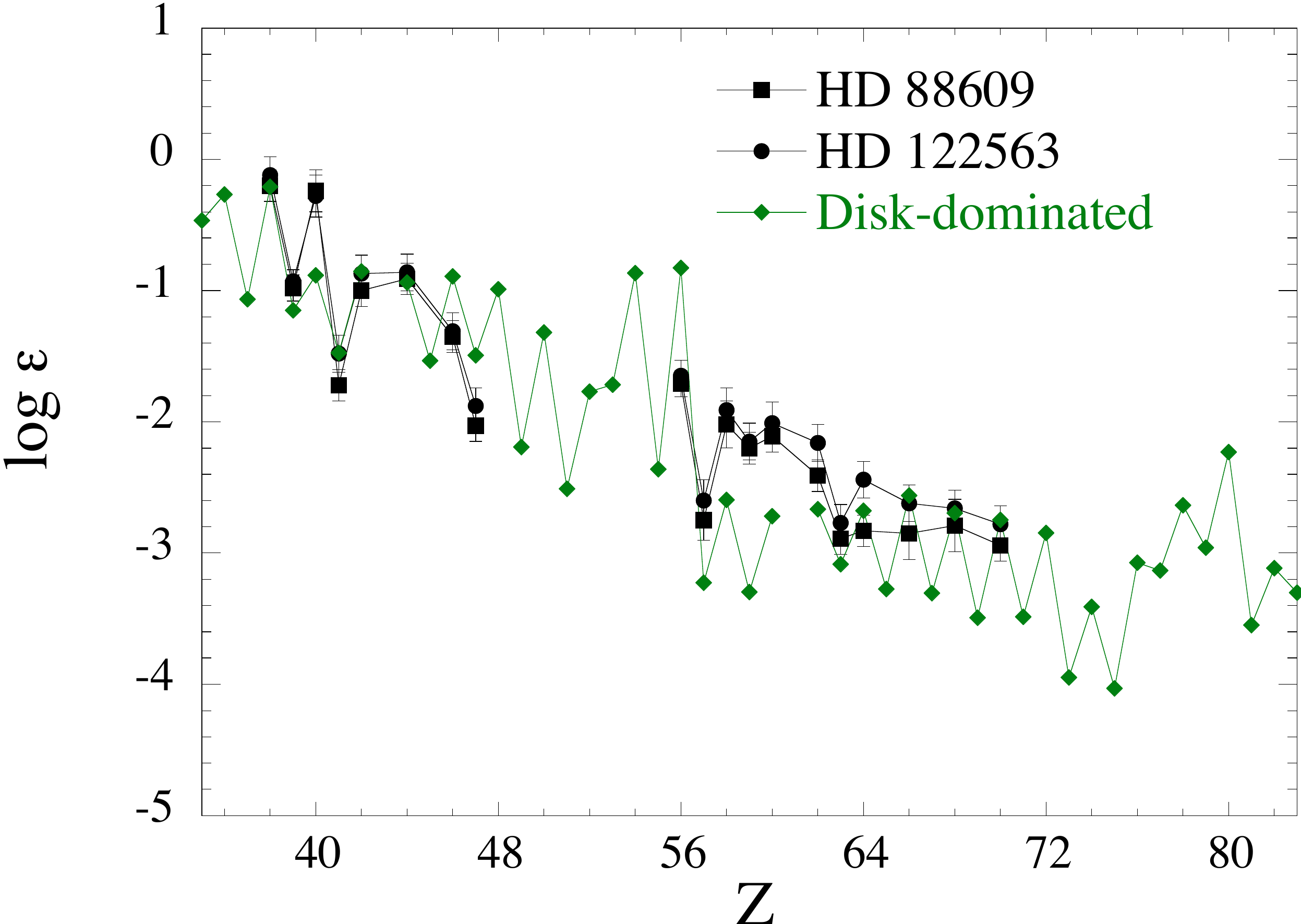}
\caption{Comparison between the HD88609 and HD122563 elemental abundances (in log\,$\varepsilon$
  scale) and those estimated for the combination of models SFHO\_1123 and M3A8m3a2, when the ejected
  mass of the dynamical component is assumed to be 100 times smaller than the one coming from the
  disk component. The calculated distribution is normalized to the Sr abundance of HD88609.}
 \label{fig_HD88609}
\end{figure}

In the previous comparison with CS22892-052 we determined the combined composition of all ejecta
components by weighting both the dynamical and disk contributions by their respective total ejected
masses. However, it cannot be excluded that the mass of the dynamical ejecta contributing to the
final composition is in fact significantly smaller, in particular when considering NS-BH
systems. Two different effects could suppress the dynamical ejecta relative to the torus ejecta. (1)
As discussed in Sect.~\ref{sec:comp-binary-merg}, the dynamical ejecta of NS-BH mergers can be
highly asymmetric, because most of these ejecta may consist of the matter shed off the outer tip of
the tidally stretched NS at its final approach to the BH. In contrast, the remnant ejecta are more
isotropic. The combination of both can therefore be strongly direction dependent. For the values of
the asymmetry parameter $B_{\mathrm{asy}}$ given in Table~\ref{table_mergers} (cf.\
Sect.~\ref{sec:comp-binary-merg}), the mass of the dynamical ejecta could be 20--100 times smaller
than the mass of the disk ejecta outside of the solid angle of the main dynamical mass stripping.
(2) Potentially, NS-BH mergers might produce little dynamical ejecta material for certain
  binary parameters while still forming a torus.  This hypothetical possibility might exist in cases
  where the NS is nearly completely and immediately accreted by the BH, in which case the tidal
  sling effect leading to mass ejection from an extremely stretched NS might be absent. In such
systems, the ejecta would consequently be essentially composed of disk material. If we assume that
the contribution of the dynamical ejecta represents only about 1\% of the total ejected mass, the
composition of the ejecta becomes strongly depleted in heavy r-process material in comparison to the
standard cases shown in Figs.~\ref{fig_dyn+disk} and \ref{fig_CS22892}. Such specific system
conditions could qualitatively explain the composition of stars like HD 88609 or HD 122563, as shown
in Fig.~\ref{fig_HD88609}. We consider this as an interesting, speculative possibility, but the
relative frequency of such possible events as well as the possible variation of the mix between the
dynamical and disk contributions will need to be assessed in a more quantitative way in future
studies.

\section{Summary, discussion, and conclusions}\label{sec:summ-disc-concl}

We have performed the first comprehensive study of r-process nucleosynthesis in merging NS-NS and
NS-BH binaries, also including the neutrino and viscously driven outflows from BH-torus systems as
remnants of the compact star mergers. Our focus was on such relic systems because they are the
generic outcome of NS-BH mergers when the BH/NS mass ratio is not too large, and for NS-NS mergers
in cases where the massive remnant cannot be stabilized by the NS EOS. We therefore considered
binary NSs that either led to BH formation directly when the two stars plunged into each other, or
produced a remnant that collapsed to a BH within less than about 10\,ms for the nuclear EOSs
employed in our work.

The binary mergers were simulated in 3D with a relativistic SPH hydrodynamics code, using the CFC
approximation for the spacetime \citep{Oechslin2007,Bauswein2010e} and four different microphysical
EOSs for non-zero temperature NS matter (all consistent with astrophysical, experimental and
theoretical bounds). The subsequent evolution of the BH-torus systems was computed over periods of
up to more than 10 seconds in 2D (axisymmetry) with a Newtonian, finite-volume Godunov-type code
\citep{Obergaulinger2008} including viscosity terms with a Shakura-Sunyaev $\alpha$-prescription for
the dynamic shear viscosity coefficient \citep{Shakura1973}. The gravitational field of the relic BH
was described by an Artemova-Bjoernsson-Novikov potential \citep{Artemova1996}, and the neutrino
transport in the accretion torus was treated by a 2D, fully energy dependent two-moment scheme with
an algebraic closure for the variable Eddington tensor (Just et al., in preparation).

Although the binary-merger and remnant modeling were not carried out continuously and not fully
consistently, we ensured compatibility of both evolution phases by connecting the investigated
BH-torus models with NS-NS and NS-BH binaries that produced remnants with the same (or at least very
similar) BH masses ($M_\mathrm{BH}\approx 3,\,4,\,6\,M_\odot$), torus masses
($M_\mathrm{torus}\approx 0.03,\,0.1,\,0.3\,M_\odot$), and BH spins ($A_\mathrm{BH}\approx 0.8$).
Starting with tori in rotational equilibrium minimized artificial transients. The action of viscous
angular momentum transport and redistribution establishes quasi-stationary accretion typically
within 10--20\,ms, and dissipative viscous heating and fast electron and positron captures on free
nucleons quickly (within $<$10\,ms) drive the initially chosen temperatures and electron fractions
in the tori to values that correspond to thermal and kinetic equilibrium conditions. Except for a
short period at the beginning of the merger-remnant simulations, the long-time evolution of the tori
therefore mostly depends on the applied value of the $\alpha$-viscosity and on the neutrino
transport, and only to a minor degree on the exact initial conditions.

Consistent with previously published conformally flat \citep[e.g.,][]{Oechslin2007, Goriely2011,
  Bauswein2013} and fully relativistic results \citep[e.g.,][]{Hotokezaka2013, Hotokezaka2013a,
  Kyutoku2011, Kyutoku2013, Deaton2013, Foucart2014, Wanajo2014a}, the ``dynamical'' ejecta of NS-NS
and NS-BH mergers, which are expelled within milliseconds of the collision of the two binary
components, were found to possess similar average properties, namely expansion velocities of
0.2--0.4\,$c$, electron fractions below $\sim 0.1$, and entropies per baryon of a few
$k_\mathrm{B}$. The considered NS-NS mergers produce $\sim$0.004--0.021\,$M_\odot$ of ejecta,
whereas the NS-BH mergers eject significantly larger masses, 0.035--0.08\,$M_\odot$, with very low
entropies ($\la 1\,k_\mathrm{B}$ per nucleon), because this matter is not shock heated as in NS-NS
collisions, but originates mostly from the outer tail of the tidally stretched NS at its final
approach to the BH. Mass lost in NS-BH mergers is also expelled much more asymmetrically than in the
case of NS-NS mergers: Corresponding hemispheric asymmetry parameters (mass difference between
dominant ejecta hemisphere and opposite hemisphere, divided by total ejecta mass) are a few per cent
for symmetric NS-NS mergers and 15--30\% for strongly asymmetric ones, but 0.93--0.98 for NS-BH
mergers.

Since the high neutron excess, thermodynamic properties, and expansion timescale are very similar,
the ejecta of NS-NS mergers as well as those of NS-BH mergers are sites of robust production of
r-nuclei with $A\ga 140$ and abundances close to the solar distribution. This result holds basically
independently of the considered nuclear EOS and the exact binary parameters and confirms the
findings of previous studies based on different kinds of relativistic merger simulations
\citep[e.g.,][]{Goriely2011, Bauswein2013, Hotokezaka2013, Wanajo2014a}.

The relic BH-torus systems lose mass in neutrino-driven baryonic winds \citep[e.g.,][]{Wanajo2012}
and in outflows triggered by viscous energy dissipation and angular momentum transport
\citep[e.g.,][]{Fernandez2013}. For the first time we quantitatively investigated these outflows by
self-consistent simulations of the accretion-torus evolution with multi-dimensional, energy
dependent neutrino transport. We found that the viscously driven ejecta amount up to, fairly model
independent, 19--26\% of the initial torus mass (for our BH spins of $A_\mathrm{BH} = 0.8$) 
and dominate the neutrino-driven ejecta by far. The neutrino-driven wind depends
extremely sensitively on the BH mass and on the torus mass, which determines the neutrino
luminosities. The neutrino energy loss rates can reach several 10$^{52}$\,erg\,s$^{-1}$ up to more
than 10$^{53}$\,erg\,s$^{-1}$ for each of $\nu_e$ and $\bar\nu_e$ for a few 100\,ms.  The
neutrino-driven ejecta carry away up to about one per cent of the initial torus mass, but their mass
can also be orders of magnitude lower. The maximum masses that can be associated with ejecta driven
by neutrino heating are 2.5--3.5$\times 10^{-3}$\,$M_\odot$ in the case of high neutrino
luminosities for typical durations of fractions of a second up to $\sim$1\,s. This number agrees
with the lower limit of the expelled mass computed by \citet{Perego2014a} for a hypermassive NS as
merger remnant, and the corresponding mass-loss rate is roughly compatible with the values obtained
for neutrino winds of proto-neutron stars in supernova cores \citep[e.g.,][]{Qian1996}.  Neutrino
heating, however, inflates the outer layers of the torus and has a positive feedback on the
viscously driven mass ejection on the level of several per cent of the torus mass (or up to
$\sim$20\% of the viscous-outflow mass).

Neutrino-driven torus winds exhibit characteristic properties which distinguish them from the
viscously triggered outflow, namely the tendency of higher mean entropies, higher electron
fractions, and larger expansion velocities (see Figs.~\ref{fig:torus_flux}, \ref{fig:torus_hist}).
Moreover, the neutrino wind is strongest at early times and at intermediate latitudes (around
45$^\circ$ away from the equatorial plane), and the entropy, electron fraction, and velocity exhibit
a strong pole-to-equator variation with higher values towards the poles
(Fig.~\ref{fig:ejecta_angledep}). In contrast, viscously driven ejecta develop on longer timescales,
are more spherical, and their properties vary little with angular direction.

Because of their greater neutron excess, viscously-driven ejecta allow for a much stronger r-process
than the neutrino wind. The combination of both components is far dominated by the viscous
contribution and matches the solar abundance pattern for all nuclear mass numbers $A \ga 90$ fairly
well. The abundance pattern for $A \le 132$ is comparatively uniform, but the strength of the third
abundance peak decreases with higher BH mass (Fig.~\ref{fig_MBH}), and the relative yields of
low-mass ($A \la 130$) and high-mass ($A \ga 130$) components depend on the value of the dynamic
viscosity and the detailed treatment of the viscosity terms in the hydrodynamics equations
(Fig.~\ref{fig_viscosity}).

A mass-weighted combination of the dynamical ejecta from the binary merger phase and the secular
ejecta from the BH-torus evolution can reproduce the solar r-abundance pattern and therefore also
that seen in ultra-metal-poor stars amazingly well in the range $90\la A < 240$ (cf.\
Figs.~\ref{fig_dyn+disk}, \ref{fig_CS22892}). In particular, the BH-torus outflows are able to well
fill the region $A \la 140$, where the prompt merger ejecta underproduce the nuclei. Since the
relative yields of the relic BH-torus systems for $A \la 130$ and $A\ga 130$ nuclei depend
sensitively on the system parameters, whereas the $A\ga 140$ species are created with a robust
pattern during the binary merging phase, we expect a larger variability in the low-$A$ regime than
for high mass numbers (see Fig.~\ref{fig_dyn+disk}).

Another interesting possibility is connected to the fact that the mass ejection during NS-BH mergers
shows extreme spatial asymmetry (corresponding to asymmetry parameters $\ga$0.95) but the mass loss
from their BH-torus remnants is much more isotropic.  This can lead to a strong suppression of the
dynamical ejecta component relative to the torus outflow in observer directions pointing away from
the hemisphere that receives most of the expelled matter of the disrupted NS.  Essentially pure
BH-torus ejecta of some of our models can reasonably well match the abundance distributions observed
in heavy-element deficient metal-poor stars like HD88609 and HD122563 (Fig.~\ref{fig_HD88609}).

Since the major part of the torus ejecta ends up in forming $A\la 130$ material, the additional mass
loss of the merger remnants does not alter event-rate estimates based on comparing yields of heavy
r-nuclei in the ejecta of the binary-merger phase with the r-process abundances in our Galaxy
(e.g. \citealp{Goriely2011}; \citealp{Bauswein2014}).

Regarding the sensitivity of the nucleosynthesis predictions with respect to the nuclear physics
input, we considered, in addition to the most recent data available experimentally, the latest
theoretical rates based as much as possible on microscopic or semi-microscopic nuclear models
\citep{Xu2013}. Such models have been shown to be able to compete with more phenomenological models
as far as the accuracy in reproducing experimental data is concerned, and additionally are believed
to be more predictive for exotic neutron-rich nuclei. Nevertheless, for such exotic nuclei,
significant uncertainties still affect the prediction of the reaction and decay rates, as well as
the fission fragment distributions \citep{Arnould2007,Goriely2013,Goriely2014,Goriely2014a}. While
we already partly investigated the sensitivity of the nucleosynthesis in the dynamical ejecta to
masses, neutron capture rates, $\beta$-decay rates, and fission probabilities
\citep{Goriely2013,Goriely2014,Xu2014}, a detailed sensitivity analysis for the composition of the
disk ejecta from the BH-torus systems remains to be performed in the future. Nuclear physics
uncertainties may affect the calculated abundance distributions, and more specifically, the precise
location and width of the second ($A\sim 130$), rare-earth ($A\sim 165$) and third ($A\sim195$)
peaks, but should not change the overall nucleosynthesis picture, and especially not the successful
r-process production of $A\ga 90$ nuclei.

Our work can represent only a beginning of the combined analysis of binary merger and post-merger
ejecta concerning their nucleosynthetic impact.  The parameter space of system properties (masses,
mass ratios, BH spin values and orientations) and internal properties (defined by the uncertain
nuclear EOS of NS matter) is huge and the involved physics is extremely complex and not fully
accounted for in our simulations.

The merger and remnant models employed in our study are still incomplete in many respects. Instead
of simulating the merger phase and remnant evolution consistently and continuously with the same
numerics in full 3D general relativity, we applied different codes and approximations for these two
evolutionary periods. Neutrino transport was only included for the BH-torus simulations, but we
ignored neutrino effects during the merger phase.  Neutrinos are unlikely to have a strong influence
in cases where NS-NS merger remnants collapse to a BH immediately and matter is ejected only on very
short timescales, and they do not play an important role in NS-BH mergers, where the material
ejected from the disrupted NS is never shock heated. However, \citet{Wanajo2014a} pointed out that
the neutrino emission and reabsorption could well have an impact on the electron fraction of NS-NS
merger ejecta for long(er)-lived massive NS-merger remnants, in particular for soft NS EOSs.
Neutrino oscillations and general relativistic effects (both ignored in our Newtonian torus models)
were found to be of relevance for an exact determination of the neutron-to-proton ratio in
neutrino-driven disk winds \citep[e.g.,][]{Malkus2012, Caballero2012}. However, these complications
may not be overly relevant because neutrino-driven ejecta are dwarfed in mass by the viscous
outflows, which are expelled at large distances from the central BH and whose electron fraction is
governed by electron and positron captures rather than neutrino captures. The low $Y_e$ of this
clearly dominant torus-ejecta component also lowers the perspective of significant nickel production
\citep{Surman2014} in these outflows.

The application of a viscosity description instead of magnetohydrodynamic modeling is a major source
of uncertainty, whose consequences may well influence the nucleosynthetic predictions, as we have
shown. A reliable numerical description of the mechanism that drives turbulent angular momentum
transport and ultimately gives rise to the (quasi-) viscous ejecta encountered for post-merger
BH-tori is beyond the scope of the present study. The $\alpha$-viscosity treatment employed here is
a computationally pragmatic approach that efficiently circumvents the task of numerically modeling
turbulent, magnetized disks. The evolution of magnetized accretion disks is an active field of
research on its own with several fundamental issues being still vividly debated, e.g., numerical
convergence \citep[e.g.][]{Fromang2007,Hawley2013,Sorathia2012}, impact of the global magnetic-field
topology \citep[e.g.][]{Beckwith2008}, or the relevance of additional instabilities besides the
magnetorotational instability (MRI) \citep[e.g.][]{Narayan2012,Yuan2012}.  Nevertheless, although
all of the aforementioned issues may have implications for the disk-outflow masses and the evolution
of a given configuration, it seems likely that the principal result of our study is robust, namely
that a significant fraction of order 10\% or more of the accretion torus become unbound with
sufficiently neutron-rich conditions to allow for the formation of r-process elements with $90\la A
\la 140$.

Our work explicitly excludes long-lived hypermassive or supermassive NSs as merger remnants, whose
disk mass loss was recently studied by \citet{Metzger2014} and \citet{Perego2014a}. Both groups
replaced the core of the massive remnant by an inner boundary condition.  Such stable objects might
be common relics of approximately symmetric NS-NS mergers with typical NS masses of
$\sim$1.35--1.45\,$M_\odot$ for EOSs that are consistent with the observed $\sim$2\,$M_\odot$ lower
limit of the maximum mass of nonrotating, cold NSs \citep{Demorest2010, Antoniadis2013}. Fully
self-consistent simulations will not only have to involve highly differential rotation and
convection and the associated amplification of magnetic fields to extreme strengths in a nuclear
environment with high sound and Alfv{\'e}n speeds, but it will also encounter all the complexities
of the high-density nuclear and neutrino physics that complicate long-time evolution calculations of
neutrino-cooling proto-neutron stars in supernovae. This ambitious undertaking is deferred to future
work.

\section*{Acknowledgments}
OJ wishes to thank Martin Obergaulinger, Bernhard M\"uller and Reiner Birkl for helpful
discussions. At Garching, this research was supported by the Max-Planck/Princeton Center for Plasma
Physics (MPPC) and by the Deutsche Forschungsgemeinschaft through the Transregional Collaborative
Research Center SFB/TR 7 ``Gravitational Wave Astronomy'' and the Cluster of Excellence EXC 153
``Origin and Structure of the Universe'' (http://www.universe-cluster.de). AB is a Marie Curie
Intra-European Fellow within the 7th European Community Framework Programme (IEF 331873). SG
acknowledges financial support from FNRS (Belgium). We are also grateful for computational support
by the Center for Computational Astrophysics (C2PAP) at the Leibniz Rechenzentrum (LRZ) and by the
Rechenzentrum Garching (RZG).


\appendix

\section{Neutrino field around a torus}\label{sec:neutr-field-around}

\begin{figure*}
\includegraphics[scale=1.1]{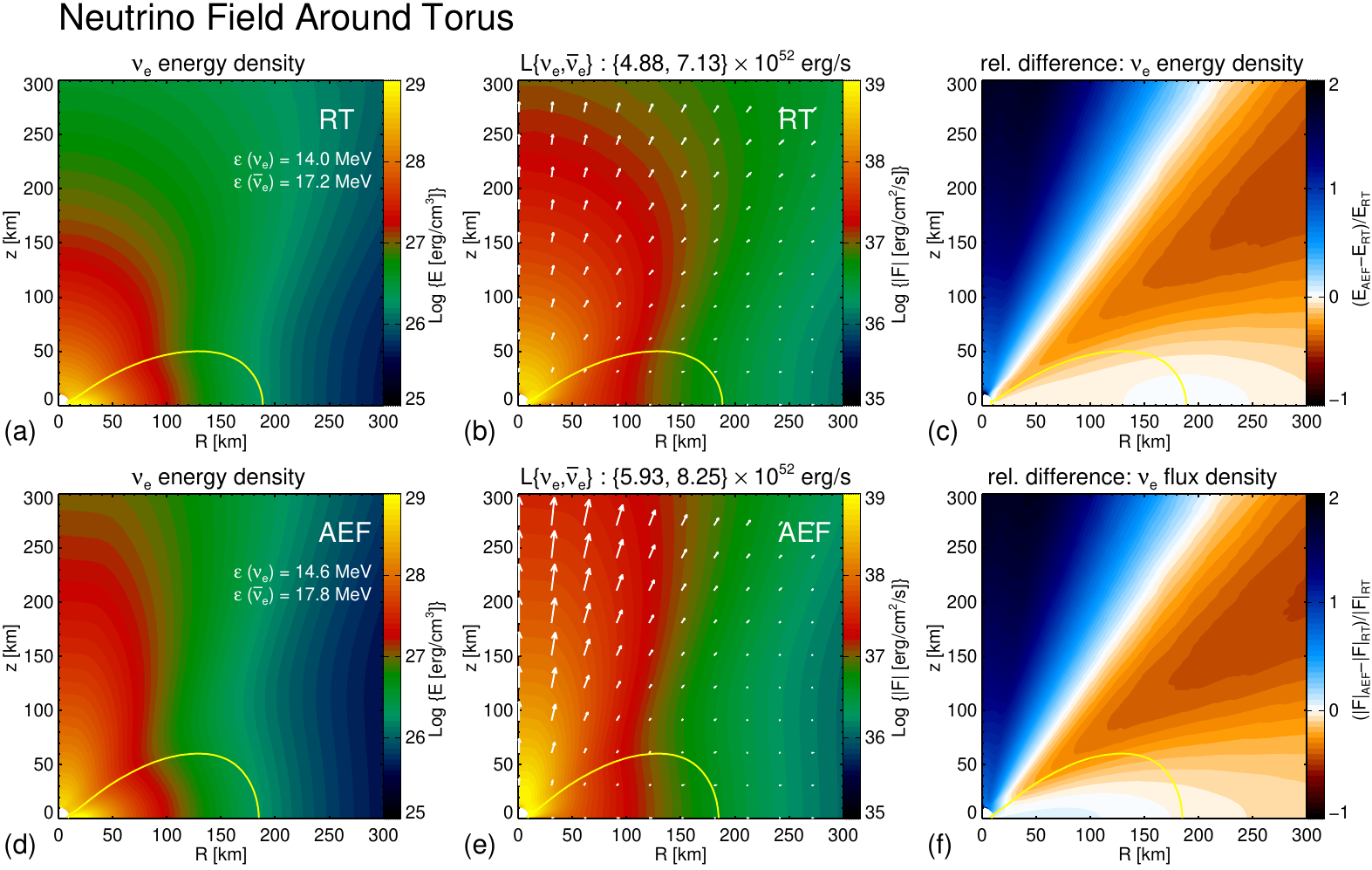}
\begin{minipage}[c]{0.65\textwidth}
\includegraphics[scale=1.1]{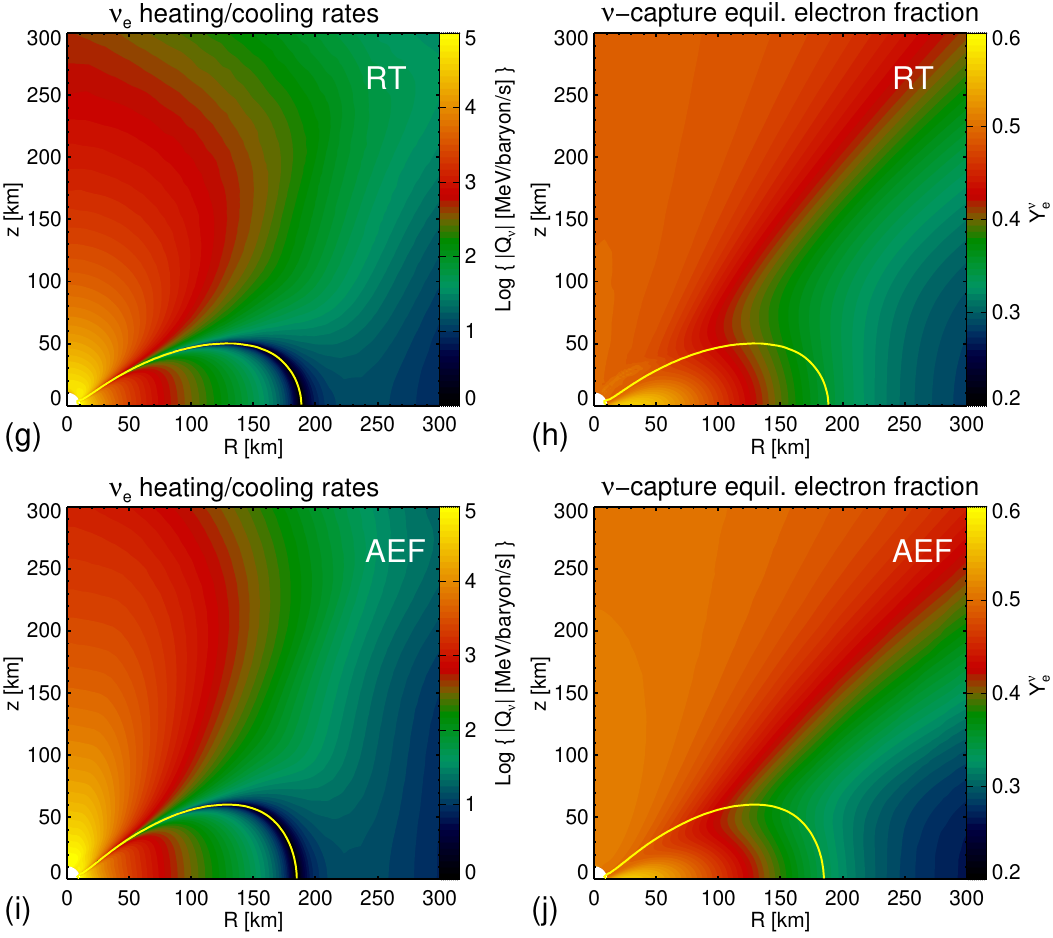}
\end{minipage}\hfill
\begin{minipage}[c]{0.32\textwidth}
  \caption{Comparison of the neutrino radiation field emerging from the configuration of model
    M3A8m3a5 at $t=50\,$ms as computed with the ray-tracing scheme (labeled RT) and with the
    algebraic-Eddington-factor method (labeled AEF). Panels~(a),~(d) show color coded the
    energy-integrated energy densities, $E$, and Panels~(b),~(e) the (absolute values of) flux
    densities, $F$, of electron neutrinos for both schemes, while in Panels~(c),~(f) the respective
    relative differences of the latter quantities between both schemes are depicted. The arrows in
    Panels~(b) and (e) indicate the flux-density vectors multiplied by $4\pi r^2$ scaled such that
    the maximum arrow length corresponds to $\sim 4\times 10^{53}$\,erg\,s$^{-1}$. The mean
    energies, $\varepsilon$, listed in Panels~(a),~(b) are computed as ratios of luminosity to total
    number flux, both calculated as integrals of the corresponding flux densities over a sphere at
    radius $r=300\,$km. The luminosities, $L$, are given on top of Panels~(b)
    and~(e). Panels~(g),~(i) show the net heating and cooling rates due to $\beta$-processes with
    both electron-neutrino species, $Q_\nu$. Finally, Panels~(h),~(j) display the electron fractions
    for local neutrino-capture equilibrium, $Y^{\nu}_e$. The yellow line in each panel demarcates
    the net cooling from the net heating region, i.e. it coincides with $Q_\nu=0$.}
\label{fig:m1test}
\end{minipage}
\end{figure*}

In order to assess the quality of the approximate neutrino-transport scheme used for the BH-torus
simulations in this paper -- denoted as AEF (algebraic-Eddington-factor) scheme hereafter -- we
present here some results of a comparison between the AEF scheme and an accurate radiative transfer
solver. For a detailed description of the AEF scheme, its implementation, and various additional
tests, we refer the reader to a forthcoming paper (Just et al., in preparation).

The hydrodynamic configuration is essentially given by the state of model M3A8m3a5 at evolution time
$t=50\,$ms (cf. Panels~(a) and (b) of Fig.~\ref{fig:torus_contour} for contour plots of some
thermodynamic quantities). However, for this test, we only evolve the neutrino radiation field while
keeping all hydrodynamic quantities fixed. Moreover, to facilitate the comparison with the reference
neutrino scheme (see next paragraph) we now ignore frame-dependent effects in the neutrino scheme,
i.e. all velocities are set to zero, and for the neutrino interactions we only include emission and
absorption by nucleons.

To judge the results of the AEF method, a reference calculation is conducted with a more accurate
(but also computationally far more expensive) neutrino scheme, namely a ray-tracing method (RT
hereafter). In the RT method, the time-independent Boltzmann equation,
\begin{equation}
  \label{eq:transfer1}
  \textbf{\em n}\cdot\nabla f(\textbf{\em x},\textbf{\em n}) = 
  \kappa_a(\textbf{\em x})\left[(f^0(\textbf{\em x}) - f(\textbf{\em x},\textbf{\em n}) \right] \, ,
\end{equation}
(where $\textbf{\em x}$ and $\textbf{\em n}$ are the spatial coordinate vector and the
neutrino-momentum unit vector, respectively) is solved for the distribution function $f(\textbf{\em
  x},\textbf{\em n})$ by direct integration along straight rays directed towards $\textbf{\em n}$,
given the absorption opacity $\kappa_a(\textbf{\em x})$ and the Fermi distribution function
$f^0(\textbf{\em x})$ (see, e.g., \citealp{Birkl2007} and \citealp{Harikae2010} for related
approaches in the context of neutrino transfer). In practice, for each species, energy group and
spatial grid point we integrate Eq.~\eqref{eq:transfer1} along $\sim 60\,000$ ray directions,
corresponding to an angular resolution of $\approx 1^\circ$. We checked that all relevant quantities
are numerically converged to better than $\sim 1\,\%$ with respect to the number of rays and the
step size used for integration of Eq.~\eqref{eq:transfer1}. The spatial and energy grids are
identical in both calculations and directly adopted from the time-dependent simulation
(cf. Sect.~\ref{sec_numerics_remnant}), except that the radial domain is now restricted to radii
$r\la 500\,$km (corresponding to $\sim 250$ radial grid points).

In Fig.~\ref{fig:m1test} we summarize the results of both calculations. A satisfactory outcome is
the fact that the luminosities with $\{L_{\nu_e}, L_{\bar\nu_e}\}_{\mathrm{RT}} = \{4.88,
7.13\}\times 10^{52}\,$erg\,s$^{-1}$ and $\{L_{\nu_e}, L_{\bar\nu_e}\}_{\mathrm{AEF}} = \{5.93,
8.25\} \times 10^{52}\,$erg\,s$^{-1}$, as well as the mean energies with $\{\varepsilon_{\nu_e},
\varepsilon_{\bar\nu_e}\}_{\mathrm{RT}} = \{14.0, 17.2\}$\,MeV and $\{\varepsilon_{\nu_e},
\varepsilon_{\bar\nu_e}\}_{\mathrm{AEF}} = \{ 14.6, 17.8\}$\,MeV (cf. Fig.~\ref{fig:m1test},
Panels~(a),~(b),~(d) and~(e)) agree fairly well, with the AEF quantities all being shifted by at
most $\sim 20\,\%$ towards higher values compared to the corresponding RT results. Also, the local
distributions of the energy-integrated energy density, $E$, and flux density, $F$, (compared in
Panels~(a)--(f) of Fig.~\ref{fig:m1test} for electron neutrinos) agree quite well in the bulk of the
torus and close to the equator.

However, the agreement of the local neutrino distribution is not everywhere perfect. For directions
closer to the polar axis the neutrino density obtained by the AEF method appears to be overestimated
by factors of up to $\sim$2--3 while being underestimated by up to few tens of percent between the
polar region and the equatorial plane. This feature might be the consequence of the constrained
ability of AEF schemes to correctly describe the superposition of oblique radiation fronts (see,
e.g., \citealp{Skinner2013, Scadowski2013} for more comments on this property, which is closely
related to what is sometimes called ``two-beam instability''). From the mathematical point of view,
this shortcoming can be understood from the fact that the moment equations in the AEF formulation
are non-linear, even in the optically thin limit, owing to the generally non-linear closure relation
for the Eddington factor, whereas the original Boltzmann equation is linear. In our case of a
radiating torus, neutrino beams that stem from different locations on the torus surface and that
should cross each other without interaction are instead slightly deflected into the radial
direction, thereby causing an artificial enhancement of the neutrino density around the polar axis
and, in turn, a corresponding shortage of neutrinos propagating along their original directions
pointing away from the axis.

As can be inferred from Panels~(g) and (i) of Fig.~\ref{fig:m1test}, the net neutrino heating and
cooling rates, $Q_\nu$, show good agreement inside the torus and near the equatorial plane but an
enhancement in the polar region compared to the corresponding RT result, very similar to the
behavior of $E$. In Panels~(h) and~(j) of Fig.~\ref{fig:m1test} we show the electron fraction
corresponding to local neutrino-capture equilibrium, $Y_e^\nu$, which is approximately computed as
$Y_e^\nu \approx \left[1+E_{\bar\nu_e}/E_{\nu_e}\times ( \varepsilon_{4,\bar\nu_e} -
  2Q_{\mathrm{np}}) / (\varepsilon_{4,\nu_e} + 2Q_{\mathrm{np}})\right]^{-1}$ where
$Q_{\mathrm{np}}$ is the mass difference between neutrons and protons and
$\varepsilon_{4,\nu_e}\equiv \int \int \epsilon^4_\nu f \,\mathrm{d}\Omega \mathrm{d}\epsilon_\nu /
\int \int \epsilon^3_\nu f \,\mathrm{d}\Omega \mathrm{d}\epsilon_\nu$ with the neutrino energy
$\epsilon_\nu$ and momentum-space solid angle $\Omega$. In contrast to the heating rates, $Y_e^\nu$
only exhibits deviations of up to $\sim 10-20\,\%$ between both calculations, which is due to the
fact that this quantity mostly depends on the ratio of the $\nu_e$ and $\bar\nu_e$ energy densities
and not on their individual values. Owing to the fact that the spatial emission characteristics of
$\bar\nu_e$ are very similar to those of $\nu_e$ in both cases, RT and AEF, the energy ratios match
rather closely between both calculations.

The test results are encouraging concerning the main conclusions drawn in this paper. Assuming the
setup is representative for all models investigated in our main study, the above results would
suggest that the global neutrino-emission properties and therefore the cooling and
(de-)leptonization behavior of the bulk of the torus are probably captured rather
accurately. Consequently, it could then be assumed that the conditions leading into the ADAF phase
and to the development of the viscous outflow are described properly. Therefore, the properties and
nucleosynthesis signature of the dominant outflow component and, hence, the main conclusions drawn
in this paper, should be fairly robust with respect to the neutrino scheme. The deposition of energy
and lepton-number by neutrinos in the surface layers of the torus might be systematically
overestimated due to the artificial overabundance of neutrinos near the axis region. This could
imply that the neutrino-driven wind masses, which are already small compared to the viscous outflow
masses, are even overestimated in our study. In contrast, the nucleosynthesis properties of the
neutrino-driven ejecta are likely to be robust because the electron fractions are driven towards
high asymptotic values of $Y_e^{\nu}\sim 0.4-0.5$ also in the RT case. However, since this test only
examines a single static torus configuration using simplified neutrino interactions, we refrain from
speculations about the quantitative error introduced by the approximations of the AEF scheme to the
time-dependent BH-torus models considered in this paper. The quantitative difference to accurate
neutrino transport in torus evolution models can ultimately only be revealed by comparing to
neutrino-hydrodynamics simulations including a Boltzmann-type transport solver.

\label{lastpage}

\end{document}